\documentclass[a4paper,fleqn,usenatbib]{mnras}

\usepackage{newtxtext,newtxmath}

\usepackage[T1]{fontenc}
\usepackage{ae,aecompl}

\usepackage{graphicx}	
\usepackage{amsmath}	
\usepackage{amssymb}	
\usepackage{todonotes}
\usepackage{bm}
\usepackage{subfig}
\usepackage{courier}
\usepackage[T1]{fontenc}
\usepackage{aecompl}
\usepackage{verbatim}
\usepackage{scrextend}
\usepackage{color}

\title[Transit Probabilities in Evolving Secular Systems]{Transit Probabilities in Secularly Evolving Planetary Systems}

\author[M. J. Read et al.]{
Matthew J. Read$^{1},$\thanks{E-mail: mjr201@ast.cam.ac.uk}
Mark C. Wyatt$^{1}$
and Amaury H. M. J. Triaud$^{1}$
\\
$^{1}$Institute of Astronomy, University of Cambridge, Madingley Road, Cambridge CB3 0HA \\
}

\date{Accepted XXX. Received YYY; in original form ZZZ}

\pubyear{2017}

\begin{document}
\label{firstpage}
\pagerange{\pageref{firstpage}--\pageref{lastpage}}
\maketitle

\begin{abstract}
This paper considers whether the population of known transiting exoplanets provides evidence for additional outer planets on inclined orbits, due to the perturbing effect of such planets on the orbits of inner planets. As such, we develop a semi-analytical method for calculating the probability that two mutually inclined planets are observed to transit. We subsequently derive a simplified analytical form to describe how the mutual inclination between two planets evolves due to secular interactions with a wide orbit inclined planet and use this to determine the mean probability that the two inner planets are observed to transit. From application to Kepler-48 and HD-106315 we constrain the inclinations of the outer planets in these systems (known from RV). We also apply this work to the so called Kepler Dichotomy, which describes the excess of single transiting systems observed by Kepler. We find 3 different ways of explaining this dichotomy: some systems could be inherently single, some multi-planet systems could have inherently large mutual inclinations, while some multi-planet systems could cyclically attain large mutual inclinations through interaction with an inclined outer planet. We show how the different mechanisms can be combined to fit the observed populations of Kepler systems with one and two transiting planets. We also show how the distribution of mutual inclinations of transiting two planet systems constrains the fraction of two planet systems that have perturbing outer planets, since such systems should be preferentially discovered by Kepler when the inner planets are coplanar due to an increased transit probability.  
		 	
\end{abstract}




\section{Introduction}
\label{sec:intro}
Over the past 20 years the number of exoplanet detections has soared most notably due to contributions from the \textit{Kepler space telescope} (Kepler herein). As of November 2016 Kepler has detected 3414 confirmed planets, with 575 existing in multi-planet systems (exoplanet.eu; \cite{2011A&A...532A..79S}). Planet multiplicity provides information on the underlying architecture of planetary systems, such as expected orbital spacing, mutual inclinations and size distributions. For the multi-planet systems observed by Kepler, super Earth/mini Neptune type objects on tightly packed orbits inside of $\sim$200 days are common (\cite{2011ApJS..197....8L}; \cite{2014ApJ...784...44L}; \cite{2016ApJ...822...86M}). Moreover such systems are observed to have small inclination dispersions of $\lesssim$5$^\circ$ (\cite{2011ApJS..197....8L}; \cite{2012ApJ...761...92F}; \cite{2012A&A...541A.139F}; \cite{2012AJ....143...94T}; \cite{2013A&A...551A..90M}; \cite{2014ApJ...790..146F}).       

How representative Kepler multi-planet systems are of a common underlying planetary architecture however is impeded by Kepler preferentially detecting objects which orbit closest to the host star. To generalise Kepler systems to an underlying population, it is therefore necessary to account for the inherent probability that transiting systems are observed. 
Taking into account such probabilities, there appears to be an over-abundance of planetary systems with a single transiting planet (\cite{2011ApJS..197....8L}; \cite{2011ApJ...742...38Y}; \cite{2012ApJ...758...39J}; \cite{2016ApJ...816...66B}). This is commonly referred to as the 'Kepler Dichotomy'. 

It is currently not known what causes this excess. Statistical and \textit{Spitzer} confirmation studies all suggest that the false positive rate for single transiting objects with R$_\mathrm{p}$ $< $4R$_\oplus$ is low at $\lesssim$15\% (\cite{2011ApJ...738..170M}; \cite{2013ApJ...766...81F}; \cite{2014AJ....147..119C}; \cite{2015ApJ...804...59D}). Perhaps then, there are populations of inherently single planet systems in addition to multi-planet systems which are closely packed and have small inclination dispersions. However there may also be a population of multi-planet systems where the mutual inclination dispersion is large, such that only a single planet is observed to transit.

The presence of an outer planetary companion may drive this potential large spread in mutual inclinations. Recent N-body simulations show that the presence of a wide orbit planet in multi-planet systems can decrease the number of inner planets that are observed to transit, either through dynamical instability or inclination excitation (\cite{2016arXiv160908058M}; \cite{2017MNRAS.tmp..186H}). Beyond a few au, planetary transit probabilities drop to negligible values. It is possible therefore that additional wide orbit planets could indeed exist in multi-planet systems observed by Kepler. Giant planets at a few au have been detected around stars in the general stellar population by a number of radial velocity (RV) surveys (\cite{2013A&A...551A..90M}; \cite{2016ApJ...817..104R}; \cite{2016ApJ...819...28W}; \cite{2016ApJ...821...89B}), with suggested occurrence rates ranging from $\sim10-50\%$ (\cite{2008PASP..120..531C}; \cite{2011arXiv1109.2497M}; \cite{2016ApJ...821...89B}). Moreover, indirect evidence of undetected giant planets has also been suggested through apsidal alignment of inner RV detected planets (\cite{2014Sci...346..212D}). As RV studies are largely insensitive to planetary inclinations, it is possible that such wide orbit planets could be on mutually inclined orbits, which may arise from a warp in the disc (\cite{2010A&A...511A..77F}) or due to an excitation by a stellar flyby (\cite{2004AJ....128..869Z}; \cite{2011MNRAS.411..859M}).

Calculating transit probabilities of multi-planet systems is complex, often requiring computationally exhaustive numerical methods such as Monte Carlo techniques (e.g. \cite{2011ApJS..197....8L}; \cite{2012ApJ...758...39J}; \cite{2016MNRAS.455.2980B}; \cite{2016arXiv160908058M}; \cite{2017MNRAS.tmp..186H}). However analytical methods can offer a significantly more efficient route for this calculation and allows for coupling with other fundamental analytical theory, such as for the expected dynamical evolution of the system from inter-planet interactions. Despite this however, analytical investigations into the transit probabilities of multi-planet systems for this purpose are relatively sparse (e.g. \cite{2010arXiv1006.3727R}; \cite{2016ApJ...821...47B}). Recently \cite{2016ApJ...821...47B} showed how differential geometry techniques can be used to calculate multi-planet transit probabilities by mapping transits onto a celestial sphere. In this paper we perform a similar analysis, however we focus on regions where pairs of planets can be observed to transit. We also give an explicit analytical form using simple vector relations to describe the boundaries of such transit regions. 

The multi-planet systems observed by Kepler appear to be mostly stable on long timescales (\cite{2011ApJS..197....8L}; \cite{2015ApJ...807...44P}). Dynamical interactions with a potential outer planet on an inclined orbit would therefore be expected to occur on secular timescales. Recent analytical work by \cite{2017AJ....153...42L} suggests that such interactions can lead to large mutual inclinations in an inner planetary system, assuming that the direction of the angular momentum vector of the outer planet is fixed. We build on this work by deriving analytical relations for the mutual inclination that can be induced in an inner planetary system by a general planetary companion. We then simplify this result specifically for when the companion is on a wide orbit. Combining this result with our robust analytical treatment of transit probabilities, we can then derive a simple relation describing how the presence of an outer planetary companion affects the transit probability of an inner system due to long term interactions. 
 
 We also complement recent N-body simulations of Kepler-like systems interacting with an inclined outer planetary companion shown in \cite{2016arXiv160908058M} and \cite{2017MNRAS.tmp..186H} by using our robust treatment of transit probabilities to consider whether an outer planet with a range of masses, semi-major axes and inclinations can reduce an underlying population of Kepler double transiting systems enough to recover the observed number of single transiting systems through long term interactions only. We also investigate whether the presence of specific wide orbit planets in multi-planet systems preferentially predicts single transiting planets with a given distribution of radii and semi-major axes.

 In \S\ref{sec:semianal} we overview our semi-analytical method for calculating the transit probability of two mutually inclined planets. In \S\ref{sec:secint} we derive a simplified form to describe the evolution of the mutual inclination between two planets due to presence of an outer planetary companion. We show how this mutual inclination affects the transit probability of the two inner planets in \S\ref{sec:combtrans}. In \S\ref{sec:realsys} we apply this work to Kepler-56, Kepler-68, HD 106315 and Kepler-48 to place constraints on the inclination of the outer planets in these systems. In \S\ref{sec:Kepdich} we investigate whether a wide orbit planet in Kepler systems can decrease the number of observed two planet transiting systems enough to recover the observed abundances of single transiting systems. We finally discuss this work in \S\ref{sec:discussion} and conclude in \S\ref{sec:conc}.

\section{Semi-analytical Transit Probability}
\label{sec:semianal}
A planet on a circular orbit with a semi-major axis $a$ and radius $R_\mathrm{p}$ subtends a band of shadow across the celestial sphere due its orbital motion. We refer to this band of shadow as the \textit{transit region} (\cite{2010arXiv1006.3727R}; \cite{2016ApJ...821...47B}). The probability that an observer will view an individual transit event of this planet, assuming that the system is viewed for long enough, is equal to the number of viewing vectors that intersect the transit region, divided by the total number of possible viewing vectors. Perhaps more intuitively, this is equivalent to the surface area of the transit region divided by the total surface area of the celestial sphere.

To calculate the area of a transit region on the celestial sphere first consider that the area of a given surface element ($S$) on a unit sphere is equal to 
\begin{equation}
S = \int^{\theta_0}_0\int^{\phi_0}_0 \sin\theta' d\theta' d\phi' = \left[1 - \cos\theta'\right]^{\theta_0}_0\left[\phi'\right]^{\phi_0}_0,
\label{eq:surfel}
\end{equation}
where $\theta'$ is the polar angle and $\phi'$ is the azimuthal angle. A given area on the celestial sphere can therefore be represented on a 2d plane of $1 - \cos\theta'$ vs. $\phi'$, from 0 $\rightarrow$ 2 and 0 $\rightarrow 2\pi$ respectively, such that the 2d plane has a total surface area of $4\pi$. Below we show how the boundaries of a given transit region traverses this 2d plane. This allows for the area contained within these boundaries and therefore the associated transit probability to be calculated.

\subsection{Single Planet Case}
 \begin{figure}
 	\includegraphics[trim={0cm 0cm 0cm 0cm}, width = \linewidth]{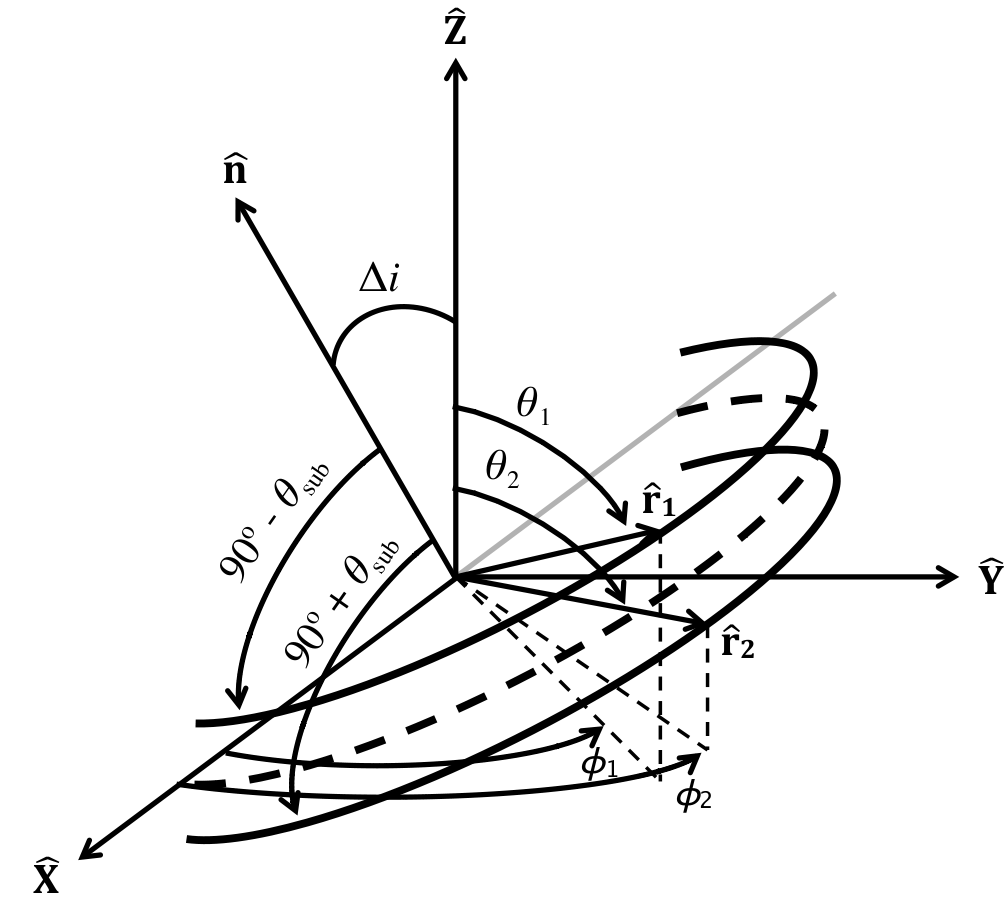}
 	\caption{The coordinate system used to show how a transit region traverses the surface of a celestial sphere. The dashed line represents an orbital plane inclined to a fixed reference plane by $\Delta i$. The direction $\hat{\mathrm{\mathbf{n}}}$ is normal to the orbital plane. The directions $\hat{\mathrm{\mathbf{r}}}$, $\hat{\mathrm{\mathbf{r}}}_1$ and $\hat{\mathrm{\mathbf{r}}}_2$ trace the central, lower and upper boundaries of a transit region respectively.}
 	\label{fig:dobcoord}
 \end{figure}
  
  \begin{figure}
  	\includegraphics[trim={0cm 1cm 0cm 0cm}, width = \linewidth]{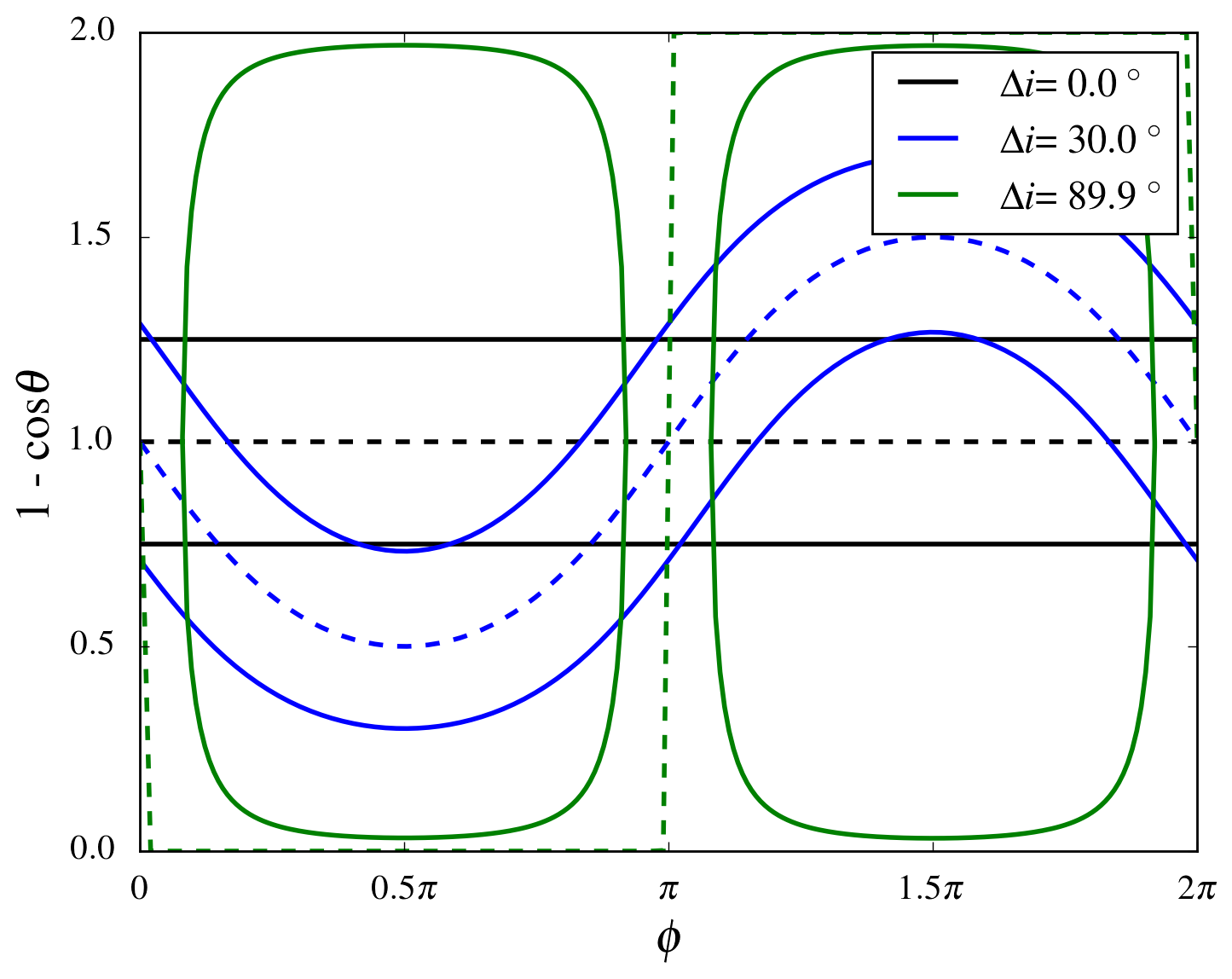}
  	\caption{The surface of a celestial sphere represented on a 2d plane. The dotted lines represent the centre of a transit region for a planet inclined to a fixed reference plane by $\Delta i$. The solid lines refer to the boundaries of such transit regions for when $R_\star/a = 0.25$. The area within these transit regions are identical, giving an identical single transit probability equal to 0.25.}
  	\label{fig:outline}
  \end{figure}
  
 Consider some fixed reference plane where [$\hat{\mathrm{\mathbf{X}}}, \hat{\mathrm{\mathbf{Y}}}$] define a pair of orthogonal directions in this plane, and $\hat{\mathrm{\mathbf{Z}}}$ defines a direction orthogonal to this plane as shown in Figure \ref{fig:dobcoord}. The fixed reference frame in Figure \ref{fig:dobcoord} is assumed to be centred on a host star with radius $R_\star$. The line of sight of an observer is considered to be randomly oriented over the surface of a celestial sphere with respect to this fixed reference plane. Now consider that the orbital plane of a planet with a semi-major axis $a$ and radius $R_\mathrm{p}$, is inclined to the fixed reference plane by $\Delta i$, with the intersection between the two planes occurring along the $\hat{\mathrm{\mathbf{X}}}$ direction. The direction of the normal of the orbital plane is given by $\hat{\mathrm{\mathbf{n}}}$. The position of a planet in the orbital plane is defined by the direction $\hat{\mathrm{\mathbf{r}}}$ which makes the angles $\theta$ and $\phi$ with the $\hat{\mathrm{\mathbf{Z}}}$ and $\hat{\mathrm{\mathbf{X}}}$ directions respectively. Hence $\hat{\mathrm{\mathbf{r}}}$ traces the centre of the transit region with respect to the fixed reference plane. As $\hat{\mathrm{\mathbf{n}}} \cdot \hat{\mathrm{\mathbf{r}}} = 0$, where $\hat{\mathrm{\mathbf{n}}} = [0, \sin\Delta i, \cos\Delta i]$ and $\hat{\mathrm{\mathbf{r}}} = [\sin\theta\cos\phi, \sin\theta\sin\phi, \cos\theta]$ it follows that 
 \begin{equation}
 - \sin\Delta i\sin\theta\sin\phi + \cos\Delta i\cos\theta = 0.
 \label{eq:cent}
 \end{equation}

Hence eq. (\ref{eq:cent}) defines how the centre of a transit region inclined to a fixed reference plane by $\Delta i$ traverses a celestial sphere. This is shown by the dashed lines in Figure \ref{fig:outline} for different values of $\Delta i$, where the surface area of the celestial sphere is shown on a 2d plane defined by eq. (\ref{eq:surfel}). We note that at the special case where $\Delta i$ = 90$^\circ$, $\phi$ can only take values of 0 or $\pi$. 

Similarly the directions that define the boundaries of the transit region can be given by $\hat{\mathrm{\mathbf{r}}}_1$ and $\hat{\mathrm{\mathbf{r}}}_2$ which makes the angles $\theta_1, \theta_2$ and $\phi_1, \phi_2$ with the $\hat{\mathrm{\mathbf{Z}}}$ and $\hat{\mathrm{\mathbf{X}}}$ directions respectively, shown in Figure \ref{fig:dobcoord}. The boundaries of the transit region also subtend an angle $\pm \theta_\mathrm{sub}$ from the orbital plane where $\sin\theta_\mathrm{sub} = R_\star/a$ assuming $R_\star \gg R_\mathrm{p}$ (\cite{1984Icar...58..121B}). As $\hat{\mathrm{\mathbf{r}}}_1 = [\sin\theta_1\cos\phi_1, \sin\theta_1\sin\phi_1, \cos\theta_1]$, $\hat{\mathrm{\mathbf{r}}}_2 = [\sin\theta_2\cos\phi_2, \sin\theta_2\sin\phi_2, \cos\theta_2]$ and $\hat{\mathrm{\mathbf{n}}} \cdot \hat{\mathrm{\mathbf{r}}}_1 = R_\star/a$ and $\hat{\mathrm{\mathbf{n}}} \cdot \hat{\mathrm{\mathbf{r}}}_2 = -R_\star/a$, it follows that 
 \begin{equation}
 -\sin\Delta i\sin\theta_{1}\sin\phi_1 + \cos\Delta i\cos\theta_{1} = R_\star/a,
 \label{eq:bound1}
 \end{equation}
 \vspace{-0.5cm}
 \begin{equation}
 -\sin\Delta i\sin\theta_{2}\sin\phi_2 + \cos\Delta i\cos\theta_{2} = -R_\star/a.
 \label{eq:bound2}
 \end{equation}
Hence eq. (\ref{eq:bound1}) and eq. (\ref{eq:bound2}) describe how the lower and upper boundaries of the transit region for a planet inclined to a fixed reference plane by $\Delta i$ traverse a celestial sphere.  The solid lines in Figure \ref{fig:outline} show these boundaries for different values of $\Delta i$, where $R_\star/a$ = 0.25. This value of $R_\star/a$ might be considered to be unrealistically large and is used for demonstration purposes only. In Appendix \ref{sec:transeq} we further discuss how the values of ($\theta_1, \phi_1$) and ($\theta_2, \phi_2$) in eq. (\ref{eq:bound1}) and eq. (\ref{eq:bound2}) respectively would be expected to change as $\Delta i$ is increased from $\Delta i = 0\rightarrow90^\circ$.
 
 An integration between the upper and lower boundaries of a transit region divided by the total surface area of the celestial sphere gives the associated \textit{single transit probability} of the planet ($R_\star/a$, \cite{1984Icar...58..121B}). All of the transit regions shown in Figure \ref{fig:outline} for different $\Delta i$ therefore contain identical areas and hence have identical single transit probabilities equal to 0.25. We note that if the planet has a non-negligible radius then the single transit probability becomes $(R \pm R_\mathrm{p})/a$ for grazing and full transits respectively. Throughout this work however we assume that $R_\mathrm{p} \ll R_\star$.

\subsection{Two Planet Case}
\label{subsec:twoplanetcase2}
Consider now a system containing two planets, both of which are on circular orbits with semi-major axes and radii of $a_1$, $a_2$ and $R_\mathrm{{p_1}}, R_\mathrm{{p_2}}$ respectively, where $a_1 < a_2$ and the orbital planes are mutually inclined by $\Delta i$ (we give an exact definition for mutual inclination in \S\ref{sec:secint}). The probability that a randomly oriented observer will view both planets to transit (assuming the system is observed for long enough) is equal to the overlap area between the transit regions of both planets, divided by the total area of the celestial sphere. We refer to this probability as the \textit{double transit probability}.

\begin{figure}
 	\includegraphics[trim={0cm 0cm 0cm 0cm}, width = \linewidth]{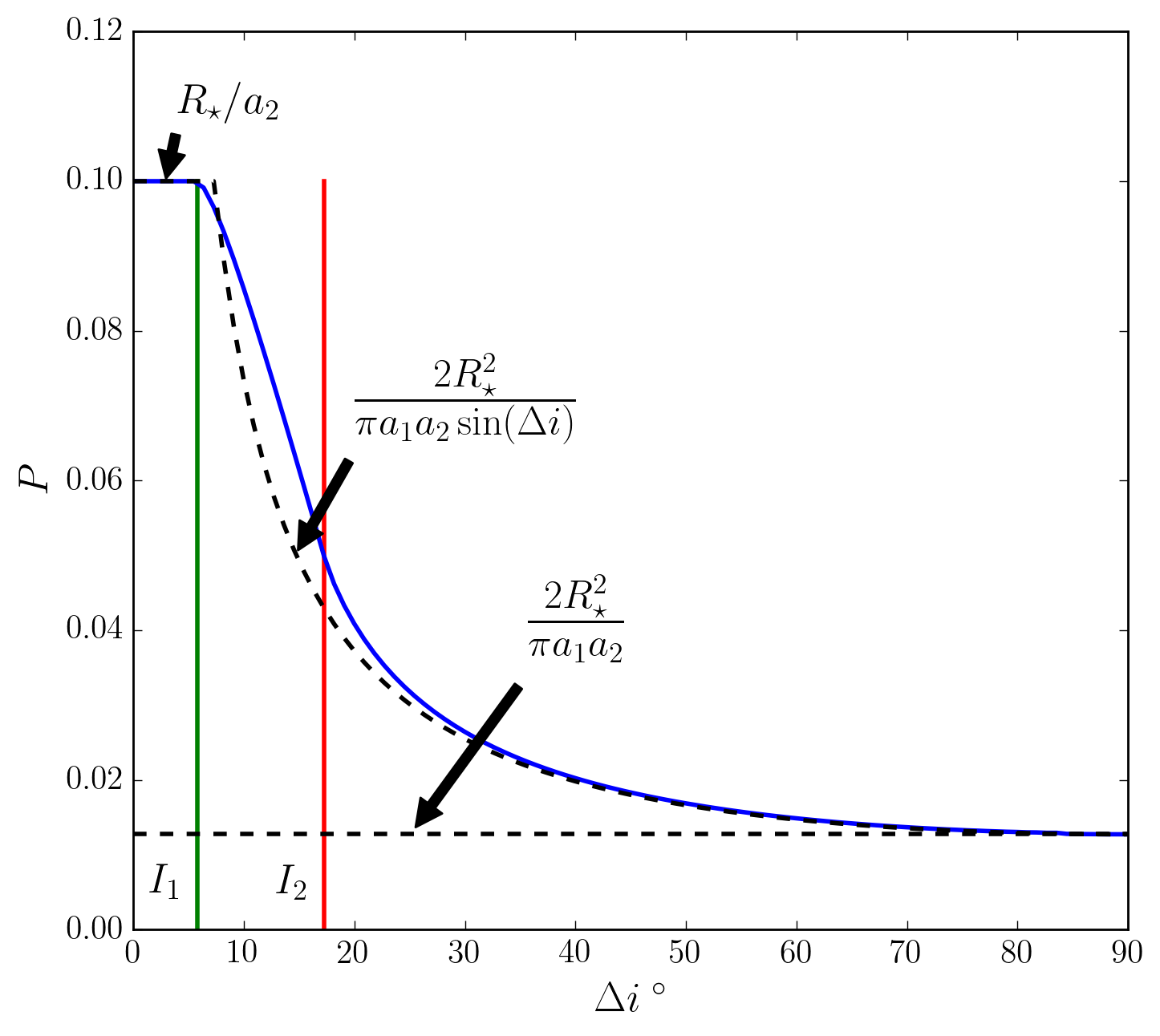}
 	\caption{The double transit probability as a function of mutual inclination between two planets from our method (blue line) for when $R_\star/a_1 = 0.2$ and $R_\star/a_2 = 0.1$. The dashed black lines represent the associated analytical estimate given by eq. (\ref{eq:doubleprob}). The green and red lines represent which inclination cause the double transit probability to go from regime 1 to 2 and regime 2 to 3, with the regimes being defined in \S\ref{subsec:twoplanetcase2}.}
 	\label{fig:pi}
 \end{figure}

Therefore, using eq. (\ref{eq:bound1}) and eq. (\ref{eq:bound2}) to find where the boundaries of the transit regions of each planet intersect, an outline of the overlap between the transit regions can be determined. The area of this overlap can subsequently be calculated by an appropriate integration, which when divided by 4$\pi$ gives the double transit probability. How the double transit probability changes as a function of $\Delta i$ is shown by the blue line in Figure \ref{fig:pi}, for when $R_\star/a_1 = 0.2$ and $R_\star/a_2 = 0.1$. We note that this result is consistent regardless of the choice of reference plane and the orientation of the orbital planes of both planets with respect to this reference plane (see \cite{2010arXiv1006.3727R} for a further discussion). That is, the double transit probability depends on the mutual inclination between the two planets only (in addition to the physical size of the respective transit regions). 

Depending on the value of $\Delta i$, the double transit probability ($P$ herein) can be split into three regimes (also discussed in \cite{2010arXiv1006.3727R}; \cite{2016ApJ...821...47B}).

(1) For low values of $\Delta i$, the transit region of the outer planet is enclosed within that of the inner planet. The double transit probability is therefore equal to $R_\star/a_2$.

(2) $\Delta i$ is large enough that the transit region of one planet is no longer fully enclosed inside the other, however there is still partial overlap for all azimuthal angles on the celestial sphere. The transition to this regime occurs for a value of $\Delta i = I_1$, which causes $\theta_1$ in eq. (\ref{eq:bound1}) for both planets to be equal at $\phi_1 = \pi/2$. Evaluating eq. (\ref{eq:bound1}) at this point gives 
\begin{equation}
\sin I_1 = -\kappa_2(1 - \kappa_1^2)^{1/2} + \kappa_1(1 - \kappa_2^2)^{1/2},
\label{eq:I1}
\end{equation}
where $\kappa_1 = R_\star/a_1$ and $\kappa_2 = R_\star/a_2$ for simplicity. We note that determining the overlap area of the two transit regions with an exact analytical expression in this regime is difficult and is commonly calculated by Monte Carlo techniques (e.g. \cite{2010arXiv1006.3727R}; \cite{2012ApJ...758...39J}; \cite{2016MNRAS.455.2980B}; \cite{2016arXiv160908058M}; \cite{2017MNRAS.tmp..186H}).
 
 (3) For large $\Delta i$, the transit regions only overlap at the intersection of the two orbital planes. The transition to this regime occurs when $\Delta i = I_2$, where $\theta_1$ for the inner planet is equal to $\theta_2$ for the outer planet at $\phi_1 = \phi_2 = \pi/2$. Evaluating eq. (\ref{eq:bound1}) and (\ref{eq:bound2}) here gives
\begin{equation}
\sin I_2 = \kappa_2(1 - \kappa_1^2)^{1/2} + \kappa_1(1 - \kappa_2^2)^{1/2}.
\label{eq:I2}
\end{equation}
The values of $I_1$ and $I_2$ are shown by the green and red lines respectively in Figure \ref{fig:pi}. If it is assumed that the transit region overlap in regime 3 can be represented as a 2d parallelogram, \cite{2010arXiv1006.3727R} showed the double transit probability can be approximated by\footnote{\label{note1}For greater accuracy, we include a 2/$\pi$ factor here that is not included in \cite{2010arXiv1006.3727R}.}

 \begin{equation}
 P = \frac{2R_\star^2}{\pi a_1a_2\sin\Delta i}.
 \label{eq:doubleprob}
 \end{equation}

 For large $\Delta i$ therefore, the double transit probability predicted by eq. (\ref{eq:doubleprob}) tends to a value of 2$R_\star^2/\pi a_1a_2$. We show eq. (\ref{eq:doubleprob}) as the black dashed line in Figure \ref{fig:pi}. We note that in \cite{2010arXiv1006.3727R} it was assumed that the double transit probability transitions straight from regime (1) to (3) at $\Delta i = \arcsin\left(\frac{2}{\pi}\cdot\mathrm{min}(R_\star/a_1,R_\star/a_2)\right)$\footref{note1}. 
 
 For $\Delta i > I_2$ our method predicts a double transit probability that agrees well with the analytical estimate from \cite{2010arXiv1006.3727R}. However there is a clear discrepancy for $I_1 < \Delta i < I_2$, for when there is partial overlap between the transit regions at all azimuthal angles. This highlights the need for semi-analytical methods like the one suggested here over purely analytical relations, to robustly calculate double transit probabilities at all values of $\Delta i$. We note that our method also agrees well with the Monte Carlo treatment of double transit probabilities shown in \cite{2010arXiv1006.3727R}.
 
Calculating transit probabilities using the method outlined here is significantly more computationally efficient than equivalent Monte Carlo methods, as it is only necessary to solve combinations of eq. (\ref{eq:bound1}) and (\ref{eq:bound2}) for different planets to find where transit regions overlap. From integrating around this overlap, the associated double transit probability is also exact and not subject to Monte Carlo noise effects from under-sampling the total number of line of sight vectors.

\section{Secular Interactions}
\label{sec:secint}
\subsection{N planet system}
\label{subsec:nplanetsys}
Consider a system of $N$ secularly interacting planets in which planet $j$ has a semi-major axis $a_j$ and mass $m_j$. The inclination and longitude of ascending node of planet $j$ are given by $I_j$ and $\Omega_j$ respectively, and can be combined into the associated complex inclination $y_j = I_je^{i\Omega_j}$. Assuming that the vector involving all the planet's orbital planes is given by $\bm{y}$ = [$y_1, y_2, ..y_N$], the evolution of complex inclinations in the low inclination and eccentricity limit can be given by Laplace - Lagrange theory in the form 
\begin{equation}
\dot{\bm{y}} = i\mathbf{{B}}\bm{y},
\label{eq:Laplace}
\end{equation}
(\cite{1999ssd..book.....M}) where $\mathbf{B}$ is a matrix with elements given by 
\begin{equation}
\begin{split}
& B_{jk} = \frac{1}{4}n_j\left(\frac{m_k}{M_\star + m_j}\right)\alpha_{jk}\tilde{\alpha}_{jk}b^{(1)}_{3/2}(\alpha_{jk}) \hspace{1cm}(j\neq k) \\
& B_{jj} = -\sum^{N}_{k=1, j\neq k}B_{jk},
\end{split}
\label{eq:matrix}
\end{equation}
$j$ and $k$ are integers associated with each planet, $M_\star$ and $m_i$ are the masses of the star and planet $i$, $n_j$ is the mean motion of planet $j$ where $n_j^2a_j^3 = G(M_\star + m_j)$, $\alpha_{jk} = \tilde{\alpha}_{jk}$ = $a_j/a_i$ for $a_j < a_k$ and $\alpha_{jk} = a_k/a_j$ and $\tilde{\alpha}_{jk} = 1$ otherwise, and $b^{(1)}_{3/2}(\alpha_{jk})$ corresponds to a Laplace coefficient given by
\begin{equation}
b^{(\nu)}_s(\alpha) = \frac{1}{\pi}\int_{0}^{2\pi}\frac{\cos(\nu x)dx}{(1 - 2\alpha\cos(x) + \alpha^2)^s} \hspace{1cm} \alpha < 1.
\end{equation}
Eq. (\ref{eq:Laplace}) can be solved to show that the evolution of $\bm{y}$ is given by a superposition of eigenmodes associated with each eigenfrequency $f_i$ of the matrix $\mathbf{B}$ 
\begin{equation}
y_j(t) = \sum^{N}_{k=1}\mathbf{\mathit{I}}_{jk}e^{i(f_kt + \gamma_k)},
\label{eq:secsol}
\end{equation}
where {\textit{{I}$_{jk}$}} are the eigenvectors of $\mathbf{B}$ scaled to initial boundary conditions and $\gamma_k$ is an initial phase term. If it is assumed that all objects are spherically symmetric, additional terms in the diagonal elements of $\mathbf{B}$ in eq. (\ref{eq:matrix}) (e.g. stellar oblateness) need not be included. A choice of reference frame for the inclination also becomes arbitrary, leading to one of the eigenfrequencies equalling zero (c.f. \cite{1999ssd..book.....M}). It is only meaningful therefore to describe a \textit{mutual inclination} between pairs of planets, with the invariable plane commonly being chosen as a reference plane. The invariable plane is defined as being perpendicular to the total angular momentum vector of a system. The mutual inclination is then the angle between individual angular momentum vectors of a pair of planets. The inclination solution described by eq. (\ref{eq:secsol}) also becomes simplified when the invariable plane is taken as a reference plane as the eigenvector associated with the zero value eigenfrequency is also equal to zero.

 \subsection{Two planet system with an inclined companion}
 \label{subsec:twoplansec}
      \begin{figure*}
      	\includegraphics[trim={0cm 0cm 0cm 0cm}, width = 0.45\linewidth]{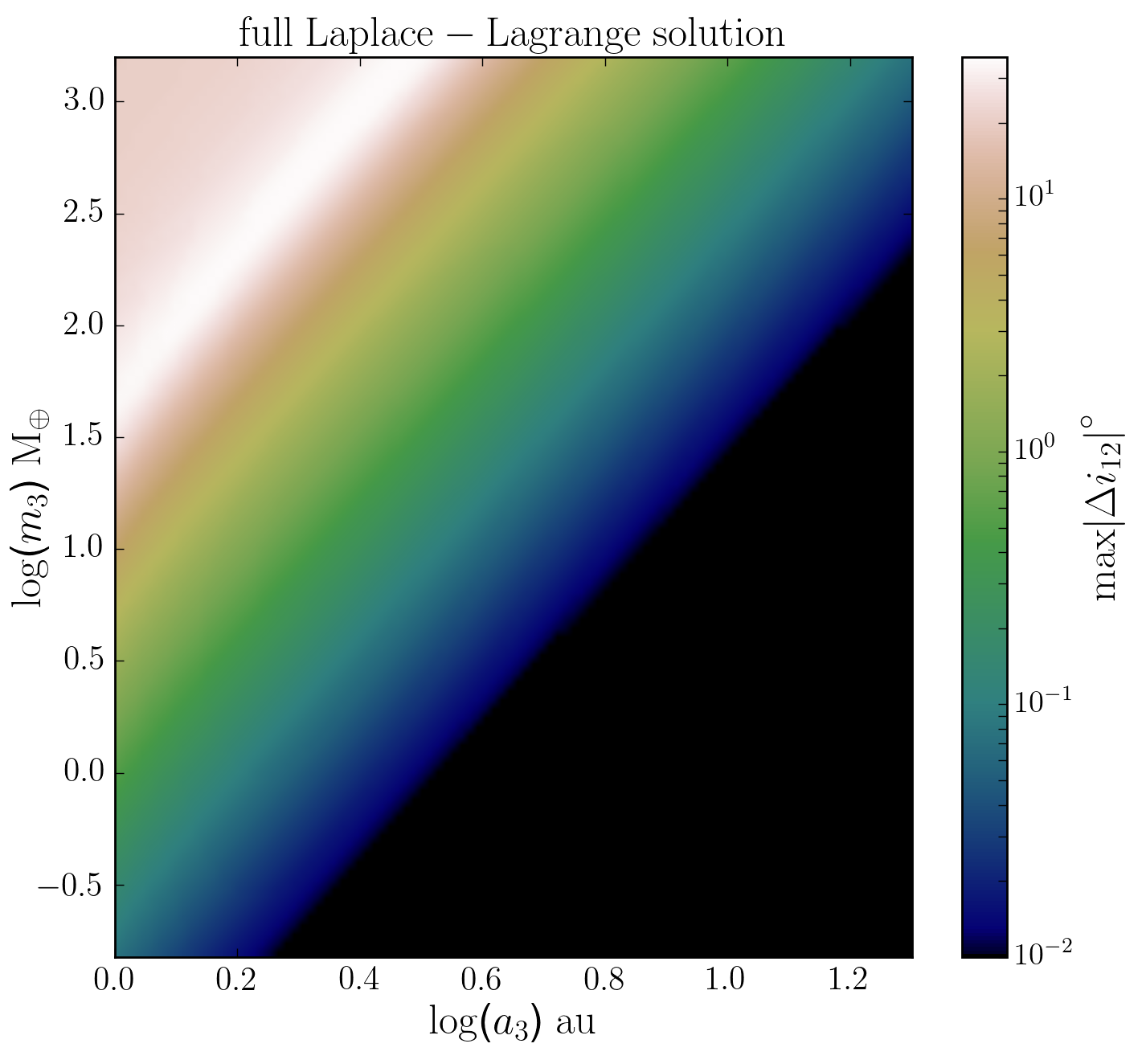}
       \includegraphics[trim={0cm 0cm 0cm 0cm}, width = 0.45\linewidth]{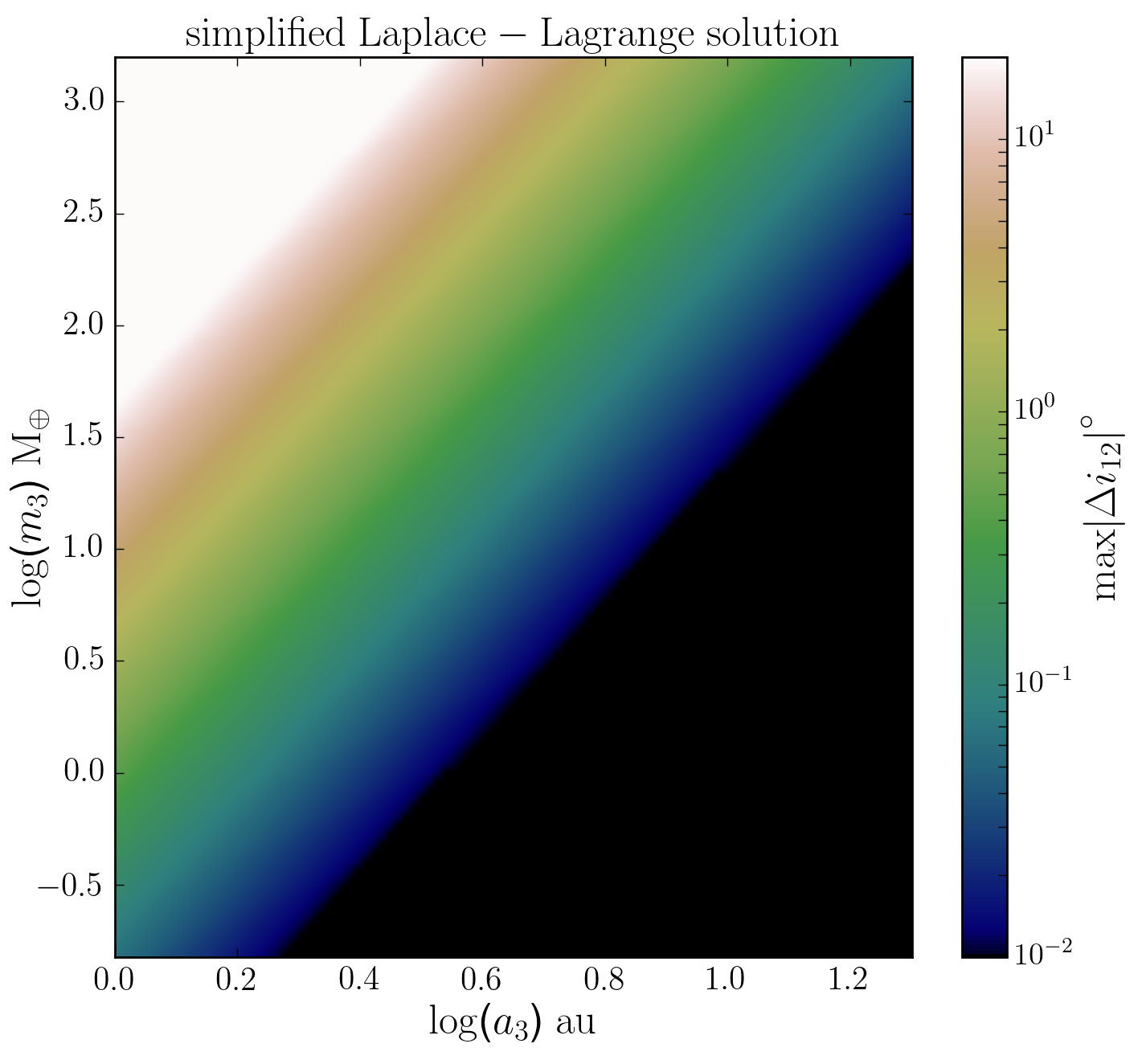}
      	\includegraphics[trim={0cm 0cm 0cm 0cm}, width = 0.45\linewidth]{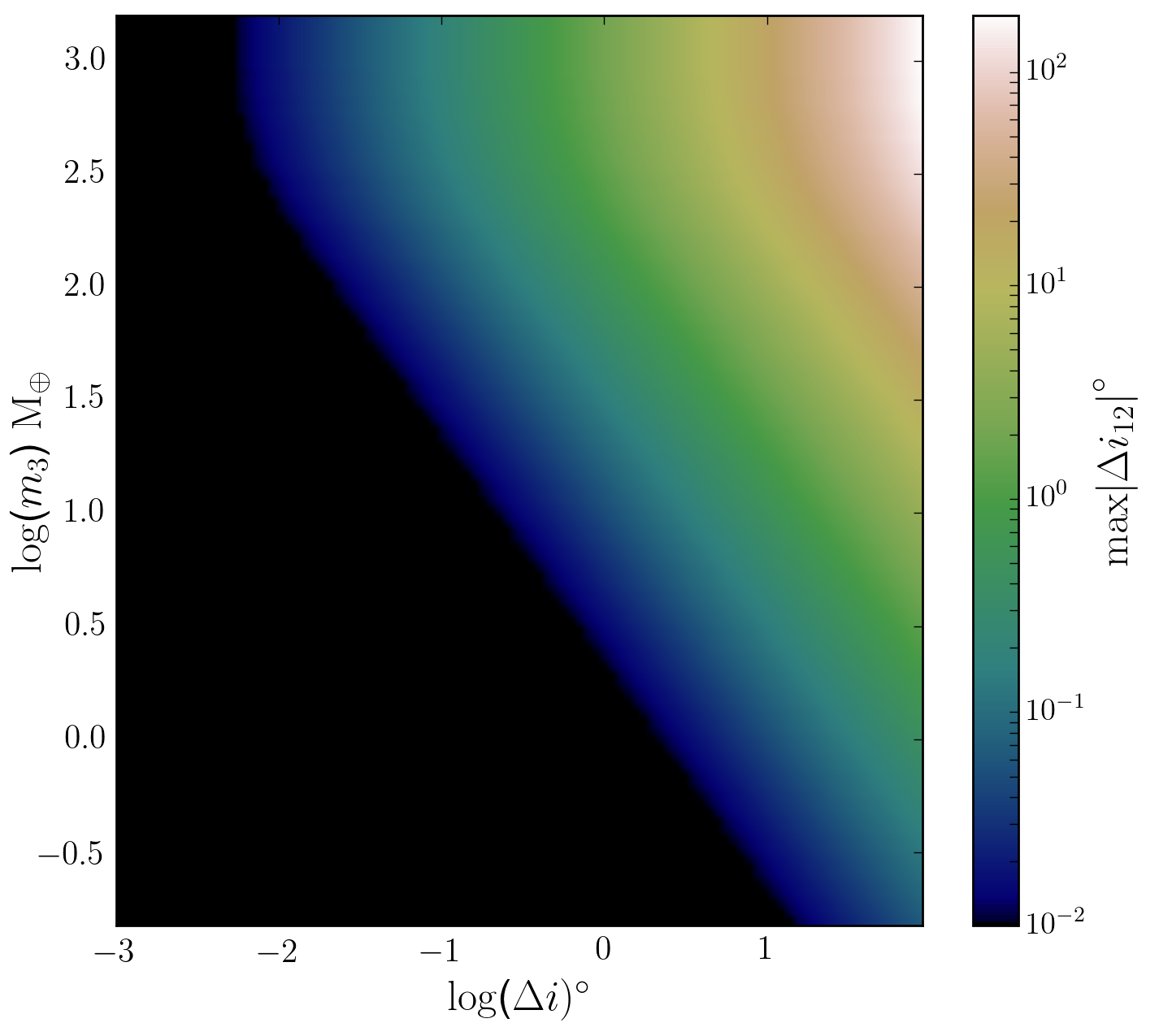}
	     \includegraphics[trim={0cm 0cm 0cm 0cm}, width = 0.45\linewidth]{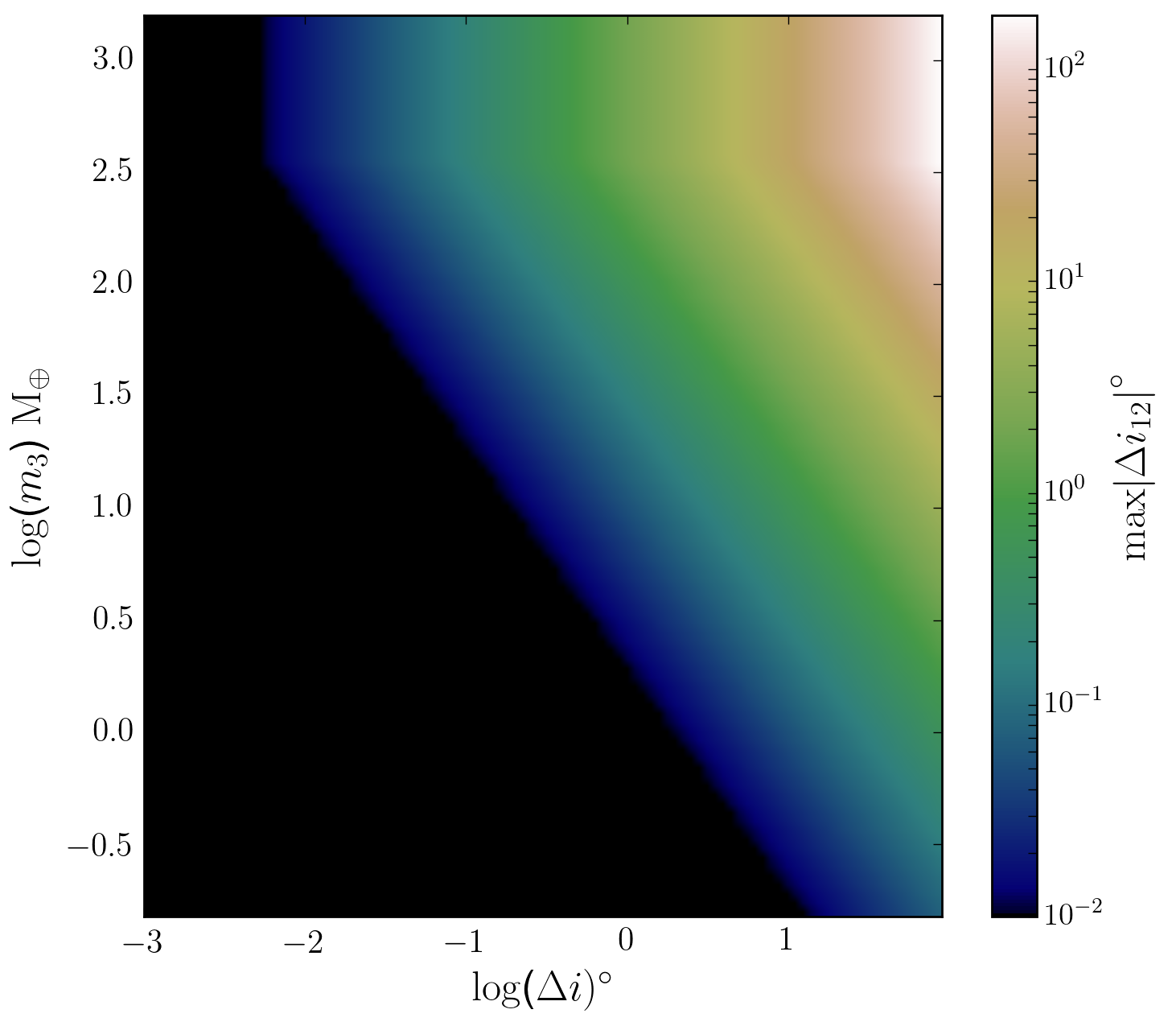}
	    \includegraphics[trim={0cm 0cm 0cm 0cm}, width = 0.45\linewidth]{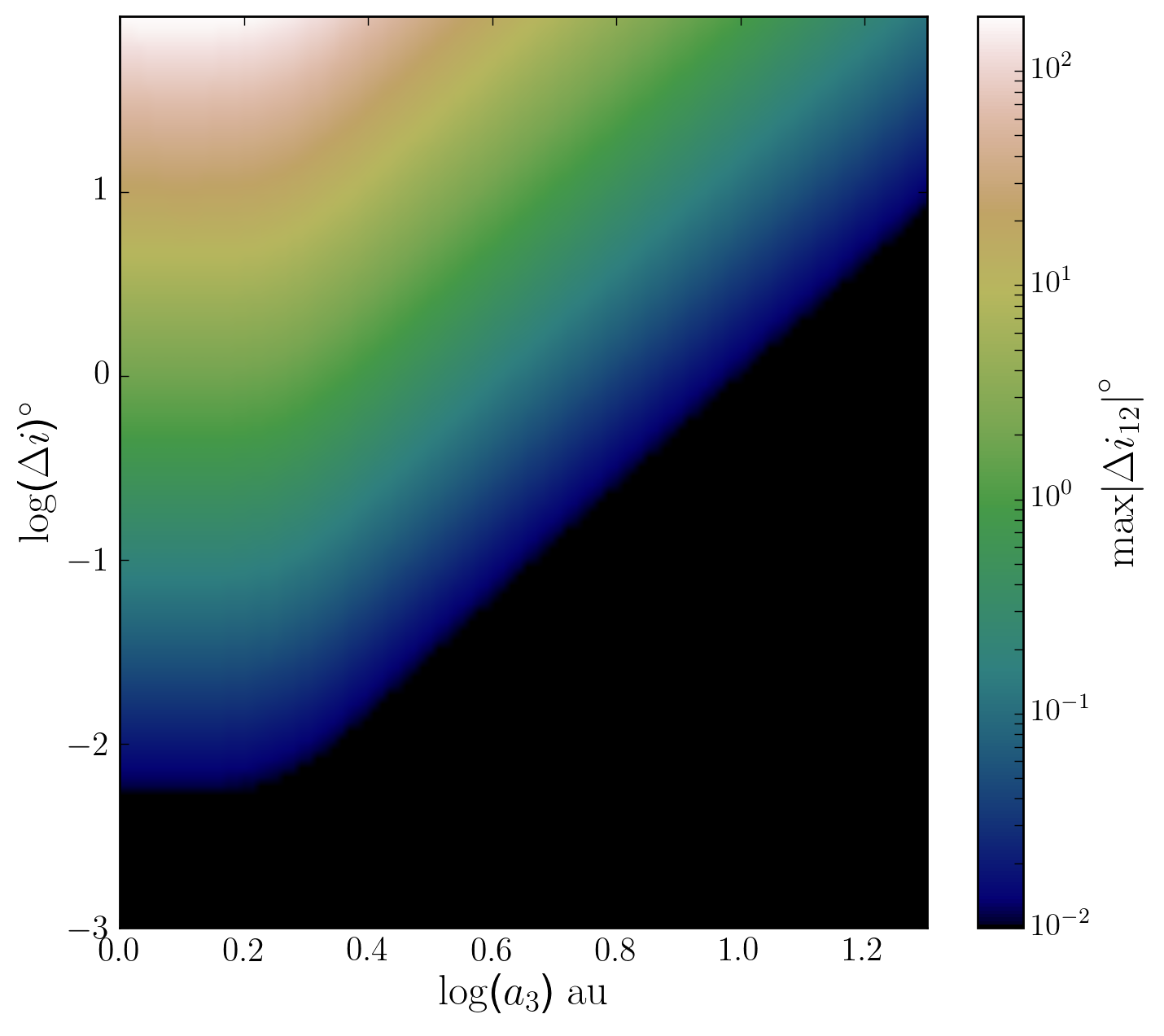}
	    \includegraphics[trim={0cm 0cm 0cm 0cm}, width = 0.45\linewidth]{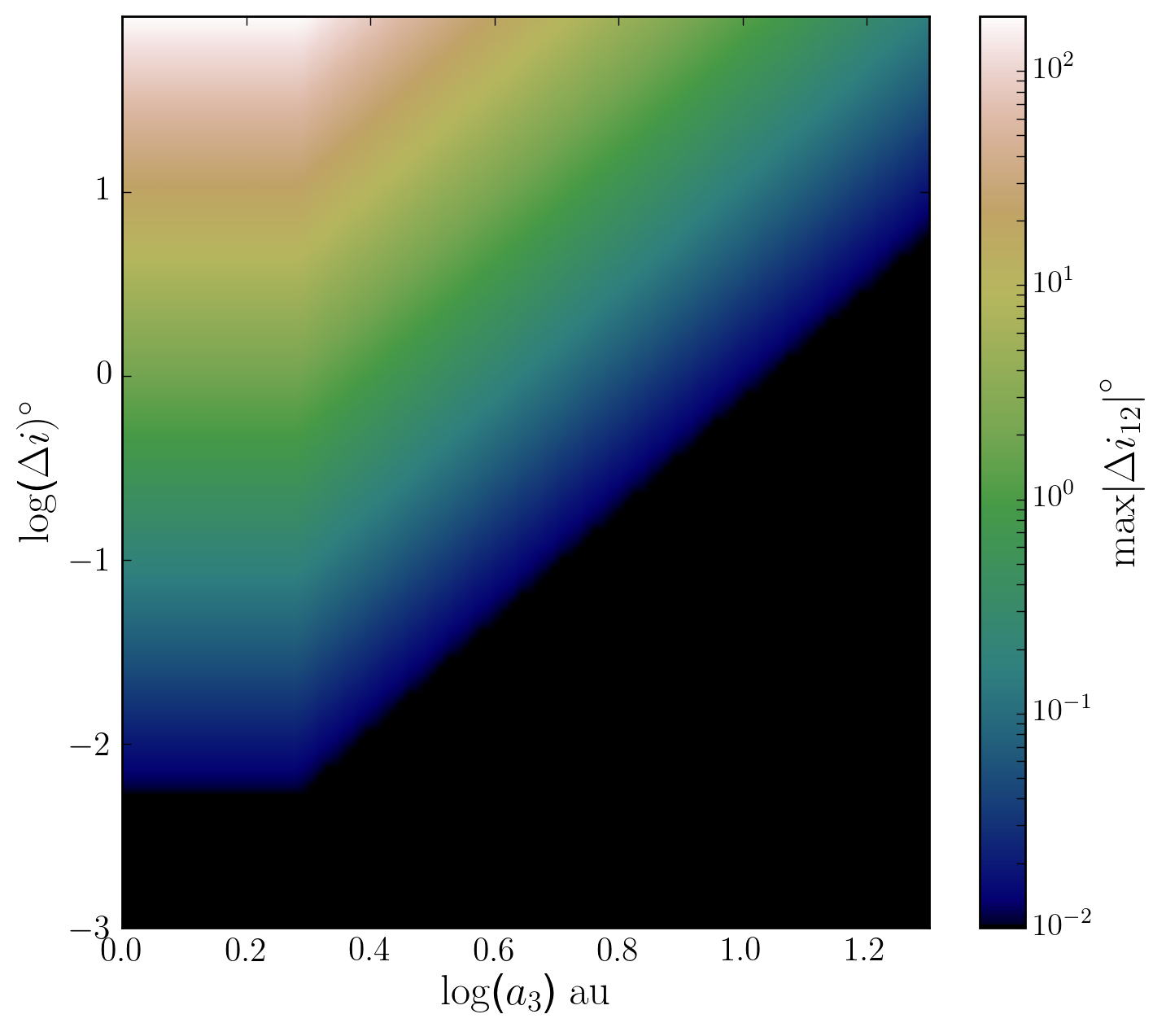}
      	\caption{The maximum mutual inclination, $\mathrm{max}|\Delta i_{12}|$, between two planets on circular, initially coplanar orbits with semi-major axis of 0.2, 0.5au and masses of 10M$_\oplus$ respectively, from the secular interaction with an outer third planet. The value of $\mathrm{max}|\Delta i_{12}|$ calculated by the full Laplace-Lagrange solution from eq. (\ref{eq:mutinc}) is given by the colour scale on the left panels. The right panel colour scales give $\mathrm{max}|\Delta i_{12}|$ calculated by the simplified Laplace-Lagrange solution for when $a_3 \gg a_1,a_2$, given by eq. (\ref{eq:mutinc_simp}) and eq. (\ref{eq:Ksimp}). For the top panels $\Delta i = 10^\circ$, for the middle panels $m_3 = 1$M$_\mathrm{J}$ and for the bottom panels $a_3 = 2$au. It is important to note that the assumptions of Laplace-Lagrange theory break down when $\Delta i \gg 20^\circ$. Larger inclinations are only included in this Figure to aid comparison between $\mathrm{max}|\Delta i_{12}|$ predicted by the full and simplified Laplace-Lagrange theory solutions.}
      	\label{fig:M_i}
      \end{figure*}
 Consider the same general two planet system from \S\ref{subsec:twoplanetcase2}. Assume that the two planets are initially coplanar. Consider now a third planet on an external circular orbit, with a mass and semi-major axis of $m_3$ and $a_3$ respectively such that $a_3 > a_2$. The orbital plane of this external planet is initially mutually inclined to the inner planets by $\Delta i$. We assume that each of the planets interact through secular interactions only and that inclinations and eccentricities remain small, allowing for application of Laplace - Lagrange theory. Assuming that the invariable plane is taken as a fixed reference plane, the initial inclination of the third planet $i_3$ is given by 
 \[ i_3 = \arctan\left[\frac{(L_1 + L_2)\sin\Delta i}{L_3 + (L_1 + L_2)\cos\Delta i}\right]\]
 where $L_j = m_ja_j^{1/2}$ and is proportional to the angular momentum in the low eccentricity limit. The initial inclination of the inner planets with respect to the invariable plane is therefore $i_1 = \Delta i - i_3$. 
  
  From eq. (\ref{eq:secsol}) the complex inclination of each of the inner two planets with respect to the invariable plane evolves in the form of   
 \begin{equation}
 \begin{split}
 & y_1 = I_{11}e^{i(f_1t + \gamma_1)} + I_{12}e^{i(f_2t + \gamma_2)}\\
 & y_2 = I_{21}e^{i(f_1t + \gamma_1)} + I_{22}e^{i(f_2t + \gamma_2)},
 \label{eq:individualinclination}
 \end{split}
 \end{equation}
 where $y_1$ and $y_2$ are the complex inclinations of the innermost and second innermost planet respectively. The evolution of the mutual inclination between the inner pair of planets is hence given by
 \begin{equation}
 y_1 - y_2 = (I_{11} - I_{21})e^{i(f_1t + \gamma_1)} + (I_{12} - I_{22})e^{i(f_2t + \gamma_2)}.
 \label{eq:mutincevofull}
 \end{equation}
 The $t = 0$ boundary conditions give $\gamma_1 = \pi$ and $\gamma_2 = 0$. Also as $y_1$($t$ = 0) = $y_2$($t$ = 0) = $i_1$, it follows from eq. (\ref{eq:individualinclination}) that $I_{11} - I_{21}$ = $I_{12} - I_{22}$. The evolution of the mutual inclination from eq. (\ref{eq:mutincevofull}) is therefore is equivalent to 
 \begin{equation}
 y_1 - y_2 = (I_{12} - I_{22})\left(e^{i(f_1t + \pi)} + e^{if_2t}\right).
 \label{eq:complexi}
 \end{equation}
 Hence the evolution of the instantaneous mutual inclination between the inner pair of planets, $\Delta i_{12} = |y_1 - y_2|$, can be calculated if the first and second elements of the eigenvector associated with the $f_2$ eigenfrequency are known.
 In Appendix \ref{sec:fullsec} we fully solve eq. (\ref{eq:Laplace}) to give $I_{12}$ and $I_{22}$ in terms of physical variables. Here we simply say that 
 \begin{equation}
 y_1 - y_2 = \Delta i K\left[e^{i(f_1t + \pi)} + e^{if_2t}\right],
 \label{eq:mutinc}
 \end{equation}
 where $K$ is dependant on the masses and semi-major axes of the three planets, shown explicitly in Appendix \ref{sec:fullsec}. We note that the maximum value of $K \approx 1$, implying that the maximum value of the mutual inclination between the inner pair of planets from eq. (\ref{eq:complexi}) is twice the initial mutual inclination with the external third planet i.e. max|$\Delta i_{12}$| = 2$\Delta i$. For given values of masses and semi-major axes of the inner pair of planets therefore, the evolution of the mutual inclination between them is dependant on three quantities, $a_3$, $m_3$ and $\Delta i$.  
 
 The left panels of Figure \ref{fig:M_i} show how max|$\Delta i_{12}$| changes as a function of different combinations of $a_3$, $m_3$ and $\Delta i$ in eq. (\ref{eq:mutinc}) for an example system where $a_1$, $a_2$ = 0.2, 0.5au and $m_1$, $m_2$ = 10M$_\oplus$ respectively. We note that the assumptions of Laplace-Lagrange theory are expected to break down when $\Delta i \gg 20^\circ$. Larger inclinations are included for demonstration purposes only. It is evident that as the third planet tends to a limit where it is on a wide orbit, with a low mass and low initial mutual inclination, the maximum mutual inclination between the inner pair of planets becomes small as one might expect.

 \subsection{Companion wide orbit approximation}
 \label{subsec:inclinedcompanion}
 In \S\ref{sec:combtrans} we look to investigate how the evolving mutual inclination between the inner pair of planets affects the associated double transit probability, for the specific case where the external third planet is assumed to be on a wide orbit. For $a_3 \gg a_1,a_2$, certain individual and combinations of $\mathbf{B}$ matrix elements from eq. (\ref{eq:matrix}) become small and we find that eq. (\ref{eq:mutinc}) can be simplified to 
 \begin{equation}
 y_1 - y_2 \approx \Delta i K_\mathrm{simp}\left[e^{i(f_1t + \pi)} + e^{if_2t}\right],
 \label{eq:mutinc_simp}
 \end{equation} 
 where
 \begin{equation}
 K_{\mathrm{simp}} = \frac{3 m_3 a_2^{7/2}}{m_2 a_1^{1/2}a_3^3}\frac{1}{b^1_{3/2}\left(\frac{a_1}{a_2}\right)\left(1 + (L_1/L_2)\right)}.
 \label{eq:Ksimp}
 \end{equation}
 Here it is assumed that as $a_3 \gg a_1,a_2$, certain Laplace coefficients from the $\mathbf{B}$ matrix elements can be simplified, specifically $b^1_{3/2}(\alpha) \approx 3(\alpha)$ (\cite{1999ssd..book.....M}). Similar simplifications can be made to each of the eigenfrequencies, for which 
 
 \begin{equation}
 \begin{split}
 & f_1 \approx -\frac{\pi m_2 a_1^{1/2}}{2 M_\star^{1/2}a_2^2}b^1_{3/2}\left(\frac{a_1}{a_2}\right)\left(1 + L_1/L_2\right),\\
 & f_2 \approx -\frac{3\pi m_3 a_2^{3/2}}{2 M_\star^{1/2} a_3^3}\frac{1}{1 + L_1/L_2}.
 \end{split}
 \end{equation}
      
 As eq. (\ref{eq:mutinc}) shows that the maximum value of the mutual inclination between the inner pair of planets cannot be larger than twice the initial mutual inclination with the wide orbit planet (max$|\Delta i_{12}| \ngtr 2\Delta i$) we assume that the maximum value of the mutual inclination between the inner two planets predicted by eq. (\ref{eq:mutinc_simp}) is    

  \begin{equation}
  \begin{split}
  \mathrm{max}|\Delta i_{12}| &\approx 2\Delta iK_{\mathrm{simp}} \hspace{1.5cm}\quad\text{for  }K_{\mathrm{simp}} < 1,\\
  &\approx 2\Delta i \hspace{2.46cm}\quad\text{otherwise.  }
  \end{split}
  \label{eq:simp_plot}
  \end{equation}
 
        \begin{figure*}
        	\centering
        	\includegraphics[trim={0cm 0cm 0cm 0cm}, width = 0.415\linewidth]{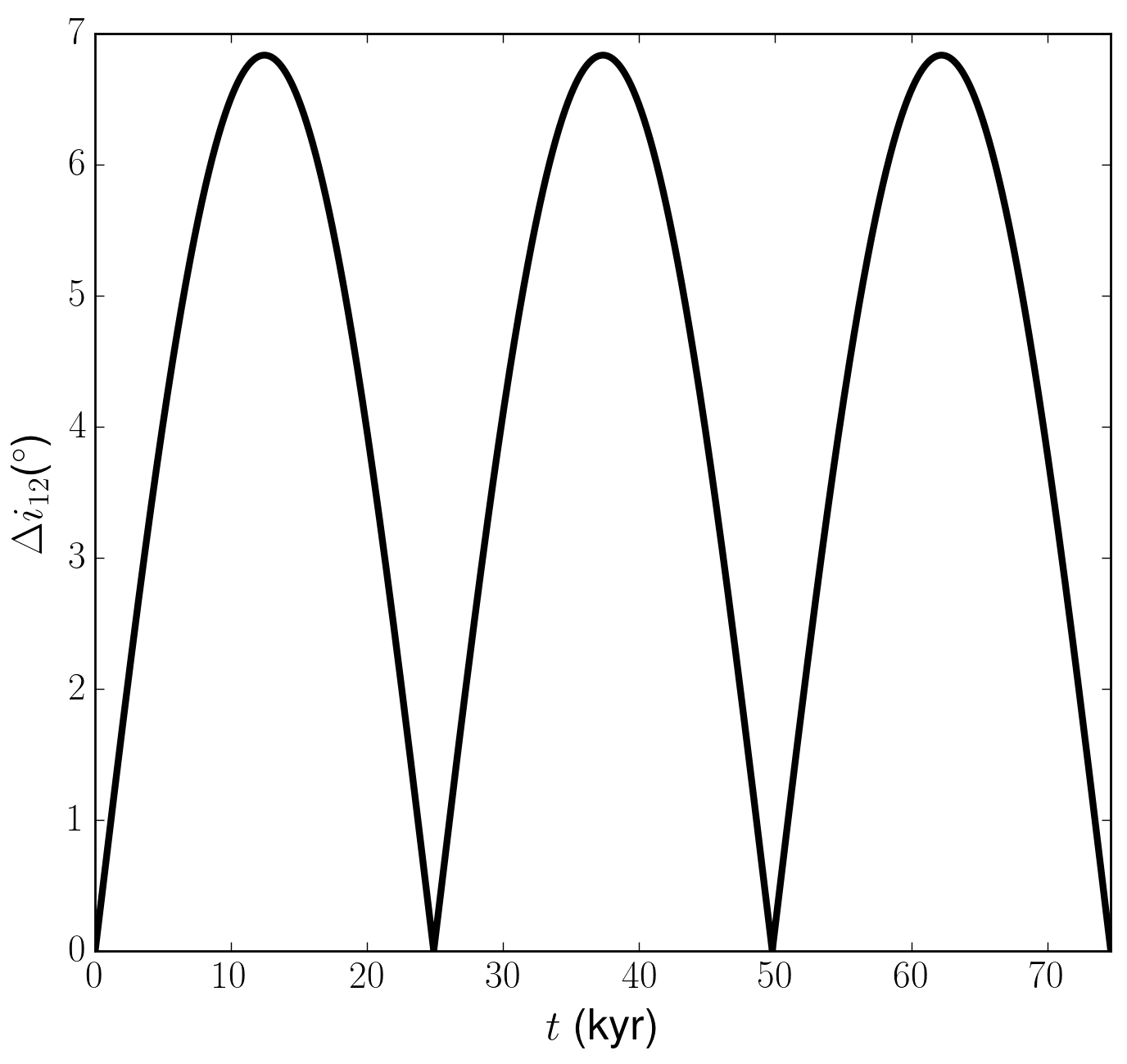}
        	\includegraphics[trim={0cm 0cm 0cm 0cm}, width = 0.425\linewidth]{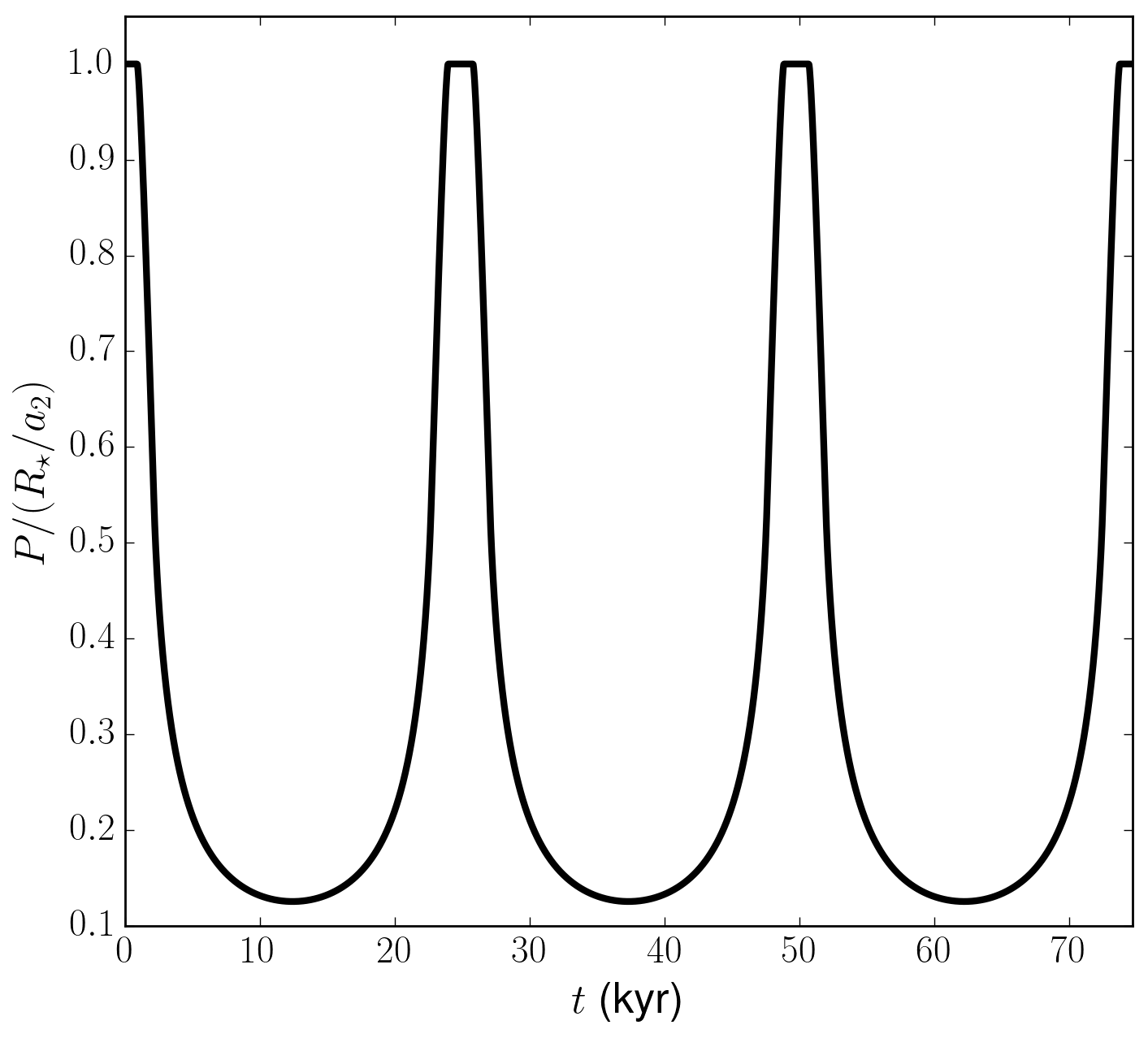}
        	\caption{(\textit{left}): The evolution of the mutual inclination of the two inner planets considered in Figure \ref{fig:M_i} due to secular interactions with a third planet with $a_3$ = 2au, $m_3$ = 1M$_\mathrm{J}$ and $\Delta i$ = 5$^\circ$. (\textit{right}): The associated evolution of the double transit probability.}
        	\label{fig:prob_time}
        \end{figure*}
 The right panels of Figure \ref{fig:M_i} show max|$\Delta i_{12}$| predicted by eq. (\ref{eq:simp_plot}) and eq. (\ref{eq:Ksimp}) using the same planet parameters as shown in the left panels. We find that when $a_3$ $\gtrsim$1.25au, the simplified form for max|$\Delta i_{12}$| from eq. (\ref{eq:simp_plot}) and eq. (\ref{eq:Ksimp}) agrees with the full Laplace - Lagrange solution to within $\sim25\%$ for all values of $m_3$ and $\Delta i$. For $a_3 \sim$1au, the simplified form of max|$\Delta i_{12}$| begins to break down and eq. (\ref{eq:simp_plot}) can underestimate max|$\Delta i_{12}$| from the full Laplace - Lagrange solution by up to a factor of 2. 
  
  This estimate is similar to the result derived by \cite{2017AJ....153...42L}, who assumed that the angular momentum vector direction of the outer inclined planet is fixed in time. They find that the maximum mutual inclination that can be induced in an inner pair of planets depends on the strength of the coupling between them (parametrized by $\epsilon_{12}$ in their eq. 12). Assuming inclinations are small we find eq. (\ref{eq:simp_plot}) agrees with the equivalent prediction of max|$\Delta i_{12}$| from \cite{2017AJ....153...42L} if $K_\mathrm{simp} = \epsilon_{12}$. Indeed, $K_\mathrm{simp}$ and $\epsilon_{12}$ are almost identical despite the different derivation techniques (e.g. we derive the full Laplace- Lagrange solution and then simplified assuming $a_3 \gg a_1, a_2$), apart from $K_\mathrm{simp}$ contains an additional factor of $a_1^{1/2}a_2^{3/2}$ whereas $\epsilon_{12}$ contains a factor of ($a_2^2 - a_1^2$). By considering different combinations of $a_1$ and $a_2$ and comparing to the value of max|$\Delta i_{12}$| given by the full solution in Appendix \ref{sec:fullsec}, we find that neither eq. (\ref{eq:simp_plot}) and (\ref{eq:Ksimp}) or the equivalent equation from \cite{2017AJ....153...42L} is favoured as a more accurate approximation, since which is closer to the full solution depends on the exact parameters.

 \section{Combining Transit Probabilities with Secular Theory}
 \label{sec:combtrans}
 Considering two inner, initially coplanar planets and an outer inclined planetary companion, we combine the analysis of transit probabilities from \S\ref{sec:semianal} with secular interactions from \S\ref{sec:secint} in two main ways. First in \S\ref{subsec:meanprob}, we assume that the outer planet is not necessarily on a wide orbit. The evolution of the mutual inclination between the inner planets is therefore assumed to be given by the full Laplace-Lagrange solution derived in eq. (\ref{eq:mutinc}). The double transit probability of the inner two planets during this evolution is then calculated through the method outlined in \S\ref{sec:semianal}. This provides the most accurate prediction for how the double transit probability of two inner planets evolves (in the low inclination limit) considering a given outer planetary companion. We make use of this method for a detailed discussion of how an outer planet affects an inner population of Kepler systems in \S\ref{sec:Kepdich}. 
 
 Second, in \S\ref{subsec:meanprobsimp} we assume that the outer planetary companion is on a significantly wide orbit. The evolution of the mutual inclination between the inner two planets is therefore given by eq. (\ref{eq:mutinc_simp}) and eq. (\ref{eq:Ksimp}). Here we look to give a simple analytical form to describe the double transit probability of two inner planets, due to secular interactions with a given outer planetary companion. We make use therefore of simple analytical relations such as eq. (\ref{eq:doubleprob}) to describe double transit probabilities. Comparing with the work in \S\ref{subsec:meanprob} allows for the accuracy of these approximations to be judged. We demonstrate in \S\ref{sec:realsys} how simple constraints can be placed on the inclination of an outer companion in specific systems using this method.

 \subsection{Two planet system with an inclined companion}
 \label{subsec:meanprob}
 From Figure \ref{fig:pi} it is clear that if the amplitude of the mutual inclination between the inner two planets is large, then the associated double transit probability, $P$, will only be at a maximum value for a small proportion of the secular evolution. The presence of an outer inclined planet may therefore result in a significant reduction in the mean double transit probability $\langle P \rangle$ on long timescales. Figure \ref{fig:prob_time} shows how both the mutual inclination and the double transit probability evolve with time for two inner planets from Figure \ref{fig:M_i}, which are perturbed by an outer planetary companion with a semi-major axis, mass and inclination of $a_3$ = 2au and $m_3$ = 1M$_\mathrm{J}$ and $\Delta i$ = 5$^\circ$ respectively. Indeed, $P$ is only at a maximum value for a small proportion of the secular evolution leading to a significant reduction in $\langle P \rangle$ compared with if the outer planet were not present. 
  
  \begin{figure*}
  	\centering
  	\includegraphics[trim={0cm 0cm 0cm 0cm}, width = 0.44\linewidth]{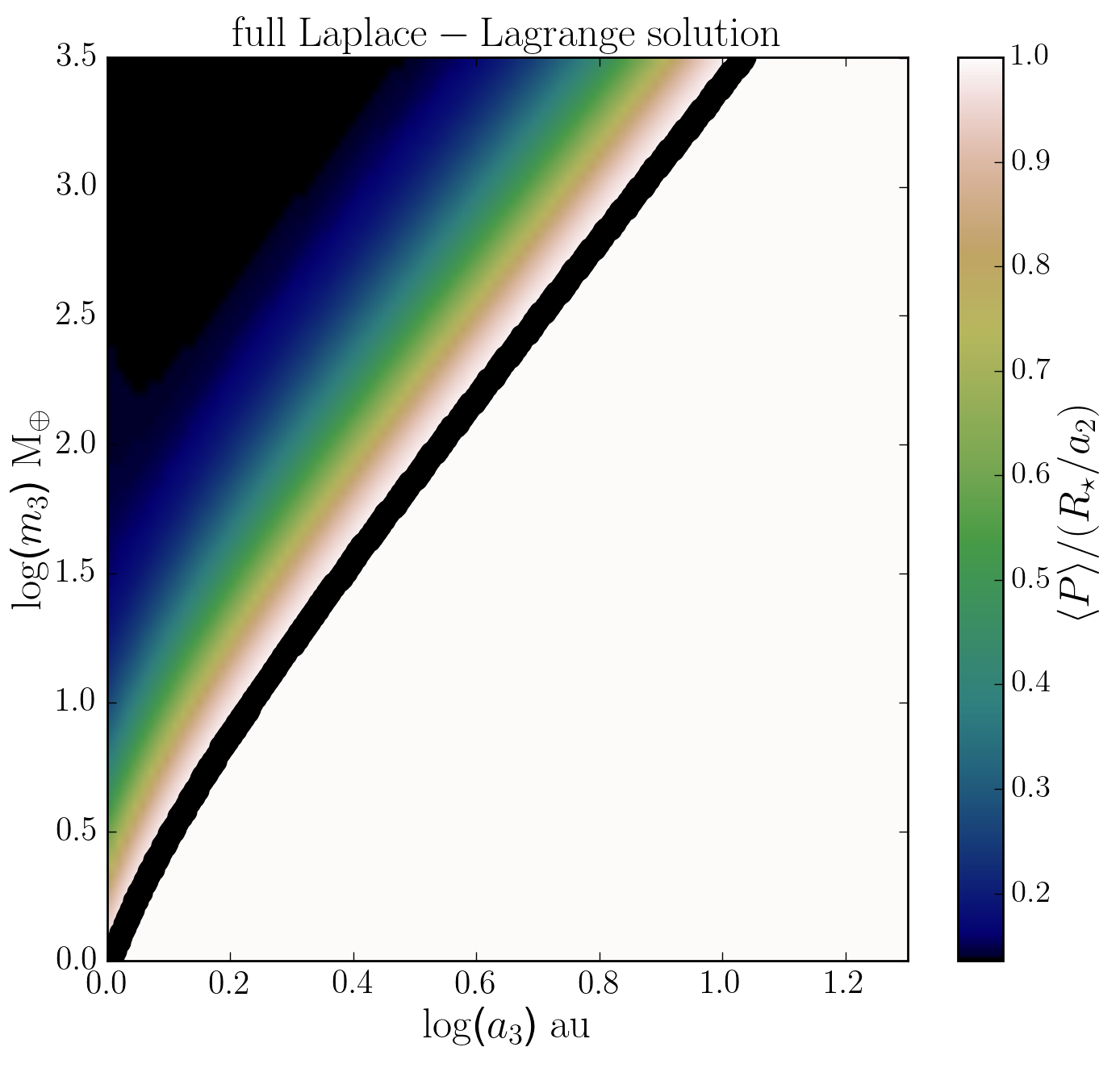}
  	\includegraphics[trim={0cm 0cm 0cm 0cm}, width = 0.44\linewidth]{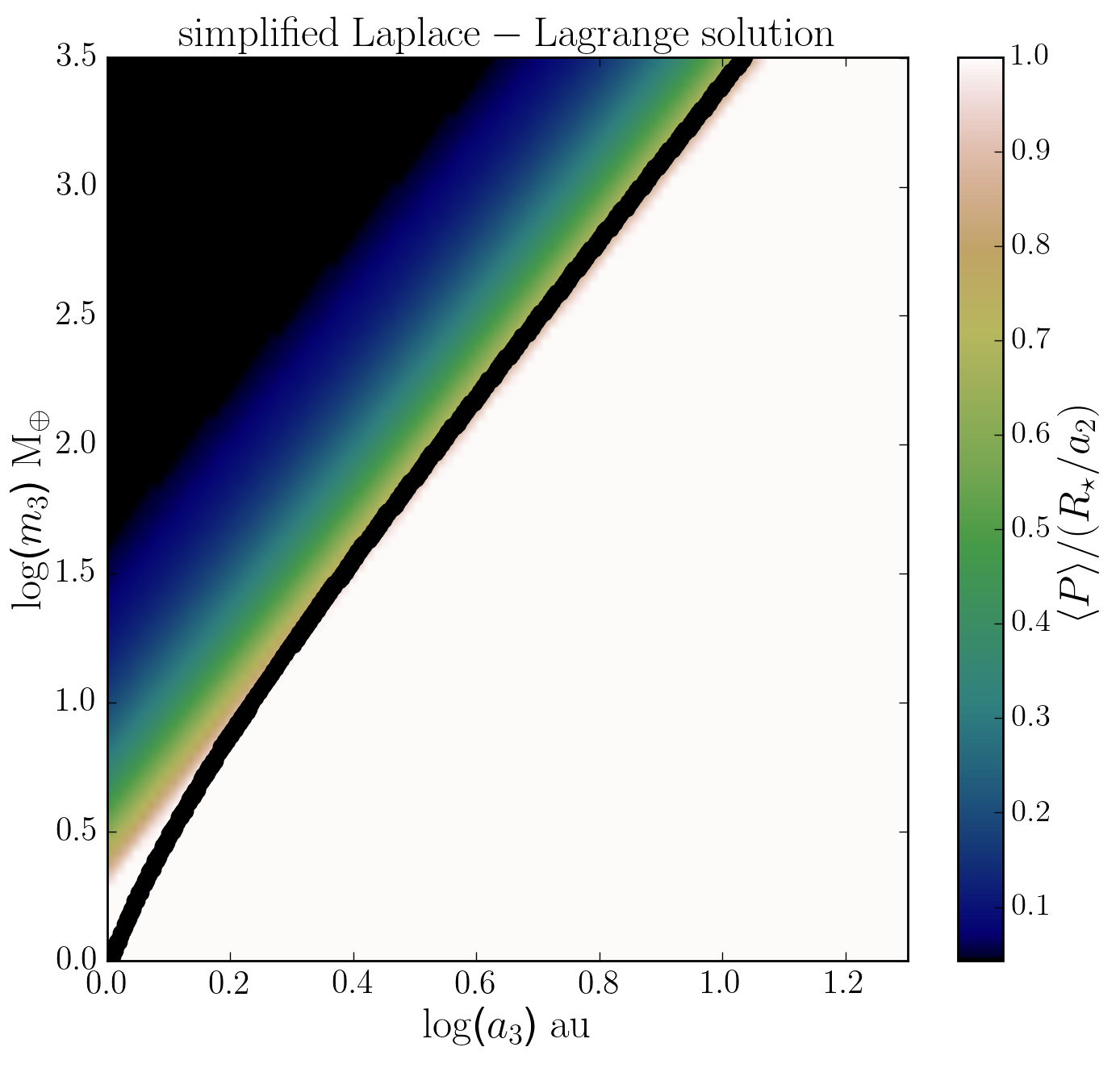}
  	\includegraphics[trim={0cm 0cm 0cm 0cm}, width = 0.44\linewidth]{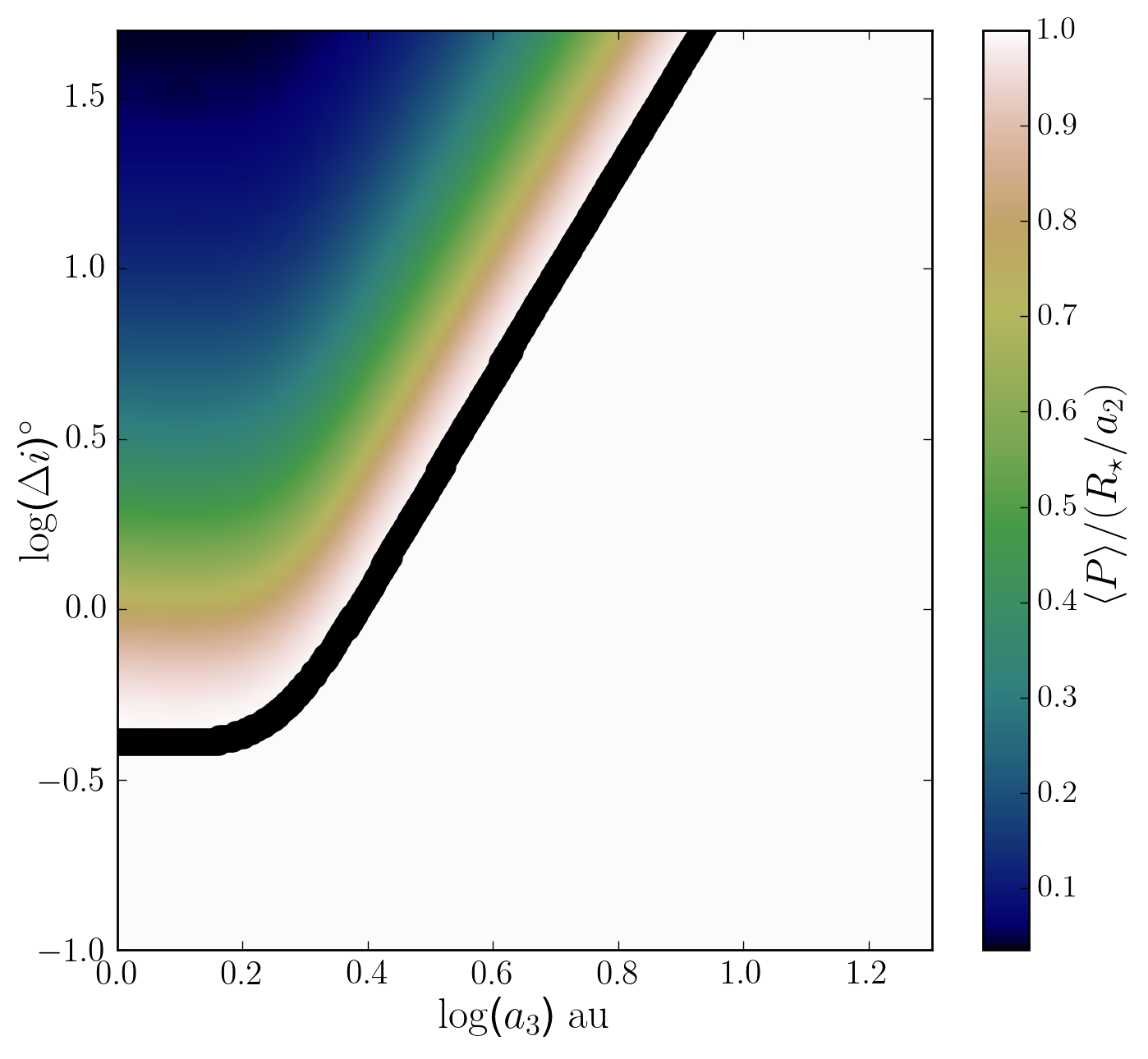}
  	\includegraphics[trim={0cm 0cm 0cm 0cm}, width = 0.44\linewidth]{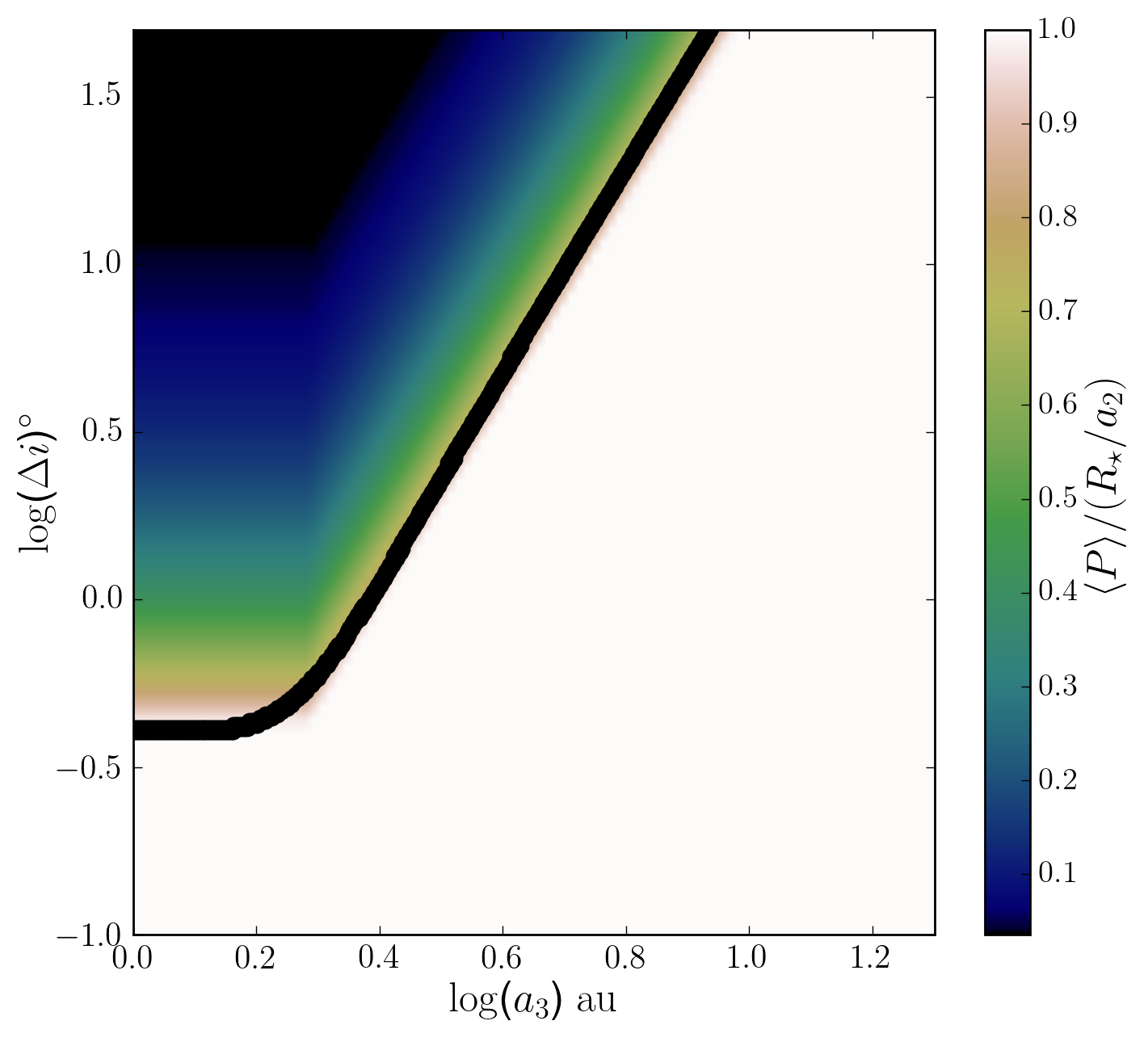}
  	\includegraphics[trim={0cm 0cm 0cm 0cm}, width = 0.44\linewidth]{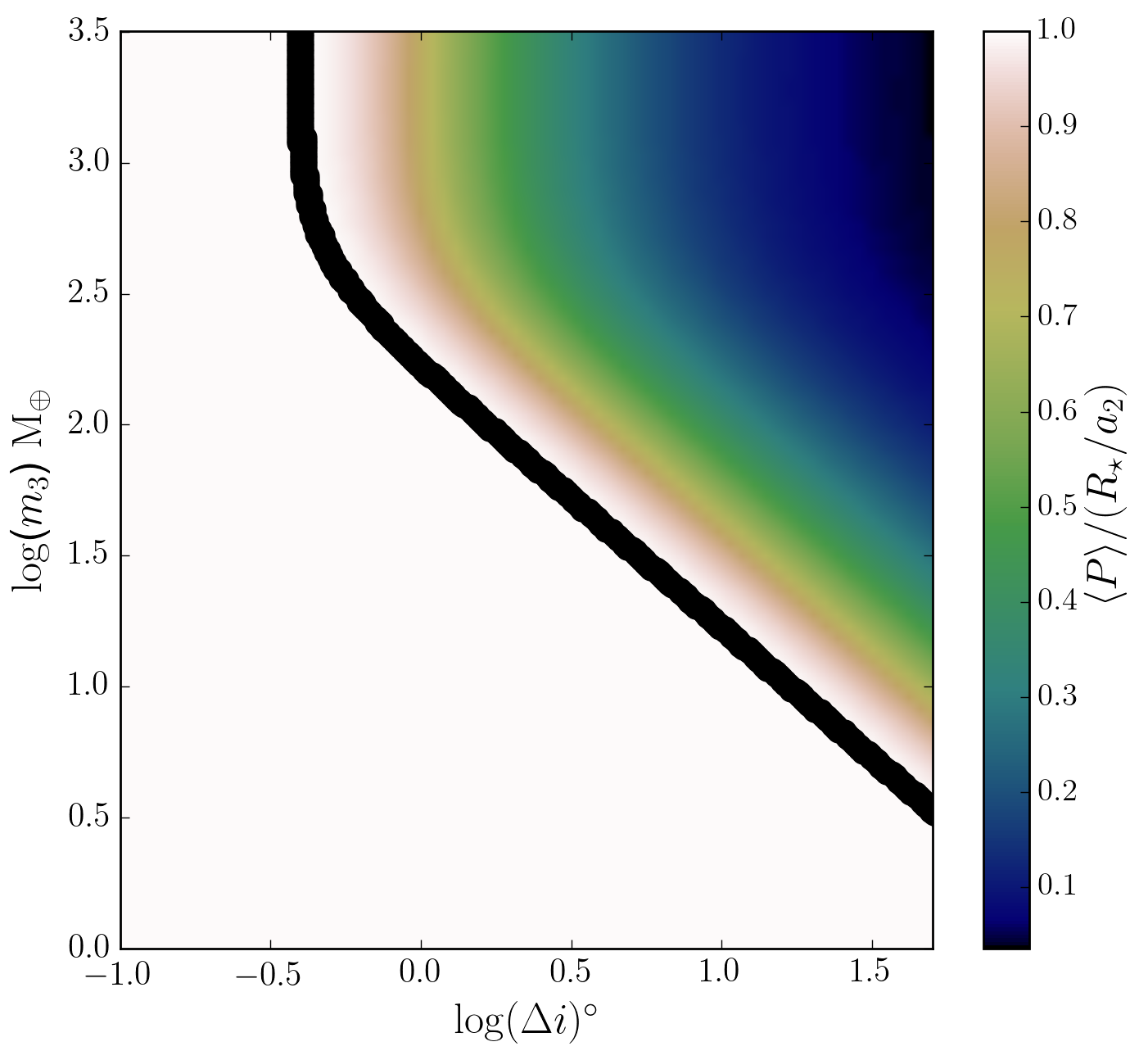}
  	\includegraphics[trim={0cm 0cm 0cm 0cm}, width = 0.44\linewidth]{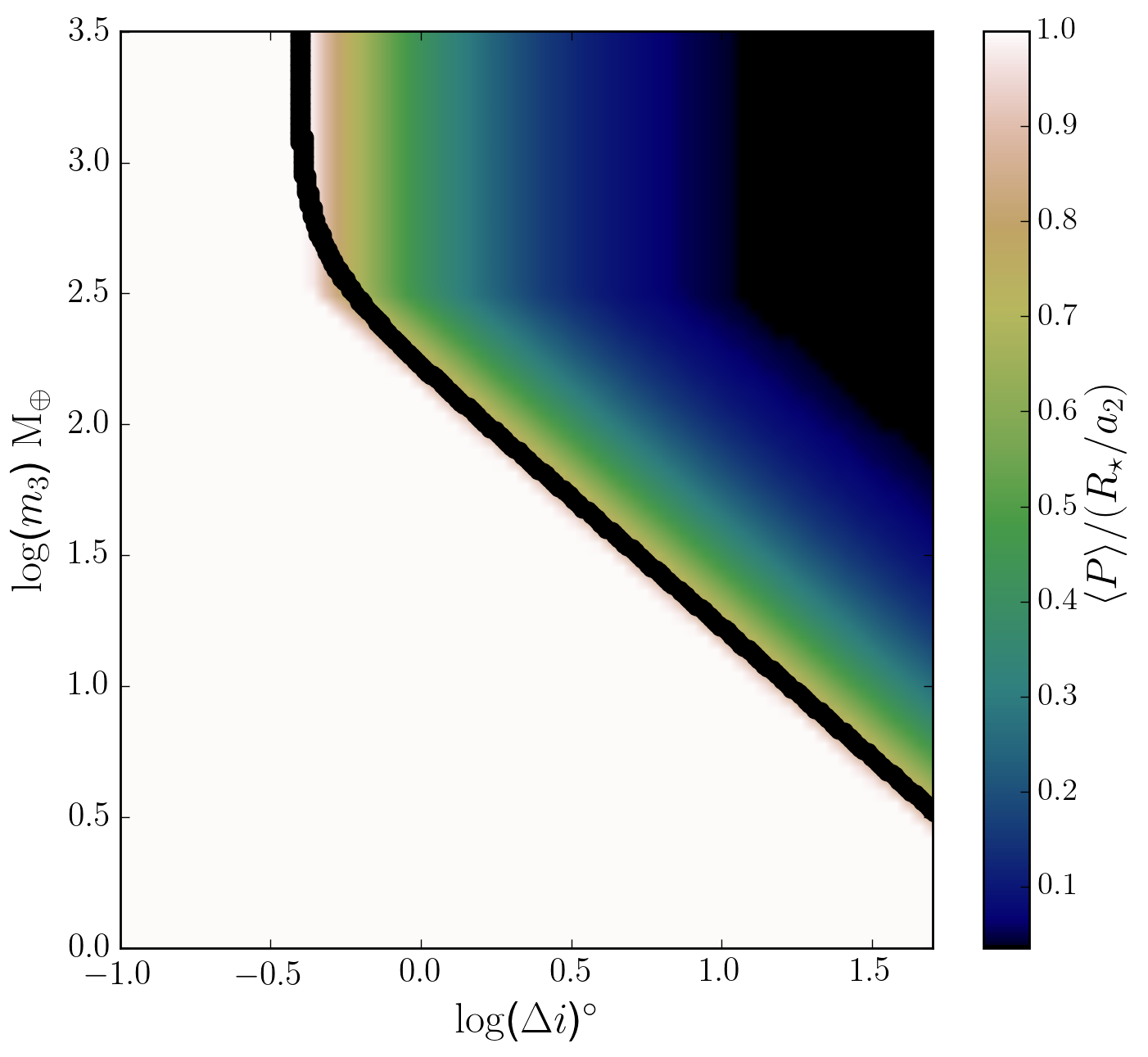}
  	\caption{The mean double transit probability of two planets $\langle P \rangle$ from Figure \ref{fig:M_i}, which are being secularly perturbed by a third planet on a mutually inclined orbit according to the full Laplace - Lagrange solution (left panels) and the simplified Laplace-Lagrange solution for when the third planet is assumed to be on a wide orbit. The black lines show the boundary where the maximum mutual inclination between the inner planets exceeds $I_1$ from eq. (\ref{eq:I1}) and $\langle P \rangle$ is assumed to be significantly reduced. The black lines on the respective left and right panels are identical and included to aid comparison. As noted in Figure \ref{fig:M_i}, Laplace - Lagrange theory is expected to break down for $\Delta i \gg 20^\circ$. Larger inclinations are only included here for demonstration purposes only.}
  	\label{fig:Pa_M}
  \end{figure*}
 
 Furthermore, the left panels of Figure \ref{fig:Pa_M} show how $\langle P \rangle$ changes due to perturbations from an outer planet with the same range of parameters considered in Figure \ref{fig:M_i}. As one may expect, through comparing the left panels of Figures \ref{fig:M_i} and \ref{fig:Pa_M}, an outer planet which induces a large value of max|$\Delta i_{12}$| also causes a significant reduction in the mean double transit probability of the inner two planets and vice versa for small values of max|$\Delta i_{12}$|. 
 
 The left panels of Figure \ref{fig:Pa_M} also suggest a clear boundary of $a_3$, $m_3$ and $\Delta i$, above which the outer planet causes $\langle P \rangle$ to be significantly reduced and below which $\langle P \rangle$ is unchanged. From Figure \ref{fig:pi}, the double transit probability of the two inner planets can be considered to be significantly reduced when $\Delta i_{12} > I_1$, where $I_1$ is given by eq. (\ref{eq:I1}). We assume therefore that the boundary where $\langle P \rangle$ is significantly reduced occurs when max|$\Delta i_{12}$| $\approx$ $I_1$. The values of $a_3$, $m_3$ and $\Delta i$ which give this boundary are shown by the black lines in the left panels of Figure \ref{fig:Pa_M}. 

 \subsection{Companion wide orbit approximation} 
 \label{subsec:meanprobsimp}
 Considering the simplified evolution of the mutual inclination from eq. (\ref{eq:mutinc_simp}) and (\ref{eq:Ksimp}) for when $a_3 \gg a_1,a_2$, here we estimate the value of the mean double transit probability itself. We assume that $\langle P \rangle$ is dominated by the maximum or minimum value of the double transit probability, $P_{\mathrm{max}}$ and $P_{\mathrm{min}}$ respectively, depending on whether max|$\Delta i_{12}$| is greater than $I_1$. We assume that $I_1 \approx R_\star/a_1 - R_\star/a_2$ from eq. (\ref{eq:I1}) for $R_\star/a_1$, $R_\star/a_2$ $\ll$ 1. From Figure \ref{fig:pi}, the value of $P_{\mathrm{max}}$ = $R_\star/a_2$, however a value of $P_{\mathrm{min}}$ is more difficult as no specific analytical estimate exists. We therefore assume $P_{\mathrm{min}}$ can be given by the estimate from \cite{2010arXiv1006.3727R} shown by eq. (\ref{eq:doubleprob}). We note that this approximation for $P_{\mathrm{min}}$ would be expected to break down if max|$\Delta i_{12}$| predicts partial overlap between the transit regions of the inner planets for all azimuthal angles (see Figure \ref{fig:pi}).   
 Assuming that the masses and semi-major axes of all the planets are known, in addition to the inclination of the outer planet and that max|$\Delta i_{12}$| is given by the simplified Laplace - Lagrange solution from eq. (\ref{eq:simp_plot}), $\langle P \rangle$ can be estimated by 
 \begin{equation}
 \begin{split}
 \langle P \rangle & \approx R_\star/a_2\hspace{2.2cm}\quad\text{for  }\mathrm{max}|\Delta i_{12}| < R_\star/a_1 - R_\star/a_2\\
 &\approx \frac{2R_\star^2}{\pi a_1a_2\sin(\mathrm{max}|\Delta i_{12}|)}\hspace{.2cm}\quad\text{otherwise,  }
 \end{split}
 \label{eq:meanpsimp}
 \end{equation}
 
 The right panels of Figure \ref{fig:Pa_M} show the value of $\langle P \rangle$ predicted by eq. (\ref{eq:meanpsimp}), using the same planet parameters as those in the left panel. The black lines are identical to those in the left panels of Figure \ref{fig:Pa_M} and are included to aid comparison between both sides of the Figure.
 
 The above assumptions bias the double transit probability toward spending a greater proportion of the secular evolution at $P_{\rm{min}}$. As such, eq. (\ref{eq:meanpsimp}) can under predict $\langle P \rangle$, by a factor of up to 4 when comparing the left and right panels of Figure \ref{fig:Pa_M}. We suggest therefore that eq. (\ref{eq:meanpsimp}) should be used as a first order approximation of $\langle P \rangle$ only.

\section{Application to specific systems}
\label{sec:realsys}
Here we consider real systems observed to have both transiting planets and an additional outer, non-transiting planet. Due to the inherent faintness of Kepler stars, follow up observations to detect non-transiting planets, namely by RV studies, are challenging. Thus the number of systems observed with such architectures are relatively sparse. We consider three of these systems: Kepler-56, Kepler-68 and Kepler-48 in addition to HD 106315. As RV surveys are largely insensitive to planetary inclinations, we apply eq. (\ref{eq:meanpsimp}) with eq. (\ref{eq:Ksimp}) to place constraints on the inclination of the non-transiting planets in these systems. 

Assume that, as the transiting planets are indeed transiting, the mean double transit probability is at a maximum. Rearranging eq. (\ref{eq:meanpsimp}) one finds
\begin{equation}
\begin{split}
\Delta i_{\mathrm{crit}} &\approx \frac{R_\star/a_1 - R_\star/a_2}{2K_{\mathrm{simp}}}\hspace{2.2cm} \quad\text{for  } K_{\mathrm{simp}} < 1\\
&\approx \frac{R_\star/a_1 - R_\star/a_2}{2}\hspace{2.2cm}\quad\text{otherwise,  }
\end{split} 
\label{eq:critinc}
\end{equation}
where $\Delta i_{\mathrm{crit}}$ is the inclination of the non-transiting planet required to significantly reduce the mean probability that the inner planets are observed to transit due to secular interactions. We note that eq. (\ref{eq:critinc}) assumes that the transiting planets are initially coplanar. However if these planets were initially mutually inclined by a small amount, a smaller secular perturbation from the outer planet would be required to significantly reduce the mean probability that the inner planets are observed to transit. In this case, $i_\mathrm{crit}$ from eq. (\ref{eq:critinc}) would be reduced.

\subsection{Kepler-56}
Kepler-56 is a red giant star with a mass and radius of M$_\odot$ = 1.32 $\pm$ 0.13 M$_\odot$ and R$_\odot$ = 4.23 $\pm$ 0.15R$_\odot$ respectively (\cite{2013ApJ...767..127H}), which is observed to host three planets. Interestingly, Kepler-56 represents one of the few red giant stars observed to host a planetary system (\cite{2014A&A...562A.109L}; \cite{2015A&A...573L...5C}; \cite{2015ApJ...803...49Q}; \cite{2016arXiv160701755P}). The two inner planets (b, c) are observed to transit with periods of 10.5 and 21.4 days respectively (\cite{2011ApJ...728..117B}; \cite{2013MNRAS.428.1077S}; \cite{2013ApJ...767..127H}; \cite{2014ApJ...787...80H}; \cite{2016ApJS..225....9H}; \cite{2016ApJ...822...86M}) and have masses of 22.1$^{+3.9}_{-3.6}$M$_\oplus$ and 181$^{+21}_{-19}$M$_\oplus$ respectively (\cite{2013ApJ...767..127H}). Keck/HIRES and HARPS-North observations have revealed a non-transiting giant planet (d) with a period of 1002$\pm$5 days and minimum mass of 5.62$\pm$0.38M$_\mathrm{J}$ (\cite{2013ApJ...767..127H}; \cite{2016AJ....152..165O}). An interesting quirk of this system is that the transiting planets, while being roughly coplanar, are misaligned to the stellar spin axis by $\sim$40$^\circ$ (\cite{2013ApJ...767..127H}). It is unclear if this large obliquity is caused by long term dynamical interactions with a highly inclined companion, such as Kepler-56d, or from the star being inherently tilted to the disk from which the planets formed (\cite{2014ApJ...794..131L}). 

Applying eq. (\ref{eq:critinc}), we find that $i_\mathrm{crit} = 704^\circ$. This unphysically large value means that, regardless of how Kepler-56d is inclined in this system, the mean double transit probability of the inner two transiting planets cannot be significantly reduced. That is, we suggest that the transiting planets in Kepler-56 are not strongly affected by the secular perturbations of Kepler-56d, regardless of its mutual inclination. This is a similar result to that found in \cite{2017AJ....153...42L} who also find that the inner planets are strongly coupled against external secular interactions. We therefore cannot place any constraint on the inclination of Kepler-56d using this method. We note however that this does not preclude that the 40$^\circ$ misalignment from the stellar spin axis comes from an inclined outer companion, since both inner planets could be inclined together without significant mutual inclination.

\subsection{Kepler-68}
Kepler-68 is a roughly solar type star with a mass and radius of 1.08$\pm$0.05M$_\odot$ and 1.24$\pm$0.02R$_\odot$ respectively (\cite{2013ApJ...766...40G}; \cite{2014ApJS..210...20M}). It hosts two transiting planets (b, c) with periods of 5.4 and 9.6 days respectively (\cite{2013ApJ...766...40G}; \cite{2014ApJS..210...20M}; \cite{2015ApJ...808..126V}; \cite{2016ApJS..225....9H}; \cite{2016ApJ...822...86M}) and fitted masses of 5.97$\pm$1.70 and 2.18$\pm$3.5M$_\oplus$ respectively (\cite{2014ApJS..210...20M}). Keck/HIRES RV follow up of this system detected a non-transiting planet (d) with a period of 625$\pm$16 days with a fitted mass of 267$\pm$16M$_\oplus$ (\cite{2014ApJS..210...20M}). 

Applying eq. (\ref{eq:critinc}) we find $i_\mathrm{crit} = 244^\circ$. Similar to Kepler-56 therefore, regardless of the mutual inclination of Kepler-68d, the mean double transit probability of the inner two transiting planets cannot be significantly reduced by secular perturbations. We therefore cannot place a constraint on the inclination of Kepler-68d using this method. We note that Kepler-68d can indeed have a large inclination without affecting the overall stability of the system according to a suite of N-body simulations, which suggest that Kepler-68d is inclined by $\Delta i < 85^\circ$ (\cite{2015ApJ...814L...9K}).

\subsection{HD 106315}
HD 106315 is a bright F dwarf star at a distance $d = 107.3 \pm 3.9$pc (\cite{2016A&A...595A...2G}) with mass and radius of 1.07$\pm$0.03M$_\odot$ and 1.18$\pm$0.11R$_\odot$ respectively (\cite{2012ApJ...761....6M}; \cite{2015PhDT........82P}; \cite{2017arXiv170103811C}). Recent \textit{K2} observations detect two transiting planets (b, c) with periods of 9.55 and 21.06 days respectively and radii of 2.23$^{+0.30}_{-0.25}$ and 3.95$^{+0.42}_{-0.39}R_\oplus$ respectively (\cite{2017arXiv170103811C}; \cite{2017arXiv170103807R}). Mass-radius relationships suggest these planets have masses of 8 and 20M$_\oplus$ respectively (\cite{2016ApJ...819...83W}; \cite{2016ApJ...825...19W}; \cite{2017arXiv170103811C}). Further Keck/HIRES RV observations also indicate the presence of a third outer companion planet (d) with a period of $P_\mathrm{d} \gtrsim 80$ days, which has a mass of $m_\mathrm{d} \gtrsim$1M$_\mathrm{J}$ (\cite{2017arXiv170103811C}). As the exact period of this outer planet is unknown we consider two possibilities where the outer planet has a period of $P_\mathrm{d} = 80$ days and $P_\mathrm{d} = 365$ days respectively. Assuming $P_\mathrm{d} = 80$ days implies a mass of $m_\mathrm{d} = 1$M$_\mathrm{J}$ (\cite{2009ApJ...700..302W}; \cite{2017arXiv170103811C}). Applying eq. (\ref{eq:critinc}) with this outer planet gives $i_\mathrm{crit} = 1.1^\circ$. This suggests that if the outer planet had a period of $P_\mathrm{d} = 80$ days, it must have an inclination of $\Delta i \lesssim 1.1^\circ$, otherwise the mean probability of observing the inner two planets to transit would be significantly reduced due to the secular interaction. Conversely, if the outer planet is assumed to be further out with $P_\mathrm{d} = 365$ days, implying a mass of $\sim$7M$_\mathrm{J}$, eq. (\ref{eq:critinc}) suggests that $i_\mathrm{crit} = 2.4^\circ$. That is, if the outer planet has a period of $P_\mathrm{d} = 365$ days, it must have an inclination of $\Delta i \lesssim 2.4^\circ$, otherwise the secular interaction would significantly reduce the mean probability that the inner planets are observed to transit.  

The mutual inclination of the outer planet might also be constrained through astrometric observations of HD 106315 with ESA's \textit{Gaia} mission (\cite{2001A&A...369..339P}; \cite{2008A&A...482..699C}; \cite{2014MNRAS.437..497S}; \cite{2014ApJ...797...14P}; \cite{2015MNRAS.447..287S}). The astrometric displacement of the host star due to the presence of a planet is defined by 
\begin{equation}
\alpha = \left(\frac{m_\mathrm{p}}{M_\star}\right)\left(\frac{a_\mathrm{p}}{1\mathrm{au}}\right)\left(\frac{d}{1\mathrm{pc}}\right)^{-1} \mathrm{arcsec},
\label{eq:alpha}
\end{equation}
with the astrometric signal-to-noise equal to $S/N = \alpha\sqrt{N_\mathrm{obs}}/\sigma$, where $N_\mathrm{obs}$ is the scheduled number of astrometric measurements ($N_\mathrm{obs}$ = 36 for HD 106315\footnote{http://gaia.esac.esa.int/gost/}) with typical uncertainties of $\sigma = 40\mathrm{\mu as}$ (\cite{2012Ap&SS.341...31D}). If $S/N > 20$, the orbital inclination can be constrained to a precision of $< 10^\circ$ (\cite{2015MNRAS.447..287S}). We find that for the example periods and masses considered above for HD 106315d that $S/N < 10$. We therefore expect that the inclination of the above examples of HD 106315d cannot be constrained using \textit{Gaia} astrometry. However if HD 106315d is outside of $\sim$1.3au, (implying a mass of $\gtrsim12\mathrm{M}_\mathrm{J}$) eq. (\ref{eq:alpha}) suggests that $S/N > 20$ such that the inclination of HD 106315d should be constrained by \textit{Gaia} astrometry. Further RV follow-up of this system will allow for greater constraints to be placed on the mass and the orbit of HD 106315d, which in turn allow for greater constraints to be placed on the inclination, either through potential astrometry measurements or through our model represented by eq. (\ref{eq:critinc}).

\subsection{Systems with three transiting planets and a wide orbit companion}
Here we generalise the affect a wide orbit planet has on the transit probabilities of three inner transiting planets. Consider Kepler-48 as an example of such a system. Kepler-48 has a mass and radius of M$_\star$ = 0.88$\pm$0.06M$_\odot$ and R$_\star$ = 0.89$\pm$0.05R$_\odot$ respectively. It hosts three transiting planets (b,c,d) with periods of 4.78, 9.67 and 42.9 days and fitted masses of 3.94$\pm$2.10, 14.61$\pm$2.30 and 7.93$\pm$4.6M$_\oplus$ respectively (\cite{2013MNRAS.428.1077S}; \cite{2014ApJS..210...20M}; \cite{2014ApJ...787...80H}; \cite{2016ApJS..225....9H}; \cite{2016ApJ...822...86M}). Keck/HIRES RV analysis also detects a non-transiting planet (e) with a period and fitted mass of 982$\pm$8 days and 657$\pm$ 25M$_\oplus$ respectively (\cite{2014ApJS..210...20M}). 
        \begin{figure}
        	\centering
        	\includegraphics[trim={0cm 0cm 0cm 0cm}, width = 0.95\linewidth]{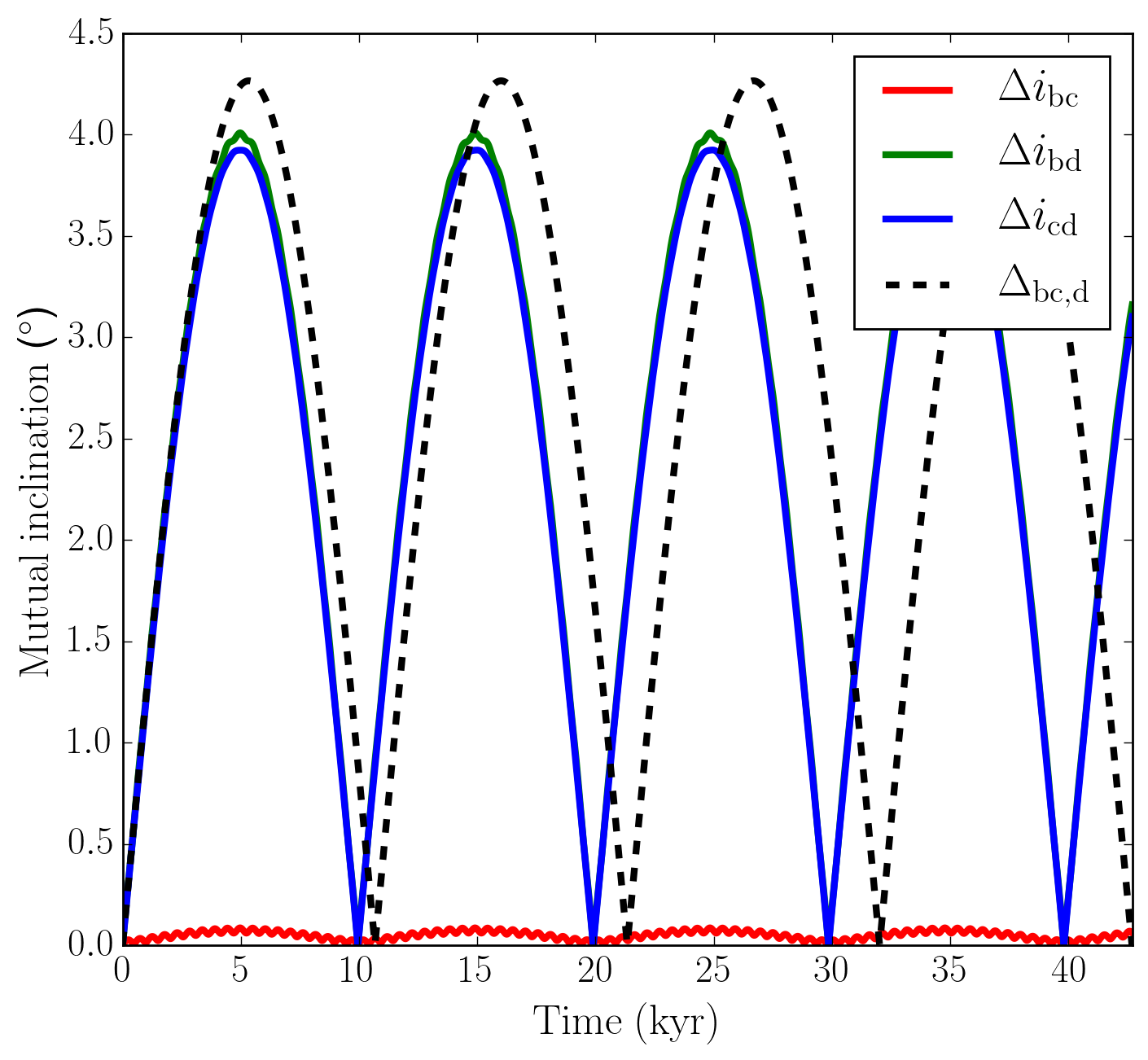}
        	\caption{The mutual inclination between the respective planets in Kepler-48, when the non-transiting planet, Kepler-48e is initially mutually inclined by $\Delta i=10^\circ$. The black dashed line shows the evolution of the mutual inclination between the inner two transiting planets with the outer transiting planet, for when the inner two planets are treated as a single body with an equal orbital angular momentum.}
        	\label{fig:Kepler48}
        \end{figure}
        
Returning to the derivation of the secular interaction in \S\ref{sec:secint}, the initial inclination of the non-transiting planet, $i_e$, with respect to the invariable plane can be generalised to 
\begin{equation}
i_\mathrm{e} = \arctan\left(\frac{\sin\Delta i\left(\sum\limits_{n=1}^{3}L_n\right)}{L_\mathrm{e} + \cos\Delta i\left(\sum\limits_{n=1}^{3}L_n\right)}\right),
\end{equation}
where $L_\mathrm{e} = m_ea_e^{1/2}$ and is proportional to the angular momentum of Kepler-48e in the low eccentricity limit and $L_n = m_na_n^{1/2}$ for either Kepler-48b, c, or d. The initial inclination of the transiting planets is therefore equal to $\Delta i - i_\mathrm{e}$. 

As the strength of the secular interaction between planets largely depends on their separation (e.g. eq. (\ref{eq:simp_plot})) we assume that Kepler-48d will be affected most by perturbations from the non-transiting planet. We demonstrate this in Figure \ref{fig:Kepler48}, which shows how the mutual inclination between each of the transiting planets evolves assuming Laplace - Lagrange theory (eq. (\ref{eq:secsol})) and that Kepler-48e is initially mutually inclined by $\Delta i = 10^\circ$. The red line shows the mutual inclination between Kepler-48b and c ($\Delta i_\mathrm{bc}$), the blue between b and d ($\Delta i_\mathrm{bd}$), and the green between c and d ($\Delta i_\mathrm{cd}$). The mutual inclination between Kepler-48b and c is largely unchanged and they remain roughly coplanar. Conversely the mutual inclination between b and d and c and d is significant and roughly equal throughout the secular evolution. It can be assumed for Kepler-48 therefore that the inner two transiting planets are largely unaffected by the secular perturbations of Kepler-48e, but both can become significantly mutually inclined to the outer transiting planet.

As such, we assume that Kepler-48b and c can be treated as a single body whose angular momentum is the sum of Kepler48-b and c, reducing the system to a total of three planets. With this approximation, the evolution of the mutual inclination between Kepler-48b and c with d ($\Delta i_\mathrm{bc,d}$) is shown by the dashed black line in Figure \ref{fig:Kepler48}. It can be seen that this way of treating Kepler-48b and c as a single body gives a good approximation for the evolution of the mutual inclination between Kepler-48b, c with d.

The initial mutual inclination of Kepler-48e which causes a significant reduction in the mean probability of the inner planets transiting, $\Delta i_\mathrm{crit}$, can therefore be approximated by eq. (\ref{eq:critinc}), where the value of $K_\mathrm{simp}$ becomes
 \begin{equation}
 K_{\mathrm{simp}} = \frac{3 m_\mathrm{e} a_\mathrm{d}^{7/2}}{m_\mathrm{d} a_\mathrm{bc}^{1/2}a_\mathrm{e}^3}\frac{1}{b^1_{3/2}\left(\frac{a_\mathrm{bc}}{a_\mathrm{d}}\right)\left(1 + (L_\mathrm{bc}/L_\mathrm{d})\right)},
 \end{equation}
with the subscripts referring to a respective planet and the subscript 'bc' to the planet which has the same total angular momentum as Kepler-48b and c. 

We find that $\Delta i_\mathrm{crit}$ = 3.7$^\circ$. This suggests therefore that the inclination of Kepler-48e $\Delta i \lesssim 3.7^\circ$, otherwise the secular interaction would cause a significant reduction in the mean probability that all three inner planets are observed to transit. Under the simpler assumption that max|$\Delta i_\mathrm{bc,d}$| $\lesssim$ $R_\star/a_\mathrm{d}$, \cite{2017AJ....153...42L} also find that the inclination of Kepler-48e, considering secular interactions only, must also be small with $\Delta i \lesssim2.3^\circ$. 

   \begin{figure*}
   	\centering
   	\includegraphics[trim={0cm 0cm 0cm 0cm}, width = 0.49\linewidth]{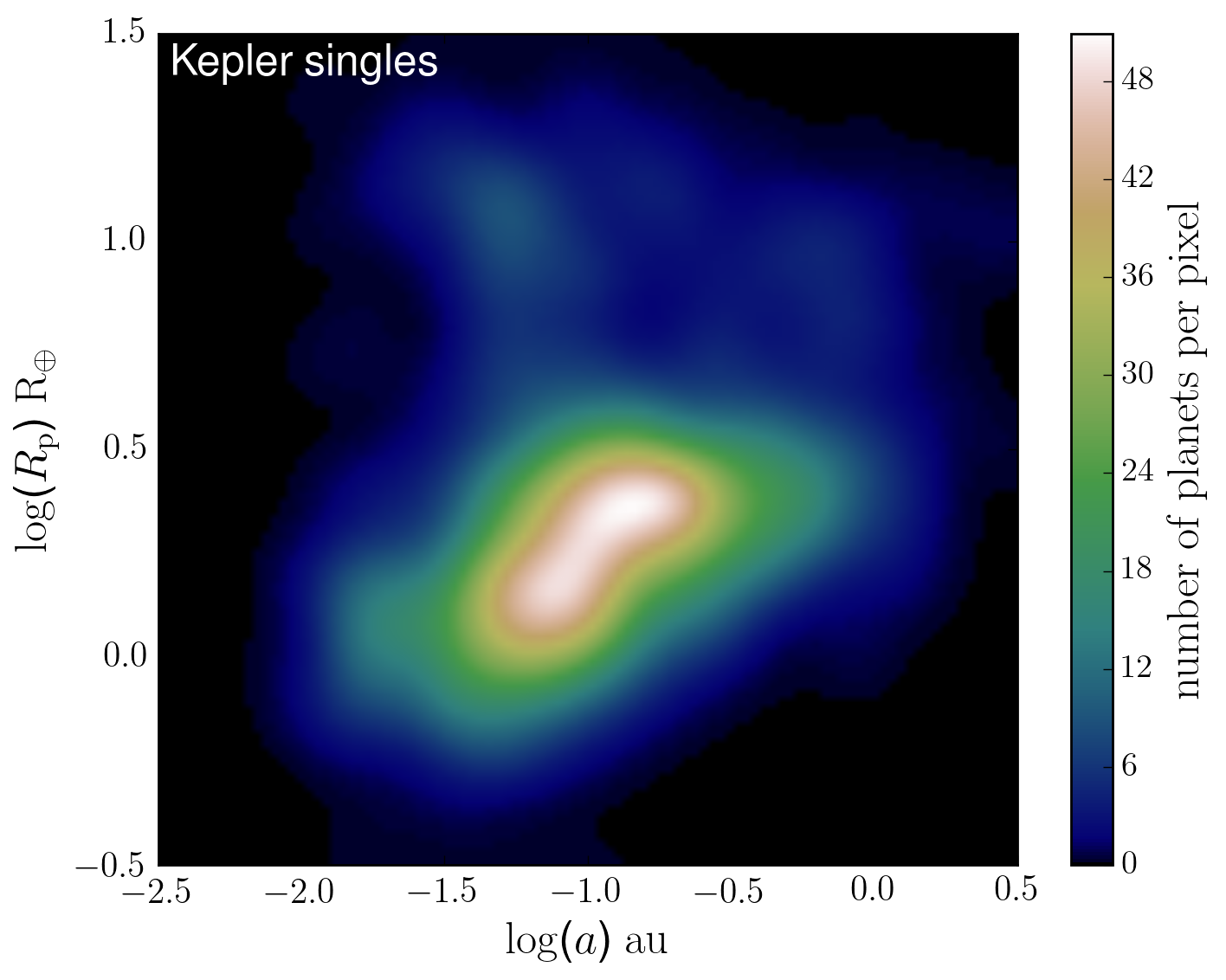}
   	\includegraphics[trim={0cm 0cm 0cm 0cm}, width = 0.49\linewidth]{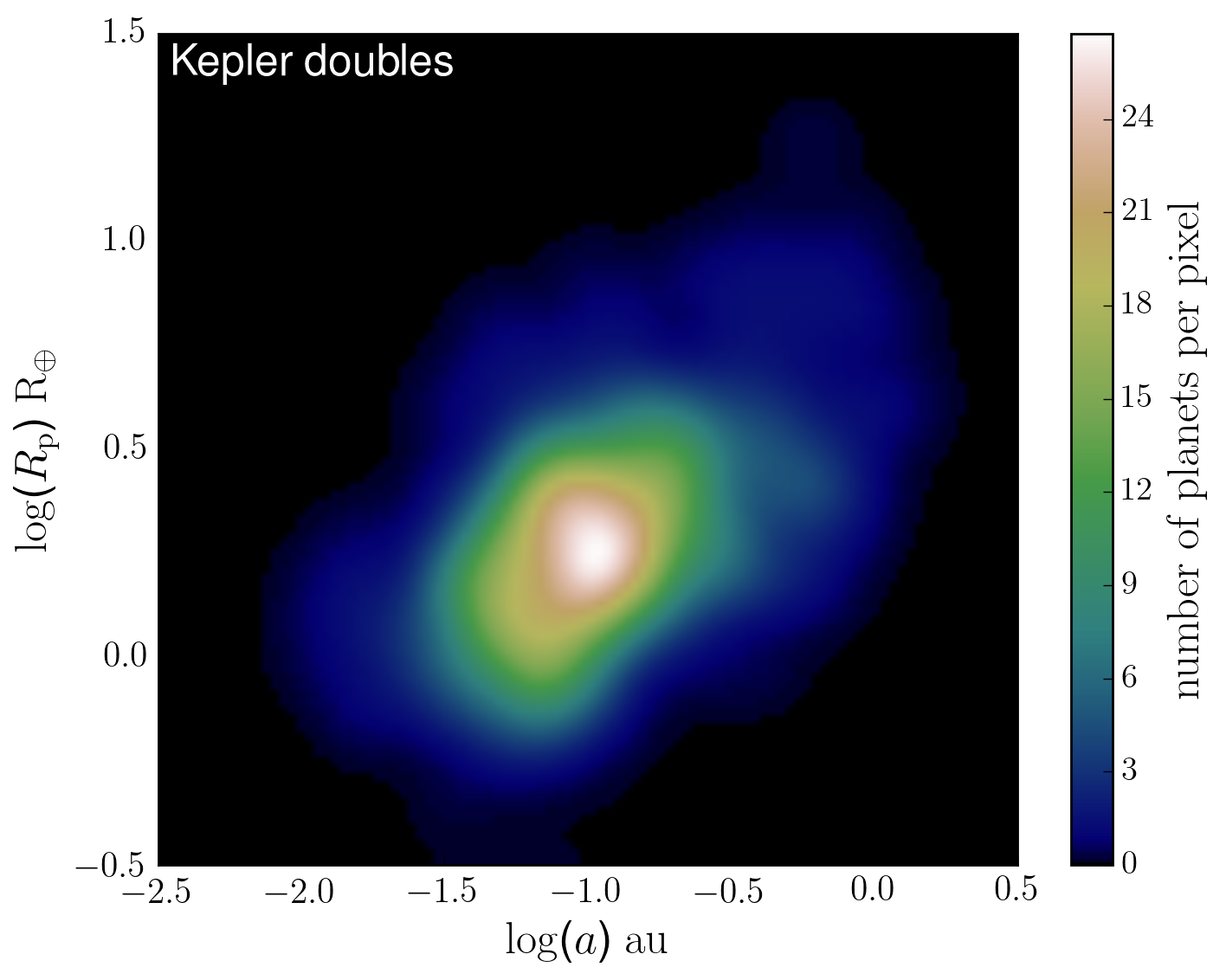}
   	\caption{The smoothed distribution of the radii and the semi-major axes of planets observed by Kepler to be in systems with a single transiting planet (\textit{left}) and in systems with two transiting planets (\textit{right}). Pixel sizes are log($a$) = 0.15 by log($R_p$) = 0.1.}
   	\label{fig:obsa_rpdist}
   \end{figure*}
   
\section{Application to the Kepler Dichotomy}
\label{sec:Kepdich}
As discussed in \S\ref{sec:intro}, Kepler has observed an excess of single transiting systems which cannot be explained by geometric effects alone, commonly referred to as the Kepler dichotomy (\cite{2011ApJS..197....8L}; \cite{2011ApJ...742...38Y}; \cite{2012ApJ...758...39J}; \cite{2016ApJ...816...66B}). This may suggest that there is a population of inherently single transiting systems in addition to a population of multi-planet systems with small inclination dispersions. However there may also be a population of multi-planet systems where the mutual inclination dispersion is large, increasing the probability that only a single planet is observed to transit. Here we investigate whether both these types of multi-planet systems can significantly contribute to the abundance of systems observed by Kepler to have one and two transiting planets respectively. 

The Kepler systems we consider are discussed in \S\ref{sec:sample}. A method for debiasing Kepler systems to a general population of planetary systems is described in \S\ref{sec:debiasing}. We consider the scenario where planets share some inherently fixed mutual inclination in \S\ref{subsec:inher}, before considering when this mutual inclination is evolving due to the presence of an outer inclined planetary companion in \S\ref{subsec:KOIwide}. We note from the outset that we do not consider Kepler systems observed to have more than two planets. Instead we look to explore what effects an outer planet might have on observables of a subset of Kepler like systems, rather than observables of the whole Kepler population. We discuss this assumption further in \S\ref{subsec:assump}.

\subsection{Kepler Candidate Sample}
\label{sec:sample}
We select planet candidates from the cumulative Kepler objects of interest (KOI) table from the NASA exoplanet archive\footnote{exoplanetarchive.ipac.caltech.edu}, accessed on 13/09/16. The vast majority of the KOIs ($\sim97\%$) that survive our cuts detailed below, to make it into our final sample are listed as being taken from the most recent Q1-17 DR24 data release. This data release is of particular note as it incorporates an automated processing of all KOIs (\cite{2016ApJS..224...12C}). 

Out of the initial 8826 KOIs we consider those which orbit solar type stars, with surface temperatures and surface gravities between 4200K $< T < $ 7000K and 4.0 $<$ log($g$) $<4.9$ respectively. This reduces the total number of KOIs to 7446. We also find the total number of unique Kepler stars within this range (discussed in \S\ref{sec:discussion}) is 164966 from the 'Kepler Stellar data' table. We next remove false positives, which refer to KOI light curves that are indicative of either an eclipsing binary, having significant contamination from a background eclipsing binary, showing significant stellar variability which mimics a planetary transit or where instrument artefacts have produced a transit like signal (see \cite{2014AJ....147..119C}; \cite{2014ApJ...784...45R}; \cite{2015arXiv150400707R}; \cite{2015ApJS..217...18S}; \cite{2016ApJS..224...12C}). This reduces our sample of KOIs (candidates herein) to 4072 objects. We subsequently remove non planetary-like objects with radii $>$22.4$R_\oplus$ (\cite{2011ApJ...728..117B}), leaving 3757 objects, after which we remove candidates with a SNR $<10$ reducing the possibility that a transit signal is caused by systematic background noise (\cite{2016ApJ...822...86M}), leaving 3327 objects. Finally we remove candidates listed as not having a satisfactory fit to the transit signal (\cite{2014ApJ...784...45R}; \cite{2015arXiv150400707R}). This gives our final sample of 3255 objects. We note that our choice of cuts means that KOI systems can become reduced in multiplicity. We find that our final sample includes systems which contain 1-6 candidates with $N_{i}$ = (1, 2, 3, 4, 5, 6) = (1951, 341, 117, 43, 15, 4) e.g. 1951 systems with a single candidate, 341 systems with two candidates etc. Herein, we consider the 1951 systems observed by Kepler to have a single transiting planet and the 341 systems observed to have two. 

The smoothed distribution of the semi-major axis and planetary radii for the single and double planet transiting systems are shown in Figure \ref{fig:obsa_rpdist}. Comparing the left and right panels of Figure \ref{fig:obsa_rpdist}, there are types of planets which are only present in single transiting systems. We briefly discuss these differences here for future reference. Large planets with short periods i.e. Hot Jupiters, are not present in Kepler systems with two transiting planets. Indeed, investigations into the formation processes of Hot Jupiters predict a lack of close companions (\cite{2009ApJ...693.1084W}; \cite{2012PNAS..109.7982S}; \cite{2015ApJ...808...14M}; \cite{2016arXiv160908110H}, see WASP-47 for an exception, \cite{2015ApJ...812L..18B}; \cite{2016A&A...595L...5A}). Long period planets are also more abundant in the population of single transiting systems. This may not necessarily indicate that long period planets inherently favour being in single transiting systems, but instead they might be the inner planet of a higher multiplicity system where the outer planets are on too long a period to produce a significant transit signal.
    \begin{figure*}
    	\centering
    	\includegraphics[trim={0cm 0cm 0cm 0cm}, width = 0.45\linewidth]{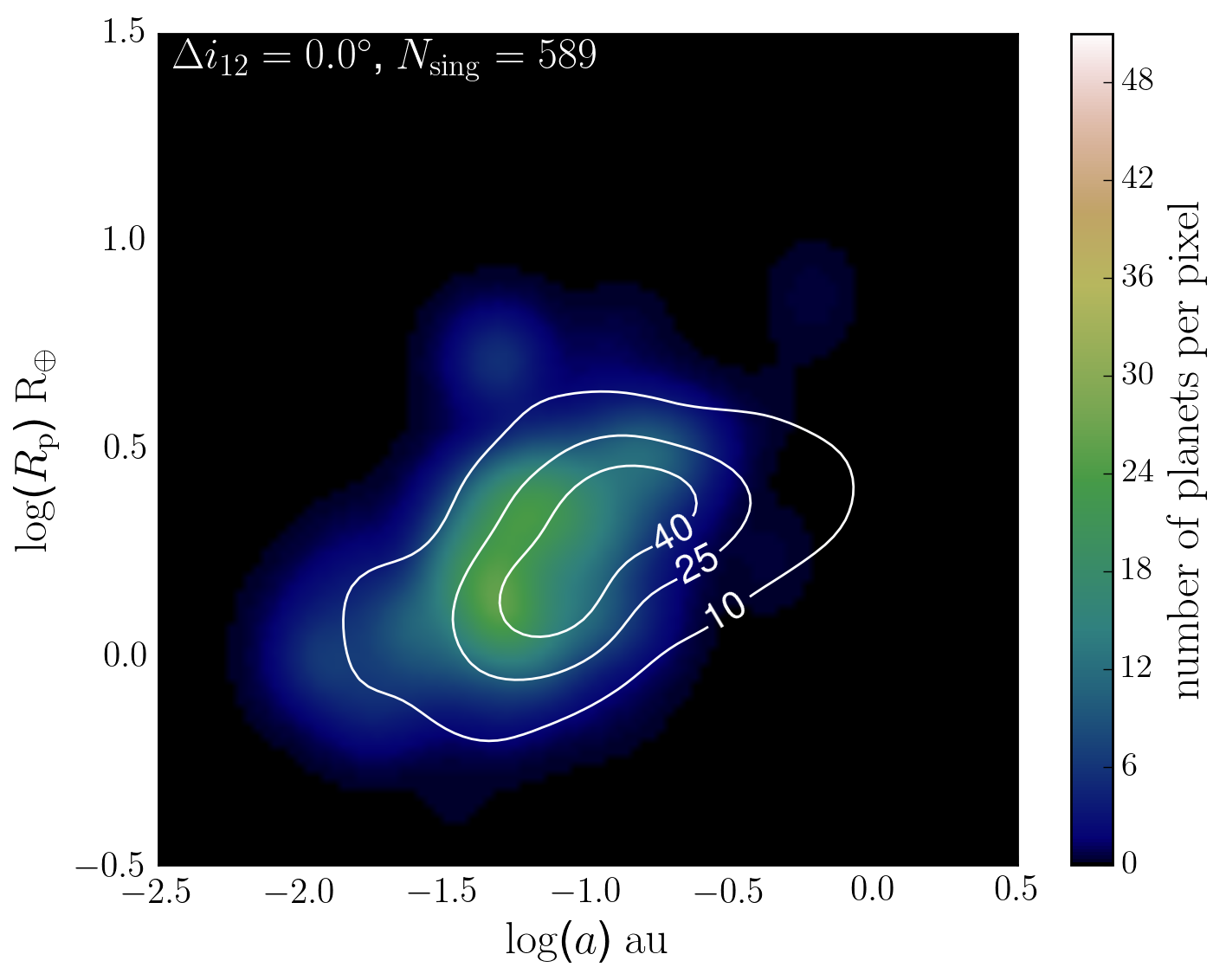}
    	\includegraphics[trim={0cm 0cm 0cm 0cm}, width = 0.45\linewidth]{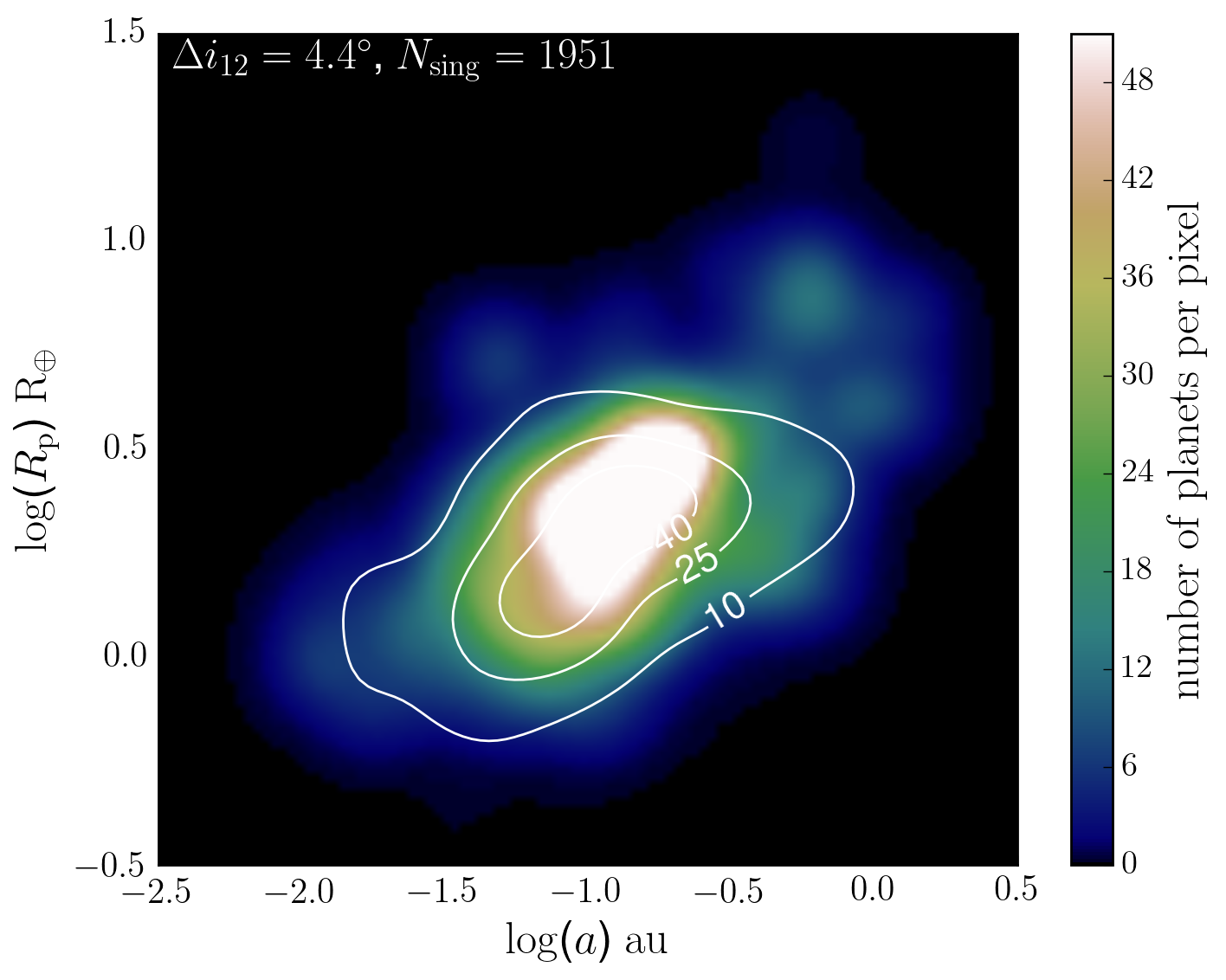}
    	\includegraphics[trim={0cm 0cm 0cm 0cm}, width = 0.45\linewidth]{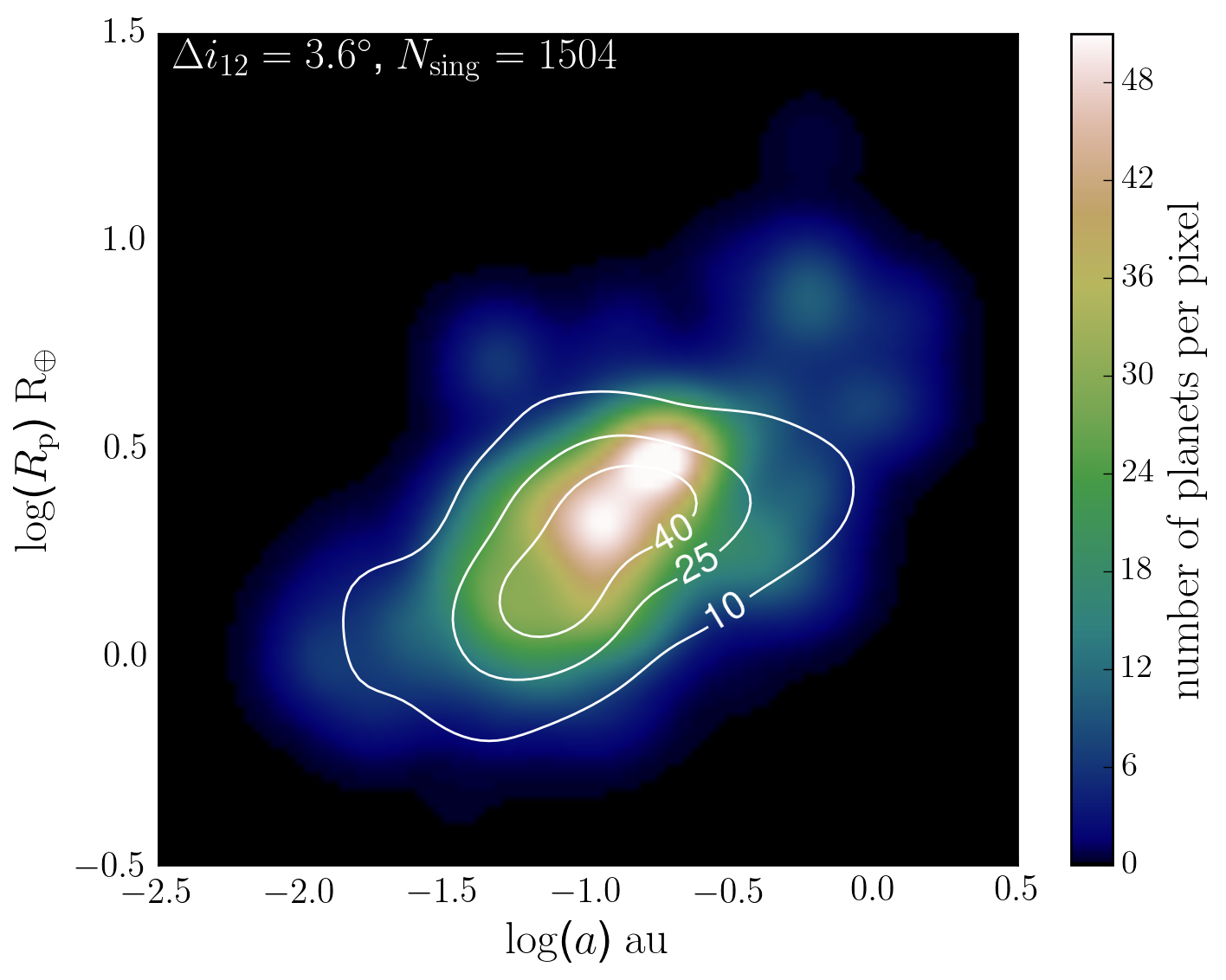}
    	\includegraphics[trim={0cm 0cm 0cm 0cm}, width = 0.45\linewidth]{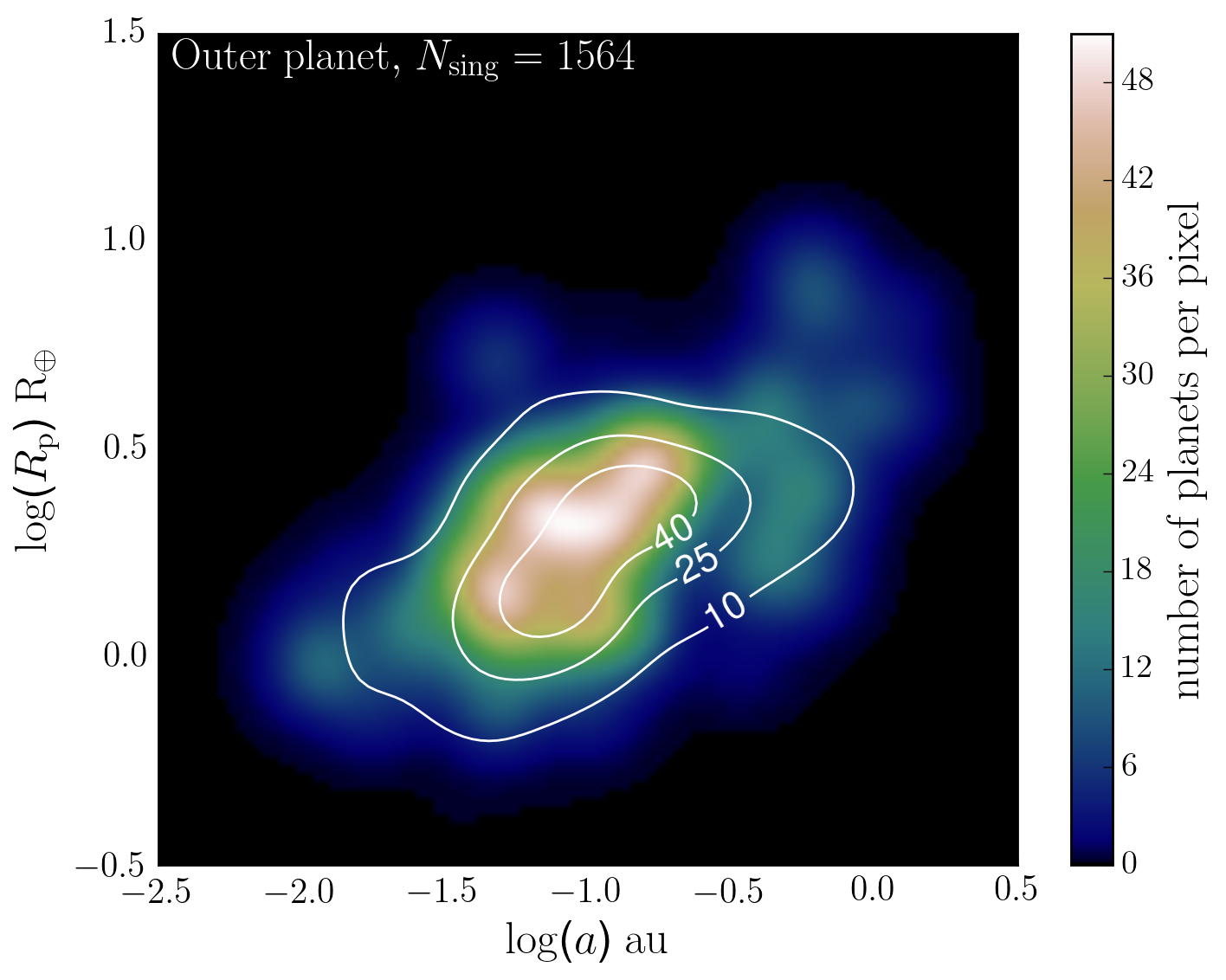}
    	\caption{The distribution of the radii and semi-major axis of single transiting planets observed from the model population with: (\textit{top}) no third planet. (\textit{middle}) A third planet with $m_3$ = 1M$_\mathrm{J}$, $a_3$ = 1.9au and $\Delta i$ = 10$^\circ$. The total number of single transiting planets predicted by the model population is equal to that observed by Kepler. The colour scale for this panel is saturated for ease of comparison. (\textit{bottom})  A third planet with $m_3$ = 24M$_\oplus$, $a_3$ = 1.07au and $\Delta i$ = 10$^\circ$. We find the 1564 single transiting planets predicted here are a best fit to those observed by Kepler (left panel of Figure \ref{fig:obsa_rpdist}). The contours show the distribution of single transiting planets from the Kepler population. Pixel sizes are log($a$) = 0.15 by log($R_p$) = 0.1.}
    	\label{fig:moda_rpdist_nothird}
    \end{figure*}
    
 Finally there appears to be an over abundance in the population of single transiting systems for planets with $R_\mathrm{p} \lesssim 2R_\oplus$ at periods $P < 10$ days ($\lesssim 0.03$au) (see \cite{2011ApJS..197....8L}; \cite{2012ApJ...758...39J}; \cite{2016PNAS..11312023S}; \cite{2016arXiv161009390L}). The formation processes which lead to these types of planets are unclear. It is also unknown if these objects are inherently rocky planets, or are the cores of Neptune sized planets whose envelopes have been irradiated (\cite{2015ApJ...807...45D}; \cite{2015ApJ...801...41R}; \cite{2016arXiv161009390L}). If these outlying systems are largely ignored, the question remains of whether the remaining planets in single transiting systems are part of the same underlying distribution of higher order planetary systems; i.e. could these single transiting systems contain similar planets which are not observed to transit? 

For our dynamical analysis it is not the radii of these planets which is of relevance, rather their masses. We estimate the masses of planets according to the following mass-radius relations. For radii less than 1.5$R_\oplus$ we use the rocky planet mass-radius relation from \cite{2014ApJ...783L...6W}, where density ($\rho_\mathrm{p}$) is related to radii ($R_\mathrm{p}$) through $\rho_\mathrm{p} = 2.43 + 3.39(R_\mathrm{p}/R_\oplus)$gcm$^{-3}$. For radii 1.5 $\leq$ $R_\mathrm{p}$ $\leq$ 4$R_\oplus$, we use the deterministic version of the probabilistic mass-radius relation for sub-Neptune objects from \cite{2016ApJ...825...19W}, where mass ($M_\mathrm{p}$) is given by $M_\mathrm{p}/M_\oplus = 2.7(R_\mathrm{p}/R_\oplus)^{1.3}$. Once radii become $R_\mathrm{p}\gtrsim$4$R_\oplus$ deterministic mass-radius relations become uncertain due to the onset of planetary contraction under self-gravity (see \cite{2017ApJ...834...17C}). From the mass-radius relations detailed in \cite{2017ApJ...834...17C}, we find their 'Neptunian worlds' deterministic relation of $M_{\mathrm{p}}/M_\oplus = (1.23R_\mathrm{p}/R_\oplus)^{1.7}$ gives the most sensible masses for all planets with $R_\mathrm{p} > 4R_\oplus$. 

\subsection{De-biasing the Kepler population}
\label{sec:debiasing}
 As previously alluded to, Kepler only observes planetary systems that have their orbital planes aligned with our line of sight. It is therefore sensible to suggest that there is a much larger, underlying population of planetary systems within which only some are observed to transit. We refer to this underlying population of planetary systems as the \textit{model population}. Conversely, we refer to the population of planetary systems actually observed by Kepler as the \textit{Kepler population}. We assume that Kepler systems are representative of planetary systems in the model population once geometrical biases have been taken into account. 
 
 To construct an underlying model population, our primary goal is for this to predict the correct number and planet parameter distribution seen in the Kepler population for systems with two transiting planets (Figure \ref{fig:obsa_rpdist} right). To achieve this we first assume that all stars either have two or zero planets. Any system which hosts two planets is assumed to be identical to one of the 341 double transiting systems observed by Kepler. We assume the abundance of a specific Kepler-like system in the model population is equal to the inverse of the mean of the double transit probability calculated by the method outlined in \S\ref{sec:semianal}. Systems with inherently low mean double transit probabilities, are therefore probabilistically assumed to be more numerous in the model population. By definition therefore, each unique system in the model population would be expected to be observed with both planets transiting exactly once and so the model population predicts the correct distribution shown in the right panel of Figure \ref{fig:obsa_rpdist}. We note that a model population generated in this way is similar to the method described in \cite{2012ApJ...758...39J}, albeit with their work predicting the correct number and planet parameter distribution seen in the Kepler population for systems with three transiting planets.    

The sum of the inversed mean double transit probabilities of all the 341 double transiting systems gives the total number of planetary systems in the model population. If we assume that all of the two planet systems are coplanar, we find the model population includes 16517 systems (the remaining 148449 systems observed by Kepler are assumed to have no planets). 

Each system in the model population can be observed to have a single transiting planet, depending on the viewing angle. The sum of the mean single transit probabilities for each of the 16517 systems in the coplanar model population gives the total number of single transiting planets, $N_\mathrm{sing}$, that would be expected to be observed. Here the mean single transit probability for a given system is equal to $R_\star/a_1 - R_\star/a_2$, where $a_1$, $a_2$ are the semi-major axes of each planet when $a_2 > a_1$ and $R_\star$ is the radius of the host star. We find $N_\mathrm{sing}$ = 589, which clearly underestimates the 1951 single transiting systems in the observed Kepler population, by a factor of $\sim3$. This is the Kepler dichotomy discussed in \S\ref{sec:intro}. We show the smoothed distribution of the semi-major axes and planet radii for these 589 predicted single transiting planets in the top left panel of Figure \ref{fig:moda_rpdist_nothird}, which when compared with the left panel of Figure \ref{fig:obsa_rpdist} clearly shows an under-prediction of the single transiting planets observed by Kepler.

\begin{figure}
	\centering
	\includegraphics[width=\linewidth]{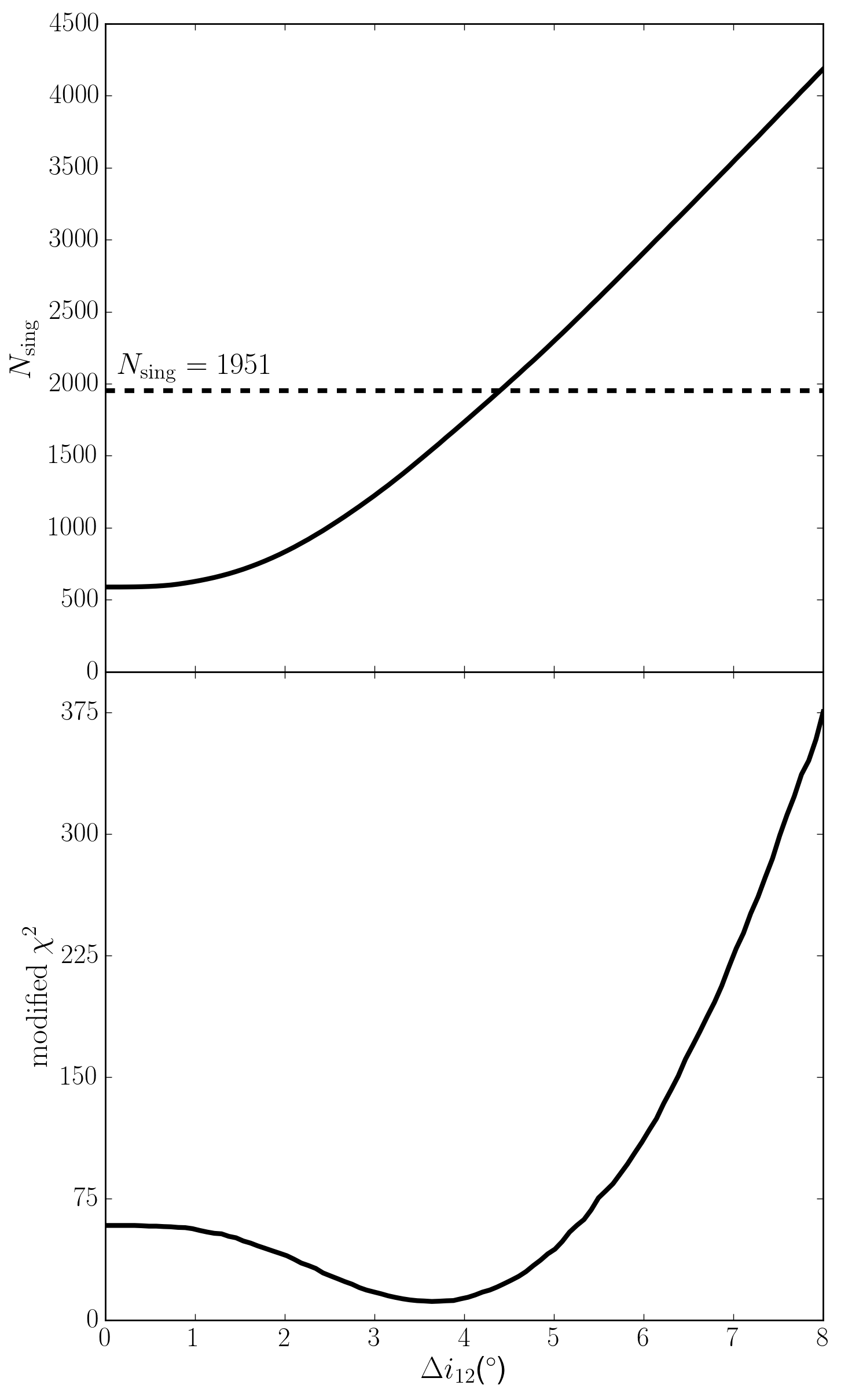}
	\caption{(\textit{top}) The expected number of single transiting planets observed from a model population generated from Kepler systems with two planets that are mutually inclined by $\Delta i_{12}$. The number of double transiting systems predicted by the model population is constant with 341 systems. (\textit{bottom}) The associated modified $\chi^2$ comparing types of single transiting planets predicted by the model population with the Kepler population. The minimum modified $\chi^2$ value corresponds to $\Delta i_{12} = 3.6^\circ$.}
	\label{fig:chis}
\end{figure}

\subsection{Inherently inclined multi-planet systems}
\label{subsec:inher}
From transit duration variation (TDV) studies, the mutual inclinations of planets in multi-transiting systems are small at $\lesssim2-3^\circ$ (\cite{2012ApJ...761...92F}; \cite{2014ApJ...790..146F}). We note that this mutual inclination also best fits the distribution of impact parameters in the Kepler population. Perhaps then, if two planets are assumed to be inherently mutually inclined by a small amount, this may account for the abundance of single transiting planets in the Kepler population. Consider a fixed mutual inclination $\Delta i_{12}$ between the two planets in each of the 341 double transiting systems. The mean single transit probability for each planet from a given system, $P_{\mathrm{sing,1}}$ and $P_{\mathrm{sing,2}}$ respectively where $P_{\mathrm{sing,1}} > P_{\mathrm{sing,2}}$, is now given by
\begin{equation}
\begin{split}
P_{\mathrm{sing,1}} = \frac{R_{\star}}{a_{1}} - P \\
P_{\mathrm{sing,2}} = \frac{R_{\star}}{a_{2}} - P,
\end{split}
\label{eq:Psing}
\end{equation}
where $P$ is the mean double transit probability and $P_\mathrm{sing,1} + P_\mathrm{sing,2}$ is the total mean single transit probability for this system. As $\Delta i_{12}$ increases, the mean double transit probability decreases (Figure \ref{fig:pi}). Therefore for a fixed population of double transiting systems considered here, the expected abundance of single transiting systems increases. Figure \ref{fig:chis} shows how $N_\mathrm{sing}$ increases with $\Delta i_{12}$ for when the number of double transiting systems is kept constant at 341 systems. If $\Delta i_{12} = 4.4^\circ$, we find $N_\mathrm{sing} = 1951$, i.e. the number of single transiting planets expected to be observed from the model population is equal to the number in the observed Kepler population. This suggests that mutual inclinations in Kepler systems observed with two planets must be less than 4.4$^\circ$, or the number of single planet systems observed by Kepler would be too large relative to the number of doubles.

 We show the distribution of the semi-major axes and radii of the expected single transiting planets for when $\Delta i = 4.4^\circ$ in the top right panel of Figure \ref{fig:moda_rpdist_nothird}. Comparing with the left panel of Figure \ref{fig:obsa_rpdist}, there is an over abundance of predicted single transiting planets with radii of $\sim2.5$R$_\oplus$ and semi-major axes of $\sim$0.15au. This is due to the model population compensating for not being able to reproduce all types of single transiting planets in the Kepler population (e.g. Hot Jupiters discussed in \S\ref{sec:sample}). Herein therefore when discussing how well a model population predicts the Kepler population of single transiting planets we refer to how well the \textit{types} of planets from each population compare, rather than the total number. That is, we look to find which value of $\Delta i_{12}$ causes the associated version of the top right panel of Figure \ref{fig:moda_rpdist_nothird} to be most like the left panel of Figure \ref{fig:obsa_rpdist}.

We judge the success of this comparison using a modified $\chi^2$ minimisation test, in which we simply sum the square of the difference between the number of singles with a given radius and semi-major axis expected from the model population, with that of the observed Kepler population. Varying $\Delta i_{12}$ we therefore look to identify a minimum in the modified $\chi^2$ space without caring for the modified $\chi^2$ value itself. We show this in Figure \ref{fig:chis}, with the modified $\chi^2$ minimum occurring for $\Delta i_{12} = 3.6^\circ$. The distribution of the single transiting planets expected from the model population for this mutual inclination is shown in the bottom left panel of Figure \ref{fig:moda_rpdist_nothird}. Comparing with the left panel of Figure \ref{fig:obsa_rpdist}, these single transiting planets share a stronger agreement with those in the Kepler population, compared with when the outer planet predicted $N_\mathrm{sing} = 1951$ (e.g. top right panel of Figure \ref{fig:moda_rpdist_nothird}). We note that the total number of single transiting planets expected from the model population for $\Delta i_{12} = 3.6^\circ$ is 1504. We assume therefore that the remaining 1951-1504 = 447 single transiting transiting planets in the Kepler population not fit by this model population are \textit{inherently single planet systems}. 

Despite the model population for $\Delta i_{12} = 3.6^\circ$ giving the lowest modified $\chi^2$ value, this mutual inclination is perhaps larger than that suggested by TDV studies. We note however that mutual inclination estimates from TDV studies consider a range of planet multiplicities. For example \cite{2012ApJ...761...92F} consider a model population of planetary systems with 1-7+ planets and predict that $\sim$50\% of observed planetary systems should contain a single planet, with the remaining systems containing multiple planets with mutual inclinations of $\lesssim3^\circ$. In order to properly predict the inherent mutual inclination in the multi-planet systems considered in this work therefore, it would be necessary to simultaneously model the TDV data directly. We consider such an analysis as part of future work. Instead in \S\ref{subsec:KOIwide} we consider the possibility that Kepler planets form coplanar, but end up mutually inclined due to perturbations from an outer planetary companion on an inclined orbit. This may provide another way to predict the correct abundance of single transiting systems observed by Kepler, and also result in a low mutual inclination for those systems with two transiting planets.

\subsection{Including an inclined planetary companion}
\label{subsec:KOIwide}
We now consider the effects of a hypothetical outer planet in each of the systems in the model population. We first amend the assumption from \S\ref{sec:debiasing} and assume that all stars either host three or zero planets. Any system which hosts three planets is assumed to be identical to one of the 341 double transiting systems from the Kepler population plus an additional outer planet. The outer planet is assumed to have the same mass and semi-major axis in all systems and starts on an inclination to the inner planets when these are coplanar, causing the mutual inclination between the inner planets to evolve according to eq. (\ref{eq:mutinc}). We assume that the outer planet satisfies the Hill stability criterion of $\Delta$ = 2$\sqrt{3}$ (\cite{1999MNRAS.304..793C}) with the outer of the inner two planets for all 341 considered systems, where $\Delta = (a_{3} - a_{2})/R_{H}$ and 
\[
R_{H}= \left(\frac{m_{2} + m_3}{3M_{\star}}\right)^{1/3}\left(\frac{a_{2} + a_3}{2}\right),
\]
where $M_{\star}$ is the stellar mass. If this criterion is not satisfied, we move the outer planet for this specific system until it is. For example, when the outer planet is assumed to have a semi-major axis and mass of 1au and 1$\mathrm{M}_\oplus$ respectively, we find 6 of the 341 systems do not satisfy this stability criterion and the outer planet needs to be moved to a mean semi-major axis of 1.2au. When the outer planet has a semi-major axis and mass of 1au and 10$\mathrm{M}_\mathrm{J}$ respectively we find 22 of the 341 systems do not satisfy the stability criterion and the outer planet needs to be moved to a mean semi-major axis of 1.4au.  

Each one of the 341 systems is again replicated enough times in the model population to be expected to be observed exactly once. That is, the inverse of the mean double transit probability of the inner two planets, gives the abundance of each of the 341 systems in the model population. The associated mean single transit probabilities for each of the inner two planets is of the same form as eq. (\ref{eq:Psing}). The sum of the mean single transit probabilities for every system in the model population therefore again gives the abundance of a given single transiting planet that would be expected to be observed from the model population that also fits the number of double transiting systems.

\begin{figure}
    	\centering
    	\includegraphics[trim={0cm 0.5cm 0cm 0cm}, width = 0.9\linewidth]{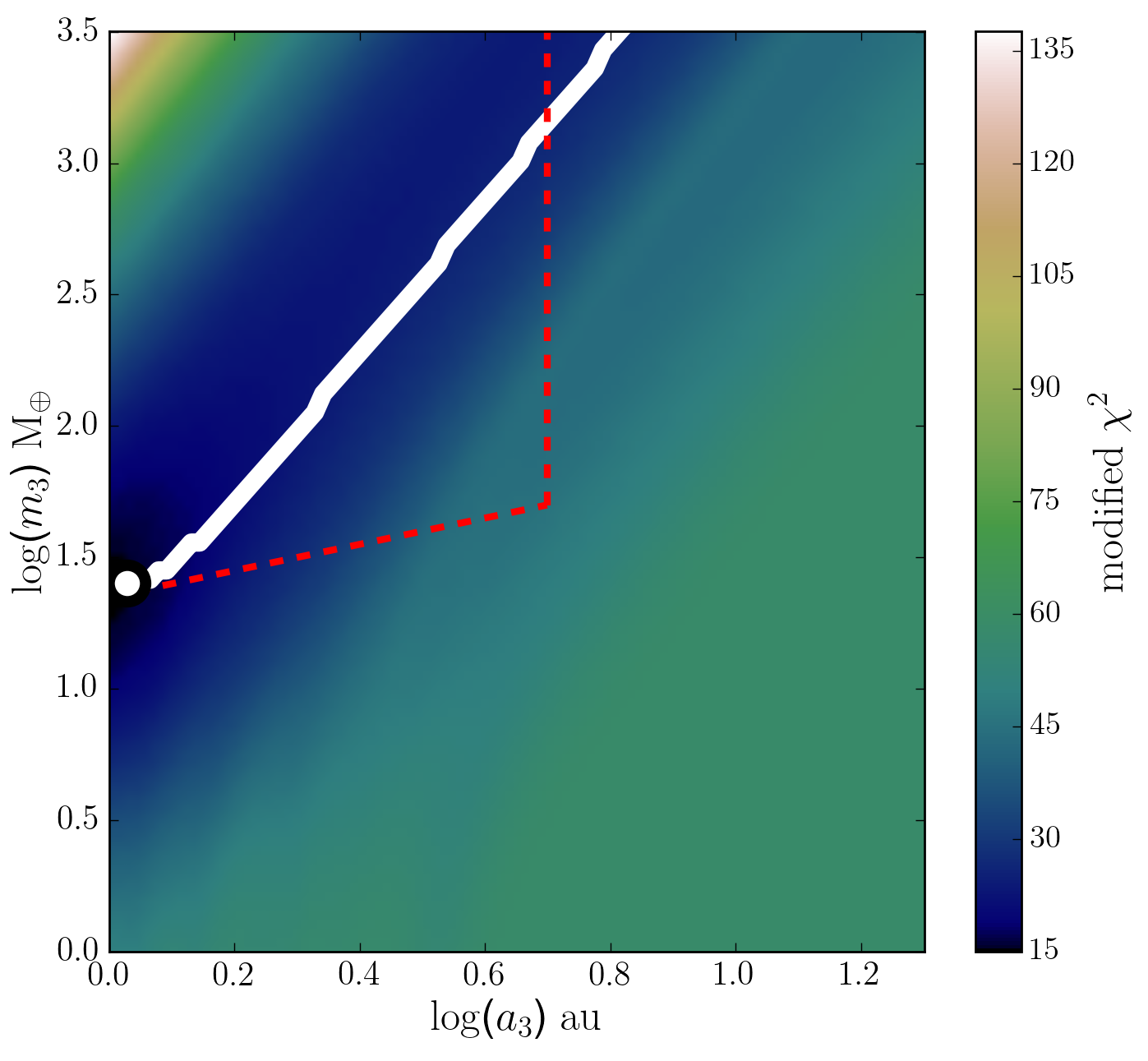}
    	\includegraphics[trim={0cm 0.5cm 0cm 0cm}, width = 0.9\linewidth]{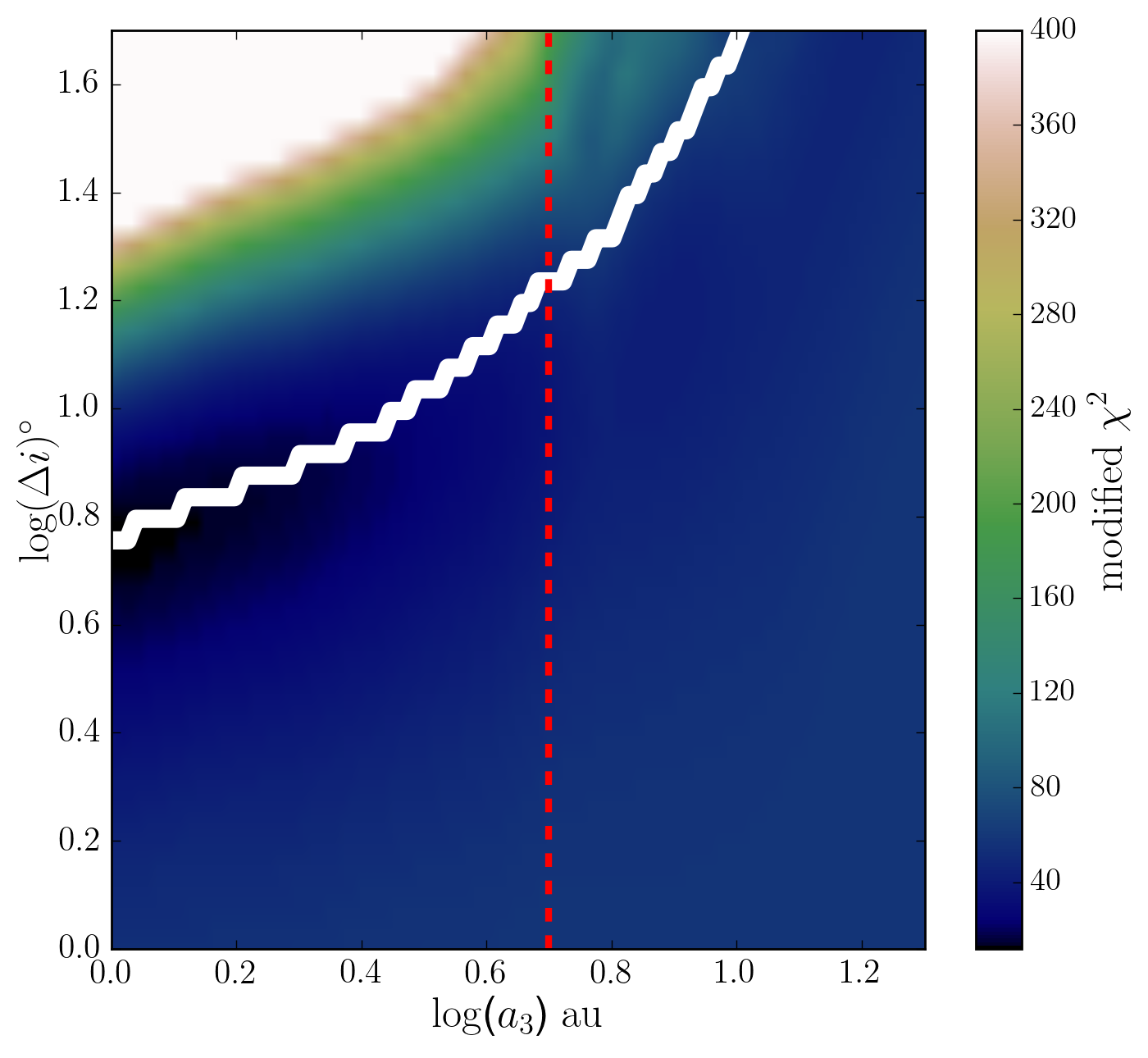}
    	\includegraphics[trim={0cm 0.8cm 0cm 0cm}, width = 0.9\linewidth]{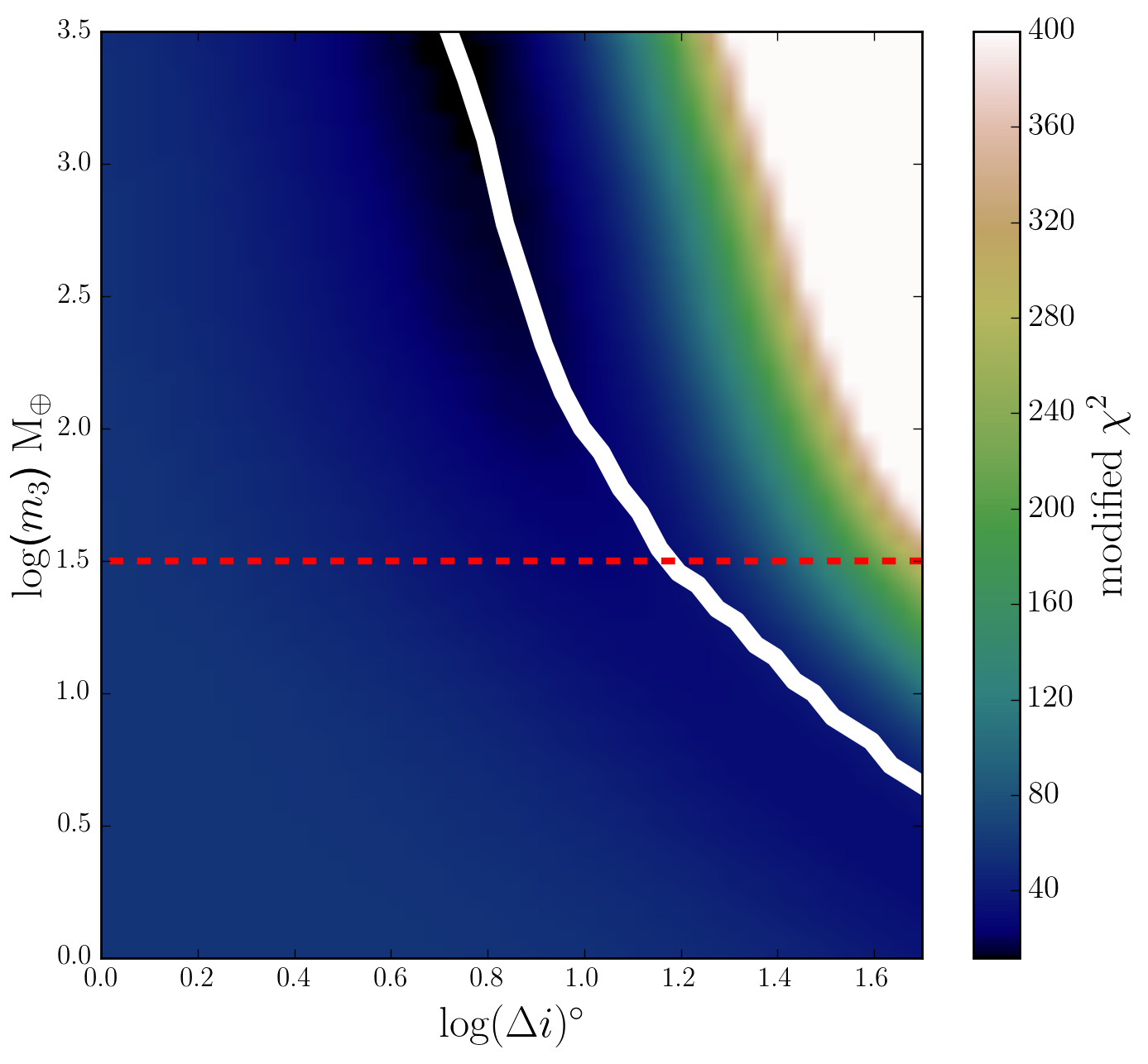}
    	\caption{Modified $\chi^2$ value comparing types of single transiting planets predicted by the model with Kepler population. For the top panel $\Delta i$ = 10$^\circ$, for the middle panel $m_3$ = 1$\mathrm{M_J}$ and for the bottom panel $a_3$ = 2au. Laplace-Lagrange theory is expected to break down for $\Delta i \gg 20^\circ$. The red dashed line refers to a rough RV detection threshold. The white line shows where the model population predicts $N_\mathrm{sing}=1564$. The white triangle and circle gives the third planet parameters used to produce the middle and bottom panels of Figure \ref{fig:moda_rpdist_nothird} respectively.}
    	\label{fig:chispace}
    \end{figure}

Similarly to the modelling approach in \S\ref{subsec:inher}, we look to identify which mass ($m_3$), semi-major axis ($a_3$) and initial inclination ($\Delta i$) of the outer planet causes the types of single transiting systems expected from the associated model population to be most like those in the observed Kepler population. For a given combination of $a_3$, $m_3$ and $\Delta i$ we therefore calculate a modified $\chi^2$ value described in \S\ref{subsec:inher}. We show these modified $\chi^2$ values in Figure \ref{fig:chispace} for an outer planet with $\Delta i$ = 10$^\circ$ (top panel), $m_3$ = 1M$_\mathrm{J}$ (middle panel) and $a_3$ = 2au (bottom panel). Inclinations of $\Delta i \gg 20^\circ$ where eq. (\ref{eq:mutinc}) is expected to break down are included for completeness.  

From the top panel in Figure \ref{fig:chispace}, it is clear that there is a 'valley' of semi-major axes and masses of the outer planet which causes a significantly lower modified $\chi^2$ value. It can be assumed therefore that such an additional planet predicts single transiting systems whose radii and semi-major axes better fit those in the Kepler population. However there is also a distinct minimum in the modified $\chi^2$ space when the outer planet has a semi-major axis of $\sim$1au for a mass of $\sim$30M$_\oplus$. Similarly in the other panels of Figure \ref{fig:chispace} there appear to be distinct minima. For the middle panel this occurs for an outer planet (of $m_3$ = 1M$_\mathrm{J}$) with a semi-major axis of 1.38au, initially inclined to the inner planets by $\Delta i = 5.7^\circ$. Finally for the bottom panel, this minimum occurs for a mass of $\sim$6M$_\mathrm{J}$ and inclination of 6$^\circ$ (where $a_3 = 2$au). Generally, we find the distribution of single transiting planets expected from the model population is more representative of those in the Kepler population for $3 \lesssim \Delta i \lesssim 10^\circ$.   

The bottom right panel of Figure \ref{fig:moda_rpdist_nothird} gives the distribution of single transiting planets expected from the model population when the outer planet exists in a minimum of the modified $\chi^2$ space with $a_3$ = 1.07au, $m_3$ = 24M$_\oplus$ and $\Delta i$=10$^\circ$ (white circle in the top panel of Figure \ref{fig:chispace}). We note that the total number of single transiting planets expected from this model population is $1564$. The outer planet parameters which predict $N_\mathrm{sing} = 1564$ are shown by the white lines in Figure \ref{fig:chispace}. This line highlights that while many outer planet parameters can predict $N_\mathrm{sing}=1564$, some predict single transiting planets which are more representative of those in the Kepler population. We note that $N_\mathrm{sing}$ predicted by the same range of outer planet parameters from Figure \ref{fig:chispace} is shown in Appendix \ref{sec:Ntot}. 
    
\section{Discussion} 
\label{sec:discussion}
 \subsection{Combining inherently mutually inclined and outer planet populations}
 \label{subsec:hybrid}
In reality it is likely that the total number of single planet transiting systems observed by Kepler ($N_\mathrm{sing, Kep} = 1951$) is contributed to by different populations of planetary systems. These may include a number of inherently single planet systems ($N_\mathrm{sing, inh}$) in addition to a number of single transiting planets observed from a population of two planet systems which have a fixed mutual inclination of $\Delta i_{12}$ ($N_\mathrm{sing, \Delta i_{12}}$). They may also include a number of single transiting planets which are observed from a population of initially coplanar two planet systems interacting with an inclined planetary companion ($N_\mathrm{sing, planet}$). Hence in general, it can be considered that 
 \begin{equation}
 N_\mathrm{sing, Kep} = N_\mathrm{sing, inh} + N_{\mathrm{sing},\Delta i_{12}} + N_\mathrm{sing, planet}.
 \label{eq:combNsing}
 \end{equation}  
 
 Here we make the assumption that the total number of double transiting systems observed by Kepler ($N_\mathrm{doub, Kep} = 341$) is made up of a fraction $f$ that are two planet systems with an inherent mutual inclination and a fraction (1-$f$) that are two planet systems with an inclined outer companion. We can thus rewrite eq. (\ref{eq:combNsing}) as 
 
 \begin{equation}
 \begin{split}
 N_\mathrm{sing, Kep} = N_\mathrm{sing, inh}  + f(N_\mathrm{sing, N doub=341})_{\Delta i_{12}} \\+ (1-f)(N_\mathrm{sing, N doub=341})_\mathrm{planet}, 
 \label{eq:combN}
 \end{split}
 \end{equation}  
 where ($N_\mathrm{sing,Ndoub=341}$)$_{\Delta i_{12}}$ is the number of singles that would have been produced from the population of two planet systems with a fixed mutual inclination of $\Delta i_{12}$, had it been numerous enough to reproduce the 341 double transiting Kepler systems (which is shown in Figure \ref{fig:chis} as a function of $\Delta i_{12}$). Conversely ($N_\mathrm{sing,Ndoub=341}$)$_\mathrm{planet}$ is the number of singles that would have been produced from the population of two planet systems which are perturbed by an outer companion, had it been numerous enough to reproduce the 341 double transiting systems. We estimate the number of inherently single planet systems to be $N_\mathrm{sing,inh} $ = 447 from \S\ref{subsec:inher}. We note that $N_\mathrm{sing,inh}$ will change for different values of $\Delta i_{12}$, however for simplicity we keep it constant at 447.

\begin{figure}
 	\centering
 	\includegraphics[trim={0cm 0cm 0cm 0cm}, width = \linewidth]{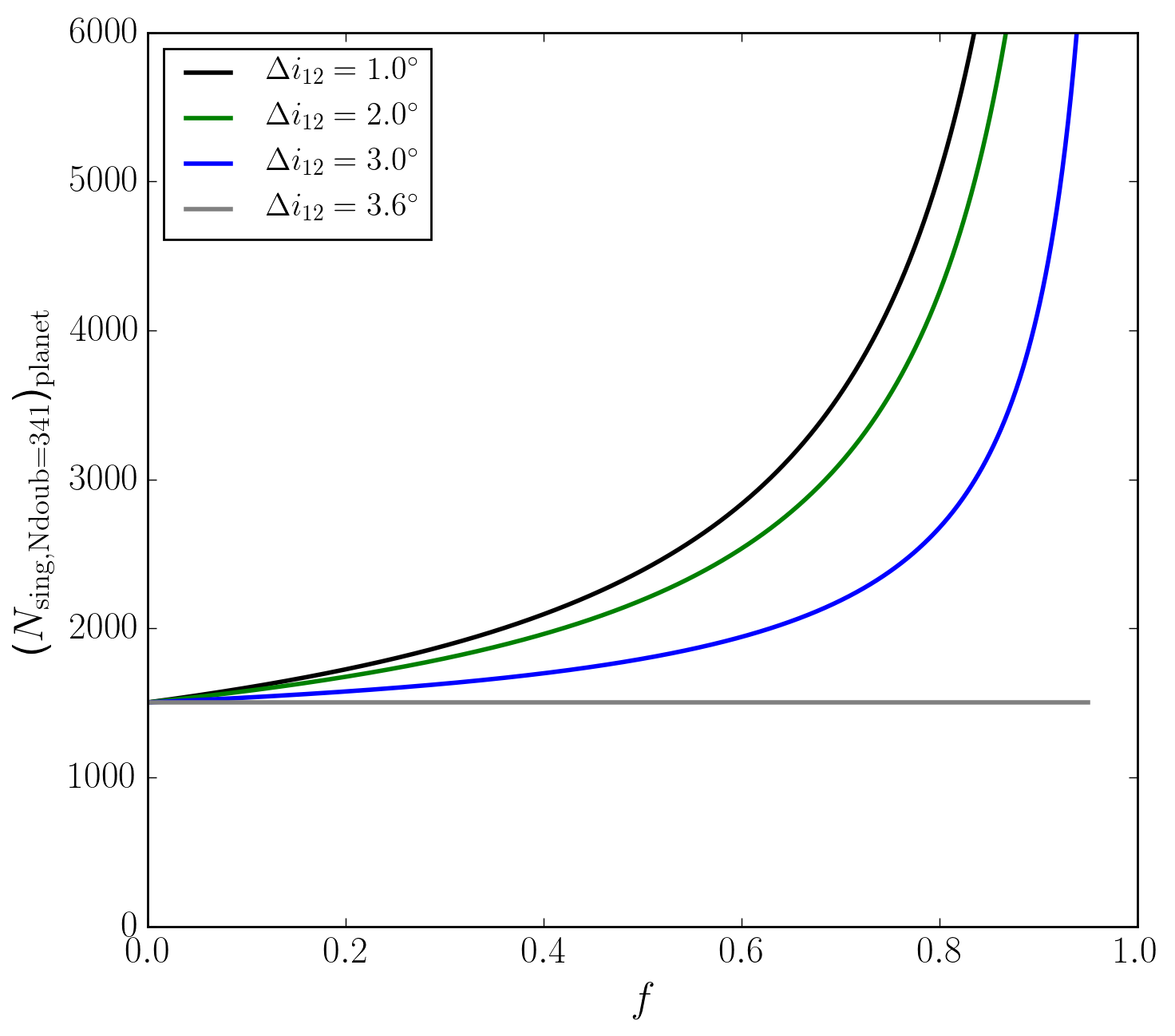}
 	\caption{The number of single transiting planets needed to be predicted by a population of two planet systems with an outer planetary companion, assuming that ($1-f$) of observed Kepler systems host such systems. The remaining fraction of observed Kepler systems are assumed to be two planet systems inherently mutually inclined by $\Delta i_{12}$.}
 	\label{fig:hybrid}
 \end{figure}

 For the assumed $N_\mathrm{sing,inh}$ and an assumed fixed mutual inclination for the fraction of the double transiting systems that are inherently inclined ($f$), eq. (\ref{eq:combN}) means that the number of single transiting systems observed by Kepler can be reproduced by specific combination with the fraction of double transiting systems that have an outer planet ($1-f$) and the properties of these planetary systems which determine the ratio of single to double transiting systems from this population (i.e. ($N_\mathrm{sing,Ndoub=341}$)$_\mathrm{planet}$). This combination is plotted in Figure \ref{fig:hybrid}, which can be read alongside Figure \ref{fig:Nsing} to determine the outer planet parameters required to reproduce the required ($N_\mathrm{sing, N doub=341}$)$_\mathrm{planet}$. For example, for $f=0.2$ and $\Delta i_{12} = 2^\circ$, ($N_\mathrm{sing, N doub=341}$)$_\mathrm{planet}$ = 1676 from Figure \ref{fig:hybrid}, which from Figure \ref{fig:Nsing} would be reproduced by an outer planet with $a_3 = 2$au, $m_3 = 132$M$_\oplus$ and $\Delta i$ = 10$^\circ$. For $f=0.5$, ($N_\mathrm{sing, N doub=341}$)$_\mathrm{planet}$ is increased to 2192 requiring the mass of this outer planet to be increased to $m_3 = 955$M$_\oplus$ (for $a_3 = 2$au and $\Delta i$ = 10$^\circ$). The outer planet parameters required to produce ($N_\mathrm{sing, N doub=341}$)$_\mathrm{planet}$ are therefore extremely sensitive to the value of $f$. However, increasing the value of $\Delta i_{12}$ for a given value of $f$ increases the value of ($N_\mathrm{sing,Ndoub=341}$)$_{\Delta i_{12}}$ and hence decreases ($N_\mathrm{sing, N doub=341}$)$_\mathrm{planet}$ as seen in Figure \ref{fig:hybrid}, requiring an outer planet which is a weaker perturber of the inner planets.

 It should be noted that $f$ and $1-f$ are not equivalent to the underlying fraction of stars that host a two planet system with a fixed mutual inclination, or a two planet system with an outer companion respectively. However if $f$ is known, such fractions for the underlying population of stars can be estimated through occurrence rate calculations. We discuss such calculations of occurrence rates in \S\ref{subsec:occurr}, however it is first necessary to estimate a value for $f$, which we discuss below.

\subsection{Comparing inherently mutually inclined and outer planet populations}
\label{subsec:disting}
From \S\ref{subsec:inher} a sole population of two planet systems which are inherently mutually inclined by $\Delta i = 3.6^\circ$ (i.e. when $f = 1$) can reproduce a population of single and double transiting systems representative of those observed by Kepler (Figure \ref{fig:moda_rpdist_nothird}). However from \S\ref{subsec:KOIwide} a sole population of two planet systems with an outer planet (i.e. $f = 0$) can also reproduce a population of single and double transiting systems representative of those observed by Kepler (Figure \ref{fig:chispace}). Here we look to differentiate between these two models by considering the predicted distribution of mutual inclinations that would be observed in the two planet populations for each model. We note that combining these two models in a way described in \S\ref{subsec:hybrid} (i.e. when $0<f<1$) would then give some intermediate distribution of mutual inclinations between the overall two planet population.     

\begin{figure}
     	\centering
     	\includegraphics[trim={0cm 0cm 0cm 0cm}, width = \linewidth, height = 7.43cm]{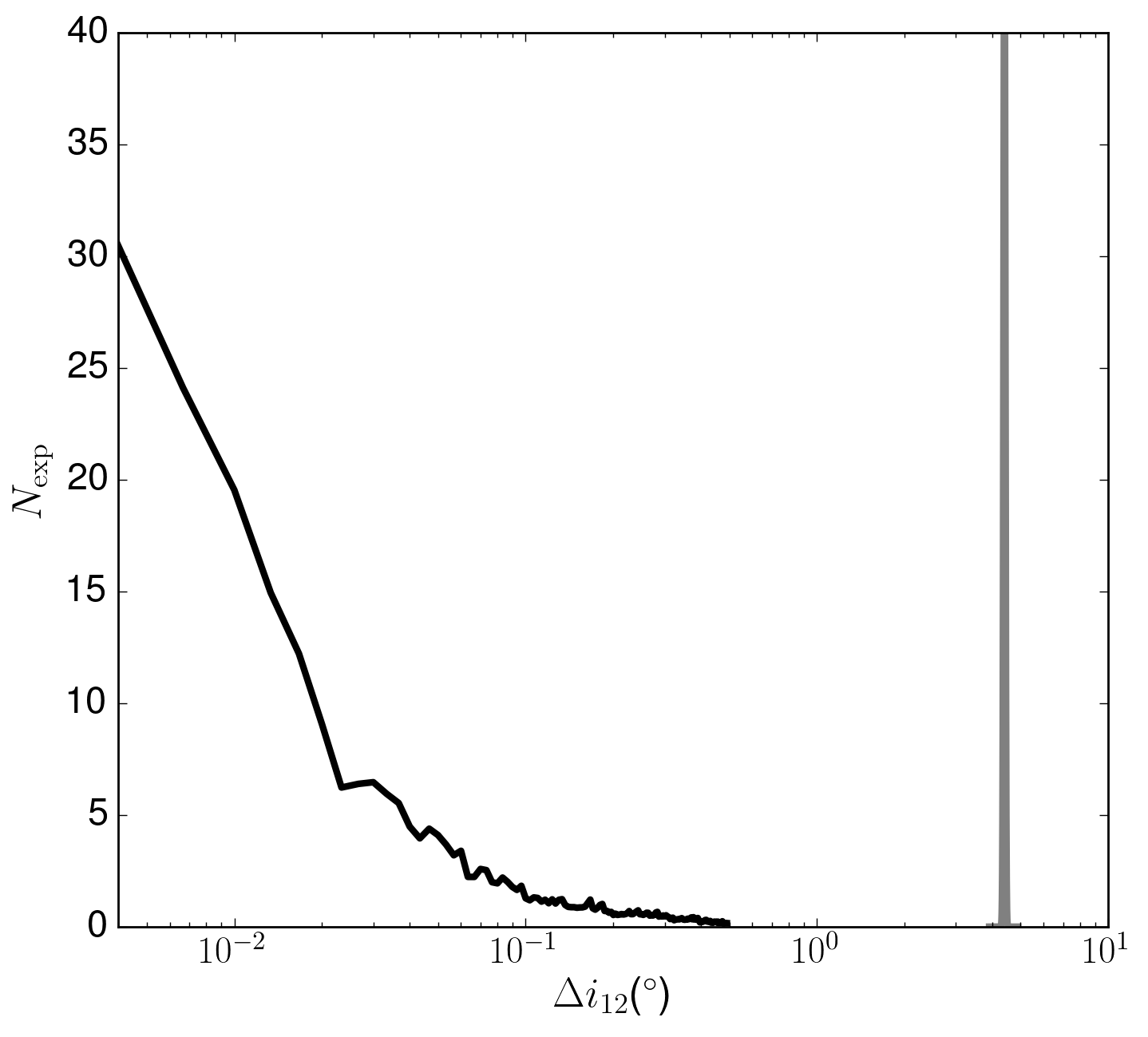}
     	\caption{Predicted distribution of mutual inclinations between the two planets in the observed Kepler double transit population for different model populations that both produce the correct number of double and single transiting systems. The grey line refers to the model where the two planet are inherently inclined by $\Delta i_{12}$=4.4$^\circ$. The black line refers to the model where two planets are secularly perturbed by a outer companion with $m_3$ = 1M$_\mathrm{J}$, $\Delta i = 10^\circ$ and $a_3$=1.9au.}
     	\label{fig:Nexp}
     \end{figure}

 For the model in which the two planets have an inherent mutual inclination of $\Delta i = 3.6^\circ$, that distribution is narrowly peaked at 3.6$^\circ$ (see Figure \ref{fig:Nexp}). In contrast, for the model in which two planets are perturbed by an inclined outer planet, the distribution of mutual inclinations is biased toward coplanar systems. This is because, while the outer planet induces a significant mutual inclination between the inner planets, as required to reproduce the correct ratio of single to double transiting systems, the inclination is not always large (see Figure \ref{fig:prob_time}) and the probability of witnessing a double transit system is much higher when their mutual inclination is low. Consider an outer companion with $m_3 = 24$M$_\oplus$, $a_3 = 1.07$au and $\Delta i = 10^\circ$, which was in a minimum of the modified $\chi^2$ space (white circle, Figure \ref{fig:chispace} top). Weighting the secularly evolving mutual inclinations between the inner two planets in the 341 considered systems by the associated double transit probability gives the predicted distribution of mutual inclinations which are most likely to be observed. This distribution is shown by the black line in Figure \ref{fig:Nexp}. It is clear that the most likely observed mutual inclination is when the inner two planets are coplanar. Moreover the number of systems expected to be observed with mutual inclinations beyond 0.5$^\circ$ drops to negligible values.  

From transit duration variation studies, the distribution of mutual inclinations between planets in multi-planet Kepler systems is peaked at $\sim 2^\circ$ (\cite{2012ApJ...761...92F}; \cite{2014ApJ...790..146F}), noting however that these works consider different planet populations to those considered here as discussed in \S\ref{subsec:inher}. Combining the two above models to produce a similar distribution in mutual inclinations may therefore allow for $f$ to be determined. We look to combine the two models in such a way to predict a value of $f$, as well as modelling the TDVs of the planetary systems considered in this work directly to predict the distribution of inherent mutual inclinations, as part of future work. For example if a fraction of two planet systems observed by Kepler are considered to have a fixed mutual inclination of $\Delta i_{12} = 4^\circ$, then in order to reproduce a distribution of mutual inclinations that peaks at $\sim$2$^\circ$ from modelling of TDVs, it might be expected that $f \sim 0.5$.  

An additional method to estimate $f$ might be to consider whether hypothetical outer planets considered in this work would have been detectable by other means. It is expected that RV studies would be most sensitive to such outer planetary companions. On Figure \ref{fig:chispace} we plot a rough constraint from RV studies, shown by the red dashed lines, assuming a detection threshold of $\sim$2m/s. Outside of 5au we assume RV studies are not sensitive to planets due to long periods. Planets above or to the left of these lines would therefore be detectable with this level of RV precision. This detection threshold suggests that a wide orbit planet located in the minima of the modified $\chi^2$ values in Figure \ref{fig:chispace} (white circle) should be just detectable by RV studies. This would assume however that all Kepler systems with two planets host this outer companion i.e. $f=0$. From Figure \ref{fig:hybrid} and highlighted in \S\ref{subsec:hybrid}, if $f> 0$ a planet with a larger mass, shorter period or larger inclination is required to reproduce the total number of single transiting systems observed by Kepler. Such outer planets should be readily detectable by RV surveys. For example, for the values of $f=0.2$ and $f=0.5$ for $\Delta i_{12} = 2^\circ$ considered in \S\ref{subsec:hybrid}, both of the outer planets in these cases would be expected to be detectable by RV surveys. Due to the inherent faintness of Kepler stars, few have been extensively studied for wide orbit planets. We suggest therefore that detailed follow-up RV studies of Kepler systems would allow for $f$ to be constrained. Generally for example, a low yield of outer planets in RV studies would suggest that $f$ is low and vice versa.

\subsection{Occurrence Rates}
\label{subsec:occurr}
Similar to that discussed specifically for Kepler systems in \S\ref{subsec:hybrid}, consider that the underlying population of planetary systems contains three possible types of planetary systems. These include inherently single planet systems, two planet systems which have a fixed mutual inclination of $\Delta i_{12}$ and two planet systems which are being perturbed by an inclined outer planet. In \S\ref{subsec:hybrid} it was shown that combining these systems with a free parameter $f$, which describes the fraction of the observed double transiting population that are two planet systems with a fixed mutual inclination, recovers the total number of single and double transiting systems observed by Kepler. 

However this value of $f$ is not the same as the fraction of the underlying population of stars that have two planets that are inherently mutually inclined. Here we define the occurrence rate of a given population to be the fraction of stars which would be expected to host such systems. Occurrence rates in this work can be estimated by taking the ratio of the number of systems in a given model population ($N_\mathrm{mod}$) to the total number of stars observed by Kepler ($N_\mathrm{Kep}$). The individual occurrence rates for the inherently single planet systems is therefore given by ($N_\mathrm{mod}/N_\mathrm{Kep}$)$_\mathrm{inh}$, for the two planet systems with the fixed mutual inclination of $\Delta i_{12}$ by ($N_\mathrm{mod}/N_\mathrm{Kep}$)$_{\Delta i_{12}}$ and for the two planet systems being perturbed by an inclined outer planet by ($N_\mathrm{mod}/N_\mathrm{Kep}$)$_\mathrm{planet}$. For example, for the population of two planet systems which were inherently mutually inclined by 3.6$^\circ$ (for when $f=1$), i.e. those which predicted a population of single transiting planets representative of those observed by Kepler (\S\ref{subsec:inher}), the number of systems in the model population was equal to 43807. From \S\ref{sec:sample} the total number of Kepler stars was 164966. Therefore the occurrence rate for this type of system, ($N_\mathrm{mod}/N{_\mathrm{Kep}}$)$_{\Delta i_{12}}$ = 27\%. Conversely, considering the population of two planet systems which were perturbed by an outer companion with $m_3=24$M$_\oplus$, $a_3=1.07$au and $\Delta i = 10^\circ$ (white circle Figure \ref{fig:chispace} top) for when $f=0$, predicted 42733 systems in the associated model population. Therefore the associated occurrence rate of this type of system ($N_\mathrm{mod}/N_\mathrm{Kep}$)$_\mathrm{planet}$ = 26\%.

The calculation of the occurrence rate for the population of inherently single planet systems is slightly different to that described above. From \S\ref{subsec:inher}, assume that there are 447 inherently single planet systems (noting that this is subject to the value of $\Delta i_{12}$). The distribution of the semi-major axes of these 447 planets is equal to the difference between the distributions of semi-major axes for the single transiting systems observed by Kepler and those predicted by the population of two planet systems with a fixed mutual inclination of $\Delta i_{12} = 3.6^\circ$, i.e. the difference between the left panel of Figure \ref{fig:obsa_rpdist} and the bottom left panel of Figure \ref{fig:moda_rpdist_nothird}. The number of inherently single planet systems in a model population is then the sum of the inverse of the single transit probabilities ($R_\star/a$) of all these 447 planets. We find this model population contains 15852 systems, predicting an occurrence rate of inherently single planet systems of 9.6\%. This is large compared with the occurrence rate of Hot Jupiters ($\sim1-2\%$ e.g. \cite{2005PThPS.158...24M}; \cite{2008PASP..120..531C}; \cite{2011arXiv1109.2497M}; \cite{2012ApJ...753..160W}; \cite{2016A&A...587A..64S}). We therefore expect that our population of inherently single planet systems is dominated by a different population, such as those described in \S\ref{sec:sample} which are poorly constrained.   

In a similar way to that described for eq. (\ref{eq:combN}), the total occurrence rate of assumed planetary systems in the underlying population of planetary systems can be estimated to be   
\begin{equation}
\left(\frac{N_\mathrm{mod}}{N_\mathrm{Kep}}\right)_\mathrm{tot} = \left(\frac{N_\mathrm{mod}}{N_\mathrm{Kep}}\right)_\mathrm{inh} + f\left(\frac{N_\mathrm{mod}}{N_\mathrm{Kep}}\right)_{\Delta i_{12}}  + (1-f)\left(\frac{N_\mathrm{mod}}{N_\mathrm{Kep}}\right)_\mathrm{planet}.
\end{equation}
 Consider the example combination of systems from \S\ref{subsec:disting} for when $f=0.2$, $\Delta i_{12} = 2^\circ$ and the outer planet parameters are $a_3 = 2$au, $m_3 = 132$M$_\oplus$ and $\Delta i = 10^\circ$. Here $f$($N_\mathrm{mod}/N_\mathrm{Kep}$)$_{\Delta i_{12}}$ $\sim$ 3\% and ($1-f$)($N_\mathrm{mod}/N_\mathrm{Kep}$)$_\mathrm{planet} \sim$21\%. We note that $f$($N_\mathrm{mod}/N_\mathrm{Kep}$)$_{\Delta i_{12}}$/($1-f$)($N_\mathrm{mod}/N_\mathrm{Kep}$)$_\mathrm{planet}$ = 3/21 = 14\%. This highlights that the occurrence rate of stars which have two planet systems with an inherent mutual inclination is similar to, but not the same as the parameter $f$. 
 
 Combining with the occurrence rate of inherently single planet systems estimated above, the total occurrence rate of planetary systems becomes 34\%. This is similar to occurrence rates of $\sim25\%-30\%$ for Kepler like planets derived from injection and recovery analysis of planet candidates from the Kepler pipeline (\cite{2013PNAS..11019273P}; \cite{2015ApJ...810...95C}).  

Estimates of occurrence rates for planets similar to the outer planets considered in this work exist from RV studies. \cite{2008PASP..120..531C} suggest an occurrence rate of 7.0 $\pm$ 1.4\% for planets with masses and semi-major axes of $m_\mathrm{p}$ = 1-10M$_\mathrm{J}$ and $\sim$1-5au respectively. Extrapolating this occurrence rate also predicts that 17-20\% of stars have gas giants within 20au. Similarly \cite{2011arXiv1109.2497M} suggest an occurrence rate of 13.9 $\pm$ 1.7\% for planets with masses and periods of $m_\mathrm{p} > 50$M$_\oplus$ and $P < 10$yrs respectively. More recently \cite{2016ApJ...821...89B} suggest that for systems with 1 or 2 RV planets, the occurrence rate of an additional companion with a mass and semi-major axis of 1-20M$_\mathrm{J}$ and 5-20au respectively is as high as 52 $\pm$ 5\%. The above example occurrence rate for the systems with an outer planet, i.e. ($1-f$)($N_\mathrm{mod}/N_\mathrm{Kep}$)$_\mathrm{planet} \sim$21\%, is then therefore not contradicted by these studies. However, this example assumed an estimated value of $f$. In addition to the methods described in \S\ref{subsec:disting}, observationally estimated occurrence rates for outer planets may also be able to constrain the value of $f$. For example if it is assumed that the occurrence rate of the types of outer planets considered in this work is 13.9\% (\cite{2011arXiv1109.2497M}), then it can be estimated that ($1-f$)($N_\mathrm{mod}/N_\mathrm{Kep}$)$_\mathrm{planet} \sim 13.9\%$. As ($N_\mathrm{mod}/N_\mathrm{Kep}$)$_\mathrm{planet} \ngtr$ 1 (i.e. it is unphysical that there are more stars in the model population than the number actually observed by Kepler), this results in an upper limit of $f\leq0.86$. We suggest therefore that combining this method of placing constraints on $f$ with those described in \S\ref{subsec:disting} might provide a strong constraint on the percentage of planetary systems which may share a fixed mutual inclination compared with systems that may host an outer inclined planet.

\subsection{Comparing with similar works}
Whether an outer planet can reduce the multiplicity of expected transiting planets in an inner planetary system in the context of N-body simulations has recently been investigated by \cite{2017MNRAS.tmp..186H}. A notable example they include is the effect of a companion with a mass of 1M$_\mathrm{J}$ at 1au, which is inclined to an inner population of planetary systems with a variety of multiplicities by 10$^\circ$. They find the ratio of the total number of double to single transiting systems that Kepler would be expected to observe is 0.184 (i.e. $\sim$5 times more expected single than double transiting systems). We find an identical outer planetary companion in our work gives this ratio to be 0.14. We suggest this difference is caused by the population of inner planetary systems used. \cite{2017MNRAS.tmp..186H} incorporate 50 model inner planetary systems with a range of multiplicities (the vast majority contained 3-6 planets at the end of their simulations), rather than the two planet Kepler systems considered in this work. Higher multiplicities increases the number of competing secular modes in the system, which can stabilise inner planets against the secular perturbations of an outer companion (e.g. \cite{2016MNRAS.457..465R}). Such an example was shown in this work in \S\ref{sec:realsys} for application to Kepler-48. Perhaps then, mutual inclinations are more easily induced between inner planets in this work, increasing the predicted number of single transiting planets that Kepler would be expected to observe, relative to a fixed population of planetary systems.    

Moreover compared with N-body simulations, our work does not allow for dynamical instability. If inclinations are large then they couple with eccentricity (\cite{1999ssd..book.....M}), potentially causing orbital crossings between neighbouring planets leading to dynamical instabilities on short, non-secular timescales. Indeed \cite{2017MNRAS.tmp..186H} find for the above mentioned outer planetary companion that roughly half of the 50 systems they consider lose at least one planet. Moreover \cite{2015ApJ...807...44P} suggest that the abundance of single and double transiting systems might be the remains of higher order planetary systems that were once tightly packed and have since undergone dynamical instability. A detailed discussion on how dynamical stability would be expected to affect our results is difficult. Our choice that all planets must be initially Hill stable is by no means a robust constraint on the long term stability of all the planetary systems we consider during the secular interaction. 
 
The effects of dynamical instability in tightly packed planet systems interacting with a wide orbit companion planet was also shown by \cite{2015ApJ...808...14M}. They find that an outer giant planet undergoing Kozai-Lidov interactions with a stellar binary (\cite{1962AJ.....67R.579K}; \cite{1962P&SS....9..719L}) can have an eccentricity which takes its orbit within the inner planets, leading to a significant reduction in planet multiplicity. Moreover more recent work in \cite{2016arXiv160908058M} suggests that these same interactions can cause $\sim$50\% of Kepler like systems to lose a planet, either through collisions or ejections. If inclination is not completely decoupled with eccentricity then, these works suggest that dynamical instability plays a significant role in sculpting an inner planetary system.  

\subsection{Metallicity Distribution}
The fraction of stars with gas giants increases with higher metal content (e.g. \cite{1996astro.ph..9148G}; \cite{2016ApJ...831...64T}). However it is unclear if this relation extends to smaller planets with $R_p \lesssim 4 R_\oplus$ (\cite{2011arXiv1109.2497M}; \cite{2016ApJ...832..196Z}). If single transiting planets are in systems which contain an outer giant companion similar to that considered in this work then the transiting planet should follow a similar metallicity relation as the giant planet. If there is an inherent population of single planet systems with $R_p \lesssim 4 R_\oplus$, in addition to a population of inherently mutually inclined double transiting systems, then these systems will follow a different metallicity relation. Therefore the population of single and double transiting systems observed by Kepler may contain a mixture of metallicity relations. If a distinction can be made between these different relations then this may place constraints on the presence of additional planets in Kepler systems with a single transiting planet.  

\subsection{Assumptions of this work}
\label{subsec:assump}
Throughout this work we have considered mutual inclinations evolve between two planets due to secular interactions with an outer planet. As stated above, increasing the multiplicity of planetary systems complicates the evolution of mutual inclinations. For application to the Kepler dichotomy, including higher multiplicity systems may cause proportionally fewer to be observed as single transiting systems. We look to investigate this as part of future work. Moreover higher multiplicity systems also allow for investigation into whether the presence of an outer planetary companion can explain the number of higher order systems observed by Kepler. This is of particular interest as \cite{2012ApJ...758...39J} find that generating a model population which predicts the number of systems observed by Kepler with three transiting planets (with small inherent mutual inclinations and no outer companion) cannot simultaneously predict the number of systems with a single and two transiting planets observed by Kepler.

We have also assumed that the inner transiting planets interacting with an outer companion were initially coplanar. However these transiting planets would most likely also have a small inherent mutual inclination (e.g. \cite{2012ApJ...761...92F}; \cite{2014ApJ...790..146F}) which in turn may affect the mean double transit probability. 

\section{Summary and Conclusions}
\label{sec:conc}
In summary, during the first part of this work we developed a semi-analytical method for the calculation of transit probabilities by considering the area a transiting planet subtends on a celestial sphere (\S\ref{sec:semianal}). Applying this method to a general two planet system, we showed how the probability that both planets are observed to transit changes as they become mutually inclined. 
 
In \S\ref{sec:secint} we discussed how the mutual inclination between two initially coplanar planets evolves due to secular interactions with an external mutually inclined planetary companion. We derived the full solution describing this evolution assuming that the mutual inclination remains small, before simplifying it under the assumption that the external planet was on a wide orbit. We found that the maximum mutual inclination between the inner two planets is approximately equal to twice the initial mutual inclination with the external planet. Below this the maximum mutual inclination between the inner two planets scales according to the mass, semi-major axis and inclination of the external planet by $\propto \Delta im_3/a_3^3$.  

How the secular interaction causes the double transit probability of the inner two planets to evolve was shown in \S\ref{sec:combtrans}. Assuming that this double transit probability is significantly reduced when the maximum mutual inclination exceeds $\approx (R_\star/a_1) + (R_\star/a_2)$ we derived an expression for the mean of the double transit probability considering a given external planetary companion. This expression was applied to Kepler-56, Kepler-68, and Kepler-48 to place constraints on the inclination of the outer RV detected planets in these systems in \S\ref{sec:realsys}. We found that the inner two transiting planets in Kepler-56 and Kepler-68 are not significantly secularly perturbed by the outer planets, regardless of their inclination. For HD 106315 we find that an outer planet inferred from recent RV analysis can cause a significant perturbation to the mutual inclination of two internal transiting planets. Moreover we find that if the outer planet is present within $\sim$1au, its inclination must be no more than $2.4^\circ$, otherwise the probability of observing both the inner planets to transit is significantly reduced. We also found that the RV detected planet in Kepler-48 needs to be inclined with respect to the inner planets by $\lesssim3.7^\circ$, otherwise the probability that all the inner planets are observed to transit is significantly reduced. We conclude therefore that using the expression for the mean transit probability between inner planets from eq. (\ref{eq:meanpsimp}) and (\ref{eq:Ksimp}) can be used to place significant constraints on the inclinations of RV detected planets, whose host systems also contain transiting planets. 

We further applied our method of calculating transit probabilities to the Kepler population in \S\ref{sec:Kepdich}. We found that relative to a fixed population of transiting systems with two planets on initially coplanar orbits, the expected number of single transiting systems can be significantly increased both by inherently inclining the two planets and by introducing an outer planetary companion. We found that an inherent mutual inclination of $\Delta i_{12} = 3.6^\circ$ predicts a population of single transiting planets most representative of those in the Kepler population. Moreover, we found that outer planets initially inclined by $\sim3-10^\circ$ to the inner planets also predict a representative population of single transiting systems. These outer planets should be detectable by RV studies. 

However it is likely that planetary systems observed by Kepler may include a combination of systems which include inherently single planet systems, two planet systems which have some fixed mutual inclination and two planet systems interacting with an inclined outer planet. For two planet systems which are perturbed by an outer planet, the distribution of the mutual inclinations between the inner planets of such systems is biased toward coplanar systems. This is due to an increased probability of observing inner planets when coplanar compared with when mutual inclinations are larger. We suggest that combining populations of inherently mutually inclined two planet systems with two planet systems which are interacting with an outer planet may be able to reproduce the observed distribution of mutual inclinations between Kepler planets. In doing so, this may provide constraints on the presence of outer planets in the Kepler population. We suggest also that detailed follow-up of RV studies in Kepler systems will provide a more direct constraints on the presence of outer planets. There should also be a dichotomy in the number of transiting systems observed by the upcoming \textit{TESS} mission (\cite{2014SPIE.9143E..20R}), however for these systems astrometry and RV techniques will be able to be used to verify the presence and influence of outer planets. 

\section*{Acknowledgements}
We thank Simon Gibbons and Grant Kennedy for useful conversations regarding this work. MJR acknowledges support of an STFC studentship and MCW acknowledges the support from the European Union through grant number 279973. We also thank the reviewer for comments which were a great help in improving this paper. This research has also made use of the NASA Exoplanet Archive, which is operated by the California Institute of Technology, under contract with the National Aeronautics and Space Administration under the Exoplanet Exploration Program.

\appendix
 \section{Further discussion of Transit equations}
 \label{sec:transeq}
 \subsection{Central Transit Line}
 The centre of the transit region is defined by eq. (\ref{eq:cent})
 \[ - \sin\Delta i\sin\theta\sin\phi + \cos\Delta i\cos\theta = 0,\]
 Assuming that $\phi = 0 \rightarrow 2\pi$ and that a corresponding value of $\theta$ for each value of $\phi$ can be in the range of $0 < \theta < \pi$, eq. (\ref{eq:cent}) can be rearranged to give
  \begin{equation}
  \begin{split}
  \theta = \arctan\left(\frac{1}{\tan\Delta i\sin\phi}\right) \hspace{2cm}&\quad\text{for  }\phi<\pi,\\
  \theta = \pi + \arctan\left(\frac{1}{\tan\Delta i\sin\phi}\right) \hspace{2cm}&\quad\text{for   }\phi > \pi.
  \end{split}
  \end{equation}
  
  \subsection{Upper Transit Boundary}
  The upper boundary of a transit region is given by eq. (\ref{eq:bound2})
  \[-\sin\Delta i\sin\theta_{2}\sin\phi_2 + \cos\Delta i\cos\theta_{2} = -\chi,\] 
   A value of $\theta_2$ for a given $\phi_2$ can be calculated through solving a quadratic of the form
   \begin{equation}
   \left(A^2 + B^2\right)x^2 + 2A\chi x + \chi^2 - B^2 = 0
   \label{eq:upquad}
   \end{equation}
   where 
   \[x = \sin\theta_2,\hspace{1cm} A = -\sin\Delta i\sin\phi_2,\hspace{1cm} B = \cos\Delta i.\]
   Depending on the value of $\Delta i$, the calculation of $\theta_2$ for a given value of $\phi_2$ can be grouped into three different regimes, (1) $\Delta i$ is small enough that the upper boundary of the transit region never crosses the fixed reference plane. (2) $\Delta i$ is large enough that the upper boundary of the transit region does cross the fixed reference plane. (3) For high values of $\Delta i$ the upper transit boundary only has values of $\theta_2$ for $0 < \phi_2 < \pi$. This can be thought of as the transit region going over the pole of the celestial sphere. 
   
   For regime (1), the value of  $\theta_2$ for a given $\phi_2$ is equivalent to that obtained from the positive root of eq. (\ref{eq:upquad}), mirrored about $\pi/2$. The transition to regime (2) occurs for when the upper transit boundary first crosses the fixed reference plane. Here $\Delta i = \arcsin(\chi)$. As $\Delta i$ is increased beyond this value the intersection between the upper transit boundary and the fixed reference plane occurs at $\phi_2 = \phi_0$ and $\phi_2 = \pi - \phi_0$, for which $\theta_2 = \pi/2$. From eq. (\ref{eq:bound2}) $\phi_0$ is given by $\phi_0 = \arcsin(\chi/\sin\Delta i)$. Therefore $\theta_2 < \pi/2$ for $\phi_0 < \phi_2 < \pi-\phi_0$ and $\theta_2 > \pi/2$ otherwise. When $\phi_0 < \phi_2 < \pi-\phi_0$, $\theta_2$ is hence obtained from the positive solution of eq. (\ref{eq:upquad}) and by the positive solution mirrored about $\pi/2$ otherwise.

   Finally the transition to regime (3) occurs when $\Delta i = \arccos(\chi)$. Similarly to regime (2) as $\Delta i$ is increased beyond this value, the upper transit boundary crosses the fixed reference plane at $\phi_2 = \phi_0$ and $\phi_2 = \pi - \phi_0$ and hence $\theta_2$ is only defined for when $\phi_0 < \phi_2 < \pi-\phi_0$. The solution from eq. (\ref{eq:upquad}) which gives the smaller value $\theta_2$ corresponds to $\theta_2 > \pi/2$ values and needs to be mirrored about $\pi/2$, with the solution giving the larger value of $\theta_2$ corresponding to $\theta_2 < \pi/2$ values.    
   
   To summarize consider that for a given value of $\phi_2$, eq. (\ref{eq:upquad}) gives two solutions for $\theta_2$, denoted as $\theta_2^{*1}$ and $\theta_2^{*2}$ respectively. For $\Delta i < \arcsin(\chi)$,
   \begin{equation}
   \theta_2 = \frac{\pi}{2} + \left(\frac{\pi}{2} - \theta_2^{*1}\right) \hspace{2.45cm}\quad\text{for  }0 < \phi_2 < 2\pi,
   \end{equation}
   where $\theta_2^{*1} > 0$ and $\theta_2^{*2} < 0$. \\
   
   \noindent For $\arcsin(\chi) < \Delta i < \arccos(\chi)$, 
   \begin{equation}
   \begin{split}
   &\theta_2 = \theta_2^{*1} \hspace{3.3cm}\quad\text{for  }\phi_0<\phi_2<\pi -\phi_0,\\
   &\theta_2 = \frac{\pi}{2} + \left(\frac{\pi}{2} - \theta_2^{*1}\right) \hspace{3.2cm}\quad\text{otherwise},
   \end{split}
   \end{equation}
   where $\theta_2^{*1} > 0$, $\theta_2^{*2} < 0$ and $\phi_0 = \arcsin(\chi/\sin\Delta i)$.
      
   \noindent For $\Delta i > \arccos(\chi)$, 
    \begin{equation}
    \begin{split}
    &\theta_2 = \text{max}\left(\theta_2^{*1}, \theta_2^{*2}\right) \hspace{2.3cm}\\
    &\text{and}\hspace{3.8cm}\quad\text{for  }\phi_0<\phi_2<\pi -\phi_0\\
    &\theta_2 = \frac{\pi}{2} + \left(\frac{\pi}{2} - \text{min}\left(\theta_2^{*1},\theta_2^{*2}\right)\right),
    \end{split}
    \end{equation}
   where $\theta_2^{*1} > 0$, $\theta_2^{*2} > 0$.
	
	\subsection{Lower Transit Boundary}
	The lower boundary of the transit region is given by eq. (\ref{eq:bound1})
	\[-\sin\Delta i\sin\theta_{1}\sin\phi_1 + \cos\Delta i\cos\theta_{1} = \chi,\]
    Depending on the value of $\Delta i$, the calculation $\theta_1$ for a given $\phi_1$ can be grouped into the same regimes as described for the upper transit boundary. However now in regime (1), $\theta_1 < \pi/2$ for $0 < \phi_1 < 2\pi$, in regime (2) the lower transit boundary crosses the fixed reference plane at $\phi_1 = \pi + \phi_0$ and $\phi_1 = 2\pi - \phi_0$ and in regime (3) $\theta_1$ is only defined for $\pi + \phi_0 < \phi_1 < 2\pi - \phi_0$. Assuming that $\theta_1^{*1}$ and $\theta_1^{*2}$ are the solutions for $\theta_1$ for a given $\phi_1$ in the modified form of eq. (\ref{eq:upquad}), then following the same discussion as for the upper transit boundary it can be shown that for $\Delta i < \arcsin(\chi)$,
	 \begin{equation}
	 \theta_1 = \theta_1^{*1} \hspace{4cm}\quad\text{for  }0 < \phi_1 < 2\pi,
	 \end{equation}
	 where $\theta_1^{*1} > 0$ and $\theta_1^{*2} < 0$. \\
   
    \noindent For $\arcsin(\chi) < \Delta i < \arccos(\chi)$, 
    \begin{equation}
    \begin{split}
    &\theta_1 = \frac{\pi}{2} + \left(\frac{\pi}{2} - \theta_1^{*1}\right)\hspace{1.cm}\quad\text{for  }\phi_0 + \pi <\phi_1<2\pi -\phi_0,\\
    &\theta_1 = \theta_1^{*1} \hspace{4.7cm}\quad\text{otherwise},
    \end{split}
    \end{equation}
    where $\theta_1^{*1} > 0$, $\theta_1^{*2} < 0$ and $\phi_0 = \arcsin(\chi/\sin\Delta i)$.
    
    \noindent For $\Delta i > \arccos(\chi)$, 
    \begin{equation}
    \begin{split}
    &\theta_1 = \text{min}\left(\theta_1^{*1},\theta_1^{*2}\right) \hspace{2.3cm}\\
    &\text{and}\hspace{3.2cm}\quad\text{for  }\phi_0+\pi<\phi_1<2\pi -\phi_0\\
    &\theta_1 = \frac{\pi}{2} + \left(\frac{\pi}{2} - \text{max}\left(\theta_1^{*1},\theta_1^{*2}\right)\right),
    \end{split}
    \end{equation}
    where $\theta_1^{*1} > 0$, $\theta_1^{*2} > 0$.

 \section{Secular Solution for Mutual Inclination Evolution}
 \label{sec:fullsec}
 From eq. (\ref{eq:secsol}) the evolution of complex inclinations according to Laplace-Lagrange theory is given by
 \begin{equation}
 y_j(t) = \sum^{N}_{k=1}\mathbf{\mathit{I}}_{jk}e^{i(f_kt + \gamma_k)},
 \end{equation}
 where \textit{I}$_{jk}$ are the eigenvectors of the matrix \textbf{B} from eq. (\ref{eq:matrix}) scaled to initial boundary conditions, $f_i$ are the eigenfrequencies of \textbf{B} and $\gamma_k$ are initial phase terms. If it is assumed that all the planets and the star are point masses and that the invariable plane is taken as a reference plane, it follows that $f_3 = 0$ and \textit{I}$_{j3} = 0$. From the initial conditions |$y_1(0)$| = |$y_2(0)$| = $i_1$. Hence the complex inclinations of the inner two planets respectively are given by 
   \begin{equation}
   \begin{split}
   y_1(t) = I_{11}\exp\left(i\left(f_1t + \pi\right)\right) +  I_{12}\exp\left(i\left(f_2t\right)\right),
   \label{eq:y_1fin}
   \end{split}
   \end{equation}
   \begin{equation}
   \begin{split}
   y_2(t) = I_{21}\exp\left(i\left(f_1t + \pi\right)\right) +  I_{22}\exp\left(i\left(f_2t\right)\right).
   \label{eq:y_2fin}
   \end{split}
   \end{equation}
   Also from the initial conditions $-I_{11} + I_{12} = i_1$ and $-I_{21} + I_{22} = i_1$. The complex mutual inclination between the inner two planets is equivalent to 
   \begin{equation}
   y_1(t) - y_2(t) = \left(I_{12} - I_{22}\right)\left[\exp(i(f_1t + \pi)) + \exp(if_2t)\right].
   \end{equation} 
     
\noindent Solving eq. (\ref{eq:secsol}), we propose a set of variables to represent the full solution of $I_{12}$ and $I_{22}$,
 
 \begin{equation}
 \begin{split}
 & K_{1m} = \frac{B_{13}B_{32}}{f_m + B_{31} + B_{32}},\\
 & K_{2m} = \frac{B_{13}B_{31}}{f_m + B_{31} + B_{32}},\\
 & K_{3m} = f_m + B_{12} + B_{13},\\
 & K_{4m} = f_m + B_{31} + B_{32},
 \end{split}
 \end{equation}
 where $m$ = 1, 2,
 \begin{equation}
 \begin{split}
 & R_{1(3-m)} = \frac{K_{3m} - K_{2m}}{B_{12} + K_{1m}},\\
 & R_{2(3-m)} = B_{31} + B_{32}R_{1(3-m)},
 \end{split}
 \end{equation}
 \begin{equation}
 \epsilon = R_{11} + \frac{R_{21}}{K_{42}}\left(R_{12} - 1\right) + \frac{R_{22}}{K_{41}}\left(1 - R_{11}\right) - R_{12}.
 \end{equation}
 Hence the components of the eigenvector associated with the $f_2$ eigenfrequency are given by
  \begin{equation}
  \begin{split}
  & I_{12} = \frac{1}{\epsilon}\left[\Delta i(1 - R_{12})\right],\\
  & I_{22} = \frac{R_{11}}{\epsilon}\left[\Delta i(1 - R_{12})\right].
  \end{split}
  \label{eq:fullsol}
  \end{equation}
 The non zero $f_1$ and $f_2$ eigenfrequencies of the matrix \textbf{B} from eq. (\ref{eq:matrix}) can be obtained by solving a quadratic of the form
 \begin{equation}
 \begin{split}
 & f^2 + f(B_{12} + B_{12} + B_{21} + B_{23} + B_{31} + B_{32}) + \\
 & [B_{12}\left(B_{23} + B_{31} + B_{32}\right) + B_{13}\left(B_{21} + B_{23} + B_{32}\right) \\
 & + B_{21}\left(B_{31} + B_{32}\right) + B_{23}B_{31}] = 0.
 \end{split}
 \end{equation}
  We note that the solution given by eq. (\ref{eq:fullsol}) recovers exactly what is predicted when solving eq. (\ref{eq:secsol}) by numerical methods. The full solution which describes how the mutual inclination between the inner two planets according to Laplace - Lagrange theory is therefore given by
 \begin{equation}
 y_1 - y_2 = \frac{\Delta i(1-R_{12})(1-R_{11})}{\epsilon}\left[e^{i(f_1t + \pi)} + e^{if_2t}\right],
 \end{equation}
 with the variable $K$ used in \S\ref{subsec:twoplansec} being equivalent to $(1 - R_{12})(1 - R_{11})/\epsilon$.

 \section{Reproducing the total number of single transiting planets observed by Kepler}
 \label{sec:Ntot}
\begin{figure}
\centering
\includegraphics[trim={0cm 0cm 0cm 0cm}, width = 0.95\linewidth]{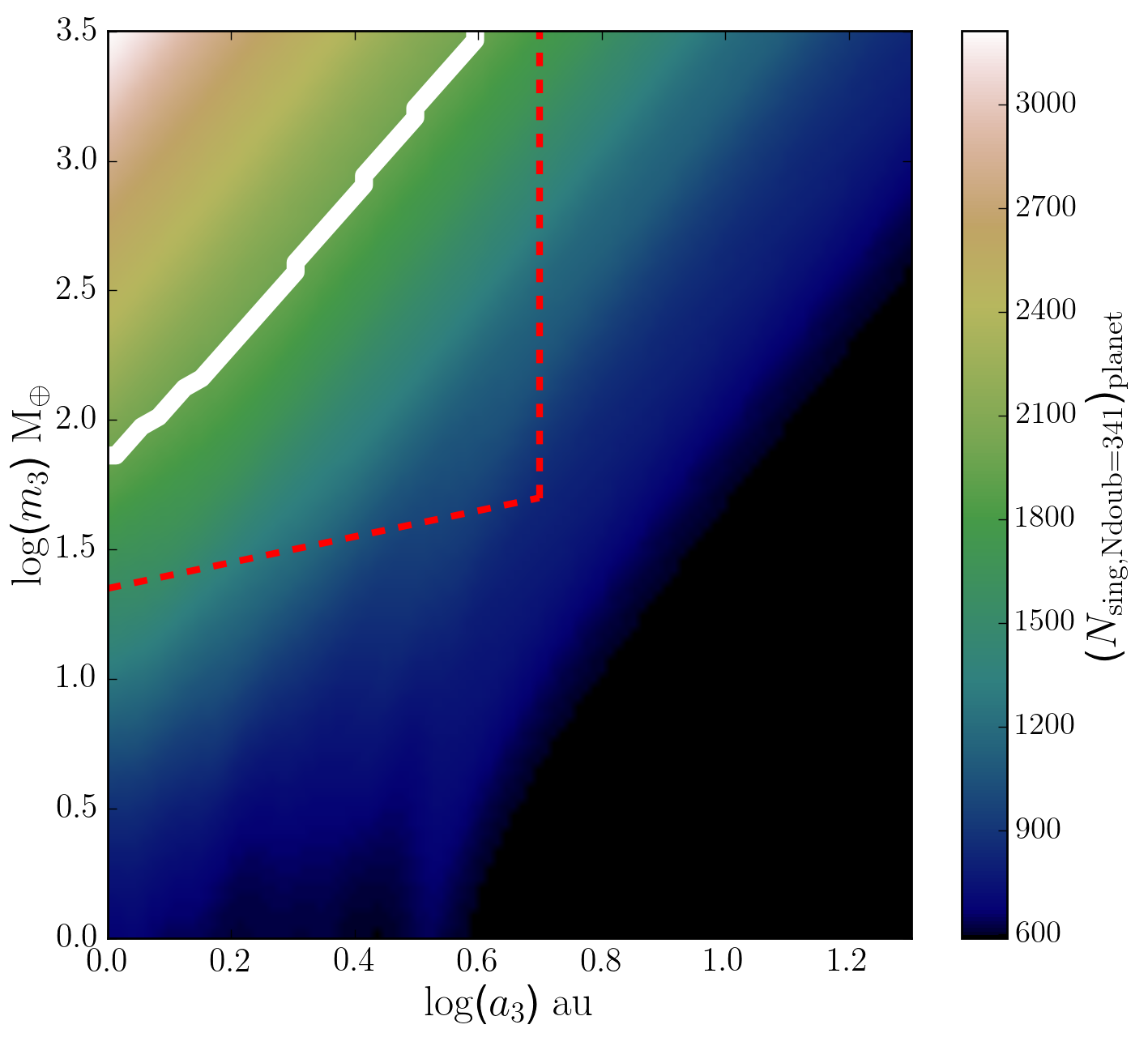}
\includegraphics[trim={0cm 0cm 0cm 0cm}, width = 0.95\linewidth]{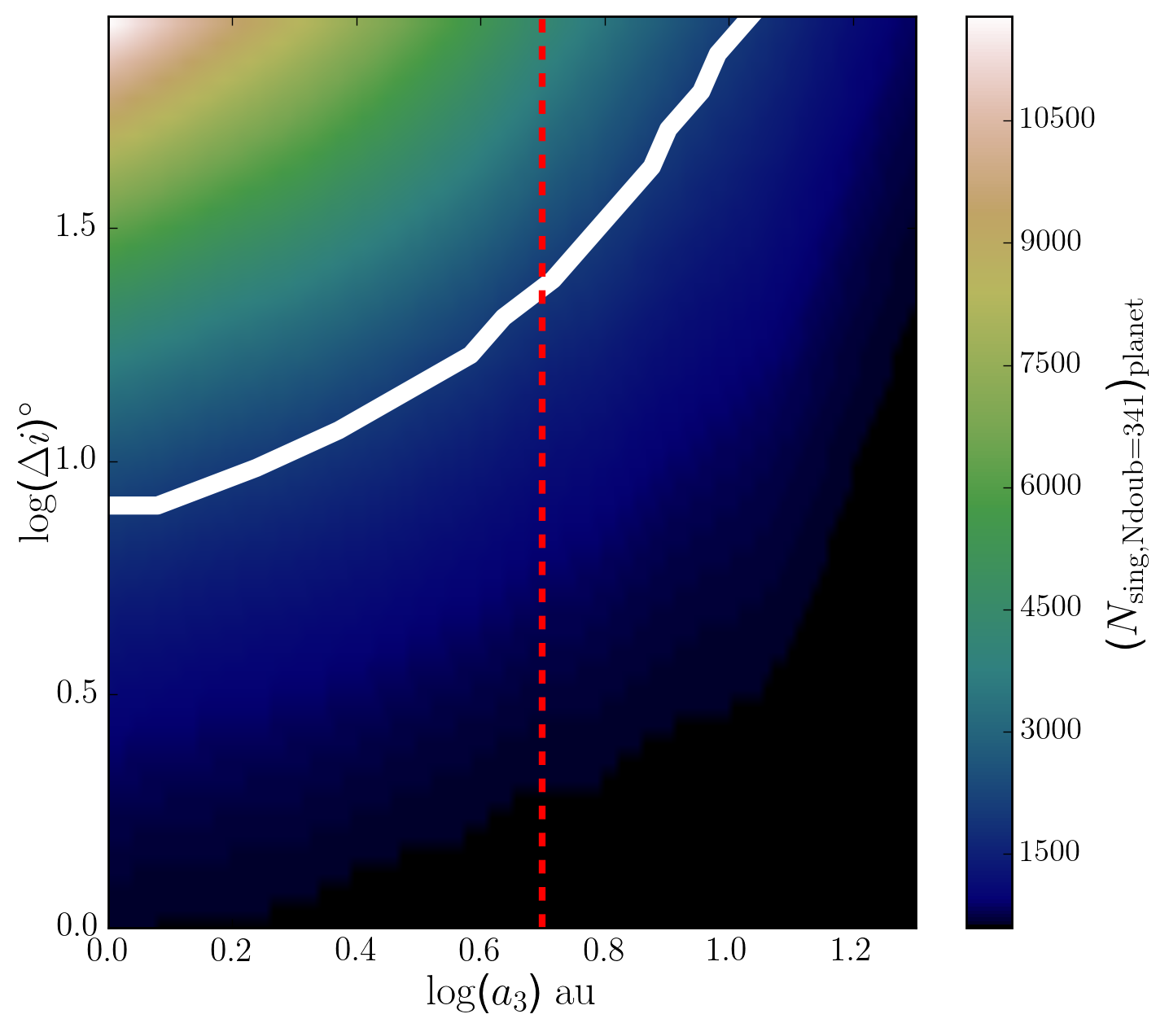}
\includegraphics[trim={0cm 0cm 0cm 0cm}, width = 0.95\linewidth]{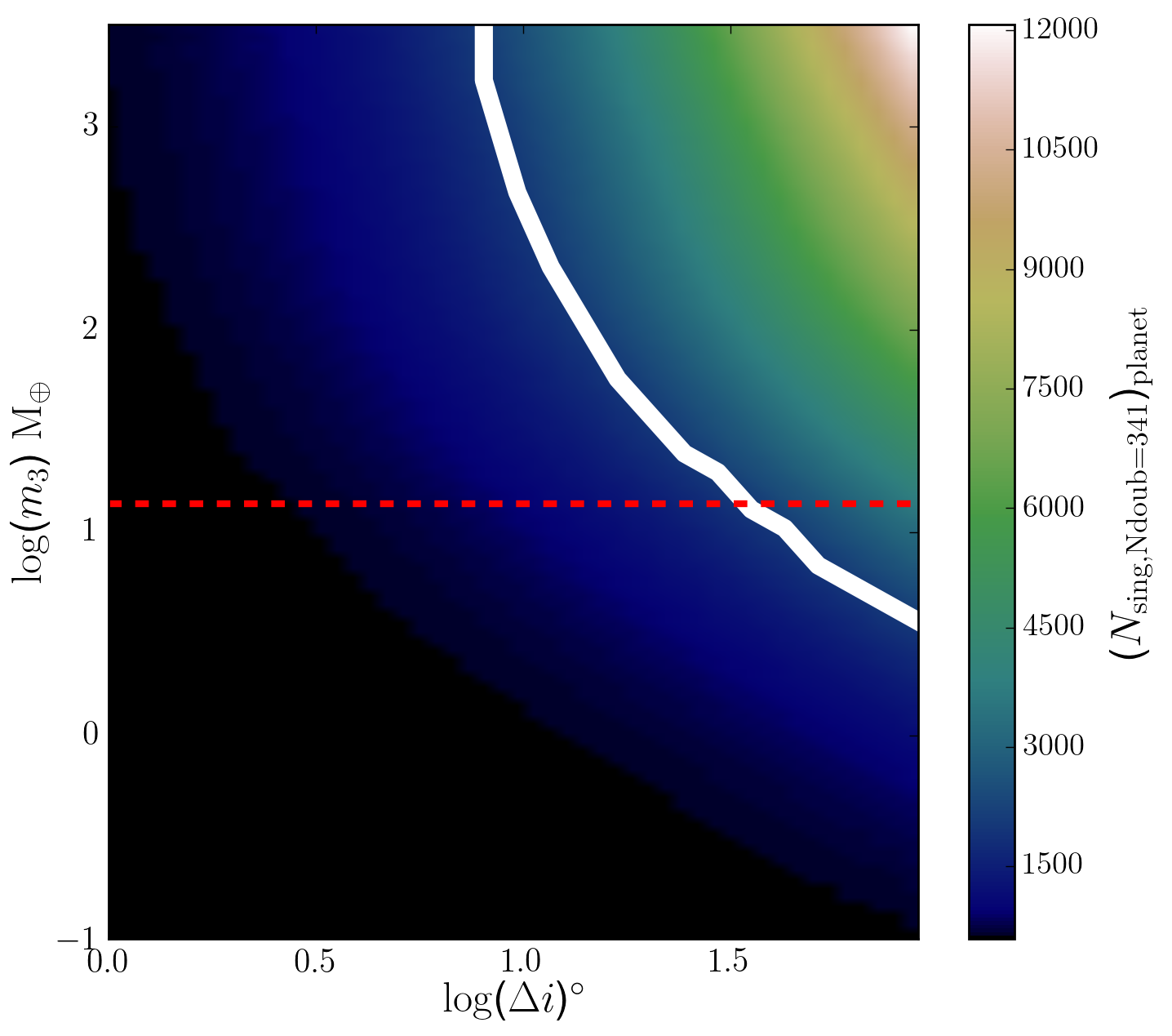}
\caption{The total number of single transiting planets Kepler would be expected to observe for given third planet parameters. The white line corresponds to the total number of single transiting systems currently observed by Kepler (1951). The red lines give an estimate for the detection threshold of RV surveys.}
\label{fig:Nsing}
\end{figure}

 In \S\ref{sec:Kepdich} we considered Kepler systems with two transiting planets which are secularly interacting with an outer planet on an inclined orbit. We found that the number of single transiting systems Kepler would be expected to observe can be dramatically increased as a result of this interaction. Figure \ref{fig:Nsing} shows the total number of single transiting objects Kepler would be expected to observe from the method outlined in \S\ref{subsec:KOIwide}, for when the outer planet has the same parameters as the respective panels of Figure \ref{fig:chispace}. Again for $\Delta i \gg 20^\circ$, Laplace-Lagrange theory is expected to break down and is included for completeness. The white line gives where the total number of single transiting planets Kepler would be expected to observe from the model population is equal to the number in the Kepler population i.e. 1951. The red dashed lines give an estimate for an RV detection threshold.
 
\bibliographystyle{mnras}
\bibliography{example} 








\bsp	
\label{lastpage}

\end{document}